\documentclass[twocolumn,superscriptaddress,nofootinbib,aps,floatfix,preprintnumbers,amsmath,amssymb]{revtex4-1}

\usepackage{slashed}
\usepackage{latexsym}
\usepackage{graphics}
\usepackage{bm}
\usepackage{graphicx} 
\usepackage{color}
\usepackage{amsmath}
\usepackage{amssymb}
\usepackage{feynmp-auto}
\usepackage{slashed}
\usepackage[dvipsnames]{xcolor}
\usepackage{xcolor}
\graphicspath{{figs/}}
\usepackage{enumitem}

\usepackage{amsfonts, amssymb,amsmath,latexsym,graphics, graphicx,epsfig,multirow,comment,
,feyn,slashed,xcolor,afterpage, makecell} 
\usepackage{booktabs, blindtext}
\usepackage{tabularx,afterpage}
\usepackage{enumerate}
\newcommand{\beq}{\begin{equation}}
\newcommand{\eeq}{\end{equation}}
\newcommand{\bea}{\begin{eqnarray}}
\newcommand{\eea}{\end{eqnarray}}

\newcommand{\sigmap}{\bar{\sigma}_n}
\newcommand{\Fmolsq}{\ifmmode {\lvert F_{\rm mol}(q) \rvert}^2 \else ${\lvert F_{\rm mol}(q) \rvert}^2$\fi}
\newcommand{\FmolvJsq}{\ifmmode {\lvert F_{{\rm mol}, v' J'}(q) \rvert^2} \else ${\lvert F_{{\rm mol}, v' J'}(q) \rvert^2}$\fi}
\newcommand{\FmolvJsqAvg}{\ifmmode {\langle\lvert F_{{\rm mol}, v' J'}(q,T) \rvert^2\rangle} \else ${\langle\lvert F_{{\rm mol}, v' J'}(q,T) \rvert^2\rangle}$\fi}
\newcommand{\FDM}{F_{\rm DM}(q)}
\newcommand{\FDMsq}{|F_{\rm DM}(q)|^{2}}
\newcommand{\DEvJ}{\Delta E_{v^\prime J^\prime}}
\newcommand{\ngemit}{n_{\gamma,{\rm em}}}
\newcommand{\ngcol}{n_{\gamma,{\rm col}}}

\begin{document}
 
\title{Direct Detection of Spin-(In)dependent Nuclear Scattering \\ of Sub-GeV Dark Matter Using Molecular Excitations} 

\author{Rouven Essig}
\affiliation{C. N. Yang Institute for Theoretical Physics, Stony Brook University, USA	}
\author{Jes\'{u}s P\'{e}rez-R\'{i}os}
\affiliation{	Fritz-Haber-Institut der Max-Planck-Gesellschaft, Faradayweg 4-6, D-14195 Berlin, Germany}
\author{Harikrishnan Ramani}
\affiliation{Berkeley Center for Theoretical Physics, Department of Physics,
University of California, Berkeley, CA 94720}
\affiliation{Ernest Orlando Lawrence Berkeley National Laboratory,
University of California, Berkeley, CA 94720, USA}
\author{Oren Slone}
\affiliation{Princeton Center for Theoretical Science, Princeton University, Princeton, NJ 08544, USA}

\date{\today}

\begin{abstract}
We propose a novel direct detection concept to search for dark matter with 100~keV to 100~MeV masses.  Such dark matter can 
scatter off molecules in a gas and transfer an $\mathcal{O}(1)$ fraction of its kinetic energy to excite a vibrational and rotational state.  The excited ro-vibrational mode relaxes rapidly and produces a spectacular multi-infrared-photon signal, which can be observed with 
ultrasensitive photodetectors.  
We discuss in detail a gas target consisting of carbon monoxide molecules, which enable efficient photon emission even at a relatively low temperature and high vapor pressure. The emitted photons have an energy in the range 180~meV to 265~meV.  
By mixing together carbon monoxide molecules of different isotopes, including those with an odd number of neutrons, we obtain sensitivity to both spin-independent interactions and spin-dependent interactions with the neutron.  We also consider hydrogen fluoride, hydrogen bromide, and scandium hydride molecules, which each provide sensitivity to spin-dependent interactions with the proton.  
The proposed detection concept can be realized with near-term technology and allows for the exploration of orders of magnitude of new 
dark matter parameter space. 
\end{abstract}

\maketitle
\tableofcontents

\section{Introduction}
\label{sec:Intro}

The evidence for the existence of dark matter (DM), which makes up about 85\% of the matter density in the Universe, is overwhelming. 
However, all the evidence is based on the gravitational interactions between the DM and ordinary matter, and we are yet to detect it in the laboratory.  
Efforts to directly detect galactic DM particles in the laboratory are crucial for developing a more 
detailed understanding of the particle nature of DM.  

The past few decades have seen tremendous progress in direct-detection searches for Weakly Interacting Massive Particles (WIMPs), which have masses above $\sim$1~GeV.  
Direct-detection experiments are typically optimized to detect WIMPs scattering elastically off nuclei, in which case the resulting nuclear recoil creates a combination of phonons, light, and/or charge, depending on the type of target material.  
Recently, however, increased attention has been given to the search for DM particles with masses a few orders 
of magnitude below the proton mass (``sub-GeV DM''), which has emerged as an important new experimental frontier, see e.g.~\cite{Essig:2011nj,Essig:2013lka,Alexander:2016aln,Battaglieri:2017aum,BRNreport}. The challenge of detecting sub-GeV DM is that very little energy and momentum is transferred from the DM to the nucleus in an elastic scattering process, and the nuclear recoil energy quickly falls below detector thresholds as the DM mass is lowered below the mass of the nucleus.  However, as emphasized in~\cite{Essig:2011nj}, inelastic processes can allow for a much larger fraction of energy to be transferred.

\begin{figure*}[t]
\centering
\includegraphics[width=2\columnwidth,keepaspectratio]{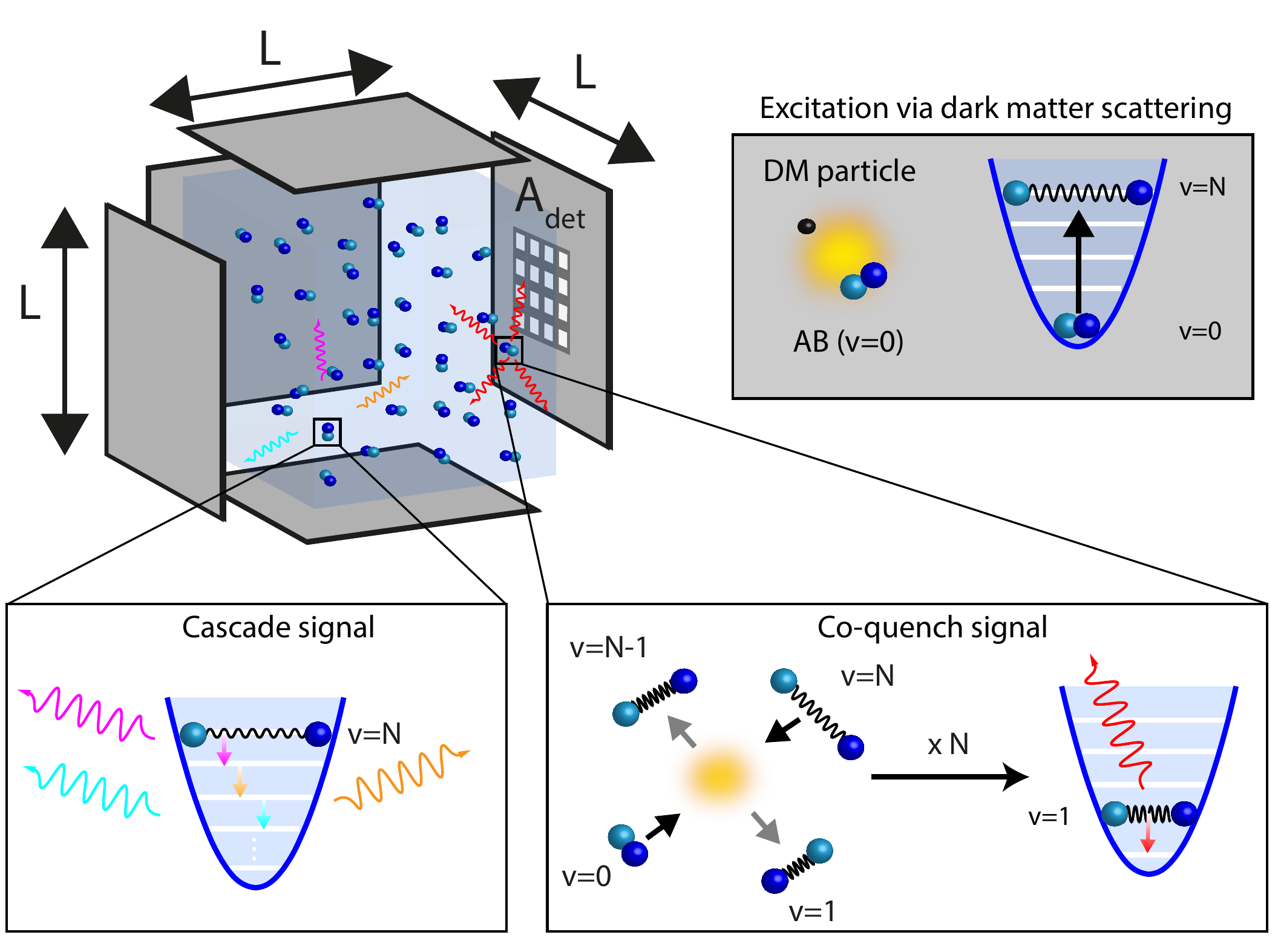}
\caption{\textbf{Top left:} Schematic showing the proposed experimental setup for detecting dark matter through ro-vibrational excitations of molecules in a gas.  A reservoir of length $L$ with reflective walls contains a gas of cold diatomic hetero-nuclear molecules (denoted AB) held at low pressure to avoid clustering.  The photodetector has surface area $A_{\rm det}$ and is shown for simplicity to be attached to one of the reservoir walls; in practice, to allow for the photodetector operating at a different temperature than the gas, the photodetector may need to be either insulated from the reservoir or the light must be transported to the photodetector by, for examples, fibers. 
\textbf{Top right:}  DM scatters off a molecule and excites a ro-vibrational mode. The excited ro-vibrational modes are short lived and relax rapidly, producing two types of infrared photons as the signal: (i) \textbf{Cascade} photons (\textbf{bottom left}), where the excited molecule cascades to a lower-lying vibrational mode (or to the ground state) emitting a single photon for every vibrational transition, which each have a large mean free path; and (ii) \textbf{Co-quench} photons (\textbf{bottom right}), where the excited molecule is resonantly quenched by scattering off and exciting neighboring molecules to their first vibrational mode, which each decay to produce a photon.  The mean free path of the co-quench photons can be enhanced by adding a buffer gas consisting of, e.g., helium, but only those produced close to the photodetector area will be measurable.}
\label{fig:Setups}
\end{figure*}

DM can be probed down to the MeV scale and below by searching for DM scattering off electrons.  This typically excites the electron to a higher energy level, and allows for the transfer of a sizable fraction of the DM's available kinetic energy.  Various target materials have been considered, including atoms~\cite{Essig:2011nj,Essig:2012yx,Essig:2017kqs}, semiconductors~\cite{Essig:2011nj,Graham:2012su,Lee:2015qva,Essig:2015cda}, scintillators~\cite{Derenzo:2016fse}, two-dimensional targets~\cite{Hochberg:2016ntt}, and superconductors~\cite{Hochberg:2015fth,Hochberg:2015pha,Hochberg:2019cyy}.  
These materials are also sensitive to the absorption of ultralight bosonic DM ($<$keV) by 
electrons~\cite{An:2013yua,An:2014twa,Bloch:2016sjj,Hochberg:2016sqx,Hochberg:2019cyy}.  
The most stringent direct-detection constraints on electron recoils from sub-GeV DM have been presented in~\cite{Essig:2012yx,Essig:2017kqs,Agnes:2018oej,Tiffenberg:2017aac,Abramoff:2019dfb,Crisler:2018gci,Romani:2017iwi,Agnese:2018col,An:2017ojc,Emken:2017hnp}, and small experiments are already being developed to probe DM down to MeV-scales~\cite{BRNreport}. 

In order to develop a full understanding of the particle nature of DM, it is essential to also probe other DM interactions besides those with electrons, such as spin-independent and spin-dependent interactions with nuclei.  Existing direct-detection constraints on DM below the GeV-scale are weak for spin-independent interactions, and are even weaker for spin-dependent interactions with the proton or neutron; only sub-GeV DM particles that interact very strongly with ordinary matter are constrained~\cite{Petricca:2017zdp,Abdelhameed:2019hmk,Abdelhameed:2019szb,Agnese:2017jvy,Amole:2017dex,Behnke:2016lsk,Fu:2016ega,Akerib:2017kat,Aprile:2019dbj,Collar:2018ydf,Bringmann:2018cvk,Gluscevic:2017ywp,Xu:2018efh,Slatyer:2018aqg,Nadler:2019zrb,Armengaud:2019kfj}. Several pathways exist in the near term that will enable significant improvement in searches for spin-independent nuclear interactions down to DM masses of $\sim$50--100~MeV~\cite{Katsioulas:2018squ,Hertel:2018aal,Kouvaris:2016afs,McCabe:2017rln,Ibe:2017yqa,Dolan:2017xbu,Bell:2019egg}; however, while there are ideas to probe spin-independent interactions for even lower DM masses~\cite{Essig:2011nj,Essig:2016crl,Schutz:2016tid,Knapen:2016cue,Budnik:2017sbu,Bunting:2017net,Rajendran:2017ynw,Knapen:2017ekk,Griffin:2018bjn,Benato:2018ijc,Trickle:2019ovy} (for an incomplete review see~\cite{Battaglieri:2017aum,BRNreport}), these usually require extensive R\&D.\footnote{For detection concepts to probe bosonic DM with various types of nuclear couplings see~\cite{Arvanitaki:2017nhi,Baryakhtar:2018doz}.}  
Spin-dependent interactions are even more challenging to probe below $\sim$1~GeV. 
There is therefore a clear need to develop new detection concepts that, with near-term technology, can probe spin-independent nuclear interactions for DM masses below 50--100~MeV and probe spin-dependent interactions for DM masses below $\sim$1~GeV.  

In this paper, we propose a novel detection concept based on DM scattering with, and subsequent excitation of, internal states of di-atomic molecules. Similar concepts were previously proposed in~\cite{Essig:2016crl} and~\cite{Arvanitaki:2017nhi}, albeit those studies considered either dissociation of the molecule or excitations following absorption of bosonic DM. Our detection concept is also distinct to that considered in~\cite{Vavra:2014nvc,Vavra:2016mll}, which used molecules, but proposed searching for excitations in liquids or ice. The proposal in this study has the features that it requires minimal R\&D and can probe both spin-independent and spin-dependent DM scattering for DM masses in the 100~keV to 100~MeV range. This is a particularly interesting mass range, since as argued above, it is below current direct-detection bounds, but above the mass range for which there are often also strong indirect limits on DM from stellar cooling~\cite{Green:2017ybv,Knapen:2017xzo,Chang:2018rso,Ramani:2019jam}.

The proposed setup is schematically depicted in Fig.~\ref{fig:Setups}. We consider DM that scatters off molecules in a gas, exciting vibrational and rotational (``ro-vibrational'') modes. The excited ro-vibrational modes are short lived and decay rapidly, producing multiple infrared signal photons. The signal photons are typically produced in two ways: (i) the excited molecule cascades down to lower vibrational modes in several steps, releasing a photon at each transition (\textbf{``cascade'' photons}), and (ii) the excited molecule is resonantly quenched by scattering off and exciting neighboring molecules to their first vibrational mode, which each decay to produce a photon (\textbf{``co-quench'' photons}).  The number of cascade and co-quench photons depends on the type of molecule and the ro-vibrational mode that is excited by the DM interaction.   Possible photodetectors could be superconducting nanowire single-photon detectors (SNSPD)~\cite{SEMENOV2001349,Goltsman2001,Natarajan2012}, superconducting transition edge sensors (TES)~\cite{Irwin:2005,Pyle:2015pya}, or microwave kinetic inductance detectors (MKID)~\cite{Mazin:12}. 
Sensitivity to spin-dependent DM interactions is obtained by working with molecules containing nuclei with an odd number of protons or neutrons, while sensitivity to spin-independent interactions is achieved with molecules containing any number of protons or neutrons. 

Much of our discussion in this paper is applicable to any molecule.  However, the ideal molecule for our purposes is a stable molecule, preferably with a $^1\Sigma$ electronic ground state, which can be cooled to relatively low temperatures while still maintaining a sufficiently high vapor pressure. The molecule should have an electric dipole moment to enable ro-vibrational transitions, and it should have a deep electronic potential well, holding a large number of vibrational levels. Among the various possible candidates, the carbon monoxide (CO) molecule appears to be very promising, due to its large dissociation energy of 11~eV and an excitation energy of 6~eV to the first electronic excited state.  By mixing together CO molecules of different isotopes, in particular, all combinations of $^{12}$C, $^{13}$C, and $^{16}$C with $^{16}$O, $^{17}$O, and $^{18}$O, one can achieve sensitivity to both spin-independent interactions and, through the isotopes with an odd number of neutrons, also to spin-dependent interactions with the neutron. Moreover, since the spectroscopy of CO is well understood, theoretical calculations of the expected DM signal are reliable. The cascade signal photons for CO will have an energy typically in the range 180~meV to 235~meV, while the co-quench photons have an energy of about 265~meV. 

In addition to the CO target, we also investigate hydrogen halides, such as hydrogen fluoride (HF), as well as a metal hydride, namely scandium hydride (HSc).  These provide sensitivity also to spin-dependent interactions with the proton.  While the molecular spectroscopy of these molecules is less well-understood than for CO, we will present several results and describe where additional theory work is required. The typical photon energies from the first five vibrational states of HSc are in the range 167~meV to 186~meV, while those of HF are in the range 416~meV to 485~meV. Detailed properties of these and other candidate molecules are given in Table~\ref{T3} in Appendix~\ref{sectMol}. 

The proposed concept has several important features: (i) The DM signal consists of \textit{multiple} photons that arrive in coincidence on a relatively short timescales of $\mathcal{O}$(0.1~s).  This allows for the use of photodetectors with non-zero, albeit small, dark counts.  Moreover, it also dramatically reduces background photons from blackbody radiation, and allows for larger gas temperatures.  (ii) Given a particular molecule, the number of detected photons depends on the DM mass and the microscopic interaction.  This allows one to infer some of the DM properties from the observed signal and also provides a handle on background discrimination. (iii) The DM scattering is an \textit{inelastic} process that excites internal degrees of freedom corresponding to ro-vibrational excitations of the molecule, which allows a sizeable fraction of the DM's kinetic energy to be transferred to the molecule. This implies that a DM particle as light as $\mathcal{O}$(100~keV) is able to excite a vibrational mode that lies $\mathcal{O}$(200~meV) above the ground state. (iv) The technological requirements for the realization of the proposed concept are expected to be available on relatively short time scales. (v) There are excellent synergies between the technological requirements needed for our proposed concept to detect DM \textit{scattering} off molecules, with concepts to probe DM \textit{absorption} by molecules~\cite{Arvanitaki:2017nhi,Baryakhtar:2018doz}, and with the use of scintillators to probe for DM-electron scattering or absorption~\cite{Derenzo:2016fse,Bloch:2016sjj}.

The remainder of the paper is organized as follows.  Sec.~\ref{sec:DM_Scattering} describes the salient features of molecules, the DM-molecular scattering kinematics, and the calculations of the molecular excitation rates.  Sec.~\ref{sec:Exp_Considerations} describes the various relaxation pathways of the excited molecule, while Sec.~\ref{sec:Sig_Gen} discusses the generation of the signal of interest (cascade and co-quench photons).  Sec.~\ref{sec:photon-collect} describes the efficiency with which signal photons can be detected, while Sec.~\ref{sec:optimization} discusses the impact of experimental parameters such a pressure and temperature on the observed signal.  Sec.~\ref{sec:bkg} contains a brief discussion of backgrounds, in particular the dark count and blackbody background.  In Sec.~\ref{sec:Projections}, we present the projected sensitivity of the experimental concept, while Sec.~\ref{sec:Conclusions} contains our conclusions.  A series of appendices provide extensive technical details on our calculations, discuss other possible molecular targets, and discuss the sensitivity to DM that interacts with ordinary matter through a dark photon.

\section{Dark Matter-molecule scattering}
\label{sec:DM_Scattering}

\subsection{Fundamental Molecular Properties}
\label{subsec:Eigenstates}

Many properties of molecules can be understood by considering the various energy scales involved in the system. Molecules are a result of two or more atoms sharing/exchanging electrons to form a bound state. Since the electrons are far lighter than the nuclei ($m_e\sim m_n/2000$), electrons move at speeds much larger than the nuclei, and hence the electronic states change only adiabatically when the nuclear states change. It is therefore possible to separate the nuclear degrees of freedom from the electronic ones. This is the core idea behind the Born-Oppenheimer approximation~\cite{Born_Oppenheimer,born2000quantum}. Within this approximation, the electronic configuration sets the potential energy of the nuclei, which in the case of diatomic molecules leads to a potential energy, $U(r)$, that depends only on the internuclear distance $r$. Neglecting the center-of-mass motion, the nuclear Hamiltonian is therefore given by
\begin{equation}
\label{eq1}
\hat{H}_{\text{mol}}=-\frac{\nabla^2_r}{2\mu_{12}} +\frac{\hat{L}^2}{2\mu_{12}r^2} + U(r)\,.
\end{equation}
Here $\mu_{12}$ is the reduced mass of the molecule, $-\nabla^2_r/2\mu_{12}$ is the radial kinetic energy operator, and $\hat{L}$ the angular momentum operator. The molecular energy depends on two distinct degrees of freedom: radial motion and molecular orientation $\Omega$. The former is associated with the vibrational modes of the molecule, while the latter is associated with the rotational modes.  

The molecular energy is obtained by solving the Schr\"odinger equation with the Hamiltonian in Eq.~(\ref{eq1}). This equation can usually be solved by separation of variables of the vibrational and rotational degrees of freedom.  As a result, the molecular wavefunction factorizes into radial and angular functions, $\Psi_{vJm}(\bm{r})=\phi_{vJ}(r) \mathcal{Y}_{J m}(\Omega)$, and the eigen-energies can be written as the sum of two contributions $E_{\text{mol}}=E_v+E_{\text{rot}}$. Here $v$ and $J$ together with $m$ are the vibrational and rotational quantum numbers, respectively, and $\mathcal{Y}_{J m}(\Omega)$ are spherical harmonics. When such a separation of variables is possible, one can often derive analytic expressions for the wavefunctions. One classic example is the Morse potential~\cite{Morse:1929zz}, which has the form 
\beq
U(r)=D_{e}\left( e^{-2 \beta (r-r_e)} - e^{-\beta (r-r_e)} \right)\,.
\label{eq:Morse_Potential}
\eeq
Here, $r_e$ is the equilibrium distance (i.e., the position of the potential minimum), $D_{e}$ is the dissociation energy of the molecule, and $\beta = \sqrt{U^{''}(r_e)/2D_{e}}$ (the values of these parameters can be found in the literature for many molecules).

Furthermore, even when an analytically solvable potential such as the Morse form is not a good approximation, one can always Taylor expand $U(r)$ around its minimum and retain arbitrary orders of the expansion. For example, at the leading order of interest, the region around the minimum is just a quantum harmonic oscillator. For many diatomic molecules, the harmonic approximation is sufficient to describe the ground state and several of the lowest lying excited modes. However, the description of higher modes requires additional anharmonic terms.  For example, including the harmonic term and the first anharmonic correction, 
the vibrational eigen-energies take the form
\beq
\label{eq4}
E_v \approx \omega_e  \left(v+\frac{1}{2}\right) - \omega_e x_e \left(v+\frac{1}{2}\right)^2\,, 
\eeq
where $\omega_e=\sqrt{U''(r_e)/\mu_{12}}$ is the harmonic frequency and $\omega_e x_e$ is the first anharmonic correction, 
where for a Morse potential $x_e = \omega_e/4 D_e$. 

To lowest order, the rotational modes of the molecule can be described by the rigid rotor approximation, i.e., the molecule rotates with a constant interatomic distance. At higher orders, the interatomic distance itself also varies. The eigen-energies associated with the rotational states are 
\beq
\label{eq:E_rot}
E_{{\rm rot},vJ}=B_v J(J+1)\,,
\eeq
where $B_v=\langle \phi_{v0} |r^{-2}| \phi_{v0} \rangle / 2\mu_{12}$ is the rotational constant associated with a vibrational state $v$. The rotational constant is often expressed as $B_v \approx B_e-\alpha_e(v+1/2)$, where $\alpha_e$ is the first anharmonic correction to the rotational constant and $B_e=(2\mu_{12}r_e^2)^{-1}$ is the equilibrium rotational constant. Finally, summing the results of Eqs.~\eqref{eq4} and~\eqref{eq:E_rot}, the molecular energy is given by~\cite{Herzberg} 
\bea
E_{vJ} & \approx & \omega_e  \left(v+\frac{1}{2}\right) - \omega_e x_e \left(v+\frac{1}{2}\right)^2 \nonumber \\
&& + B_eJ(J+1)-\alpha_e(v+1/2)J(J+1)\,.
\label{eq6}
\eea
For the case of a Morse potential, the fundamental frequency of the oscillator, $\omega_e$, the first anharmonic correction to the vibrational modes, $\omega_e x_e$, and the anharmonic correction to the rotational energy, $\alpha_e$, are given by 
\bea
\omega_e & = & \sqrt{2 D_e \beta^2/\mu_{12}} \nonumber \\
\omega_e x_e & = & \beta^2/2 \mu_{12} \nonumber \\
\alpha_e & = &\omega_e^{-1} \left(6\sqrt{\omega_e x_e B_e^3}-6B_e^2\right)\,.
\eea

Taking Eqs.~\eqref{eq4} and~\eqref{eq:E_rot} and using the fact that the electronic excitation energy is of order the nuclear dissociation energy, one can easily estimate the typical energy scales of the system. Specifically, since the dissociation energy is $D_e \sim \alpha_{\rm EM}^2 m_{\rm e}$ ($\alpha_{\rm EM}$ is the fine structure constant and $m_{\rm e}$ is the electron mass), the equilibrium distance is typically the Bohr radius, $r_e \sim (\alpha_{\rm EM} m_{\rm e})^{-1}$, and $U^{''}(r_e) \sim D_{e}/r_e^2$, one finds the approximate scaling of excitation energies to be $\mathcal{O}\left[\alpha_{\rm EM}^2 m_{\rm e} (m_{\rm e}/\mu)^\gamma\right]$, with $\gamma=0$, $0.5$, and $1$ for electronic, vibrational, and rotational modes, respectively. As expected, there is a large separation of energy scales between the various modes. 

For concreteness, this study presents results for carbon monoxide, hydrogen halides, and a metal hydride, for which $D_e\approx 3-11$ eV and $\omega_e\approx 0.2 -0.5$~eV, corresponding to IR wavelengths for transitions between consecutive vibrational states (for more details see Appendix~\ref{sectMol}). Rotational transitions typically correspond to wavelengths about an order of magnitude larger.  Since individual photons from the 
rotational transitions are very challenging to detect, we will be predominantly interested in vibrational transitions.  However, the various rotational states will play an important role in calculating the excitation and de-excitation probabilities. 
For example, calculating these probabilities requires understanding the properties of the molecular gas in the initial state before scattering events occur. 

The proposed experiment would operate at temperatures low enough to avoid multiple-photon backgrounds from blackbody radiation. For the explored molecules, this turns out to be in the range $45-115$~K and depends on the molecule (see Sec.~\ref{sec:bkg} for details). The temperature sets the distribution of thermally excited ro-vibrational states according to a Maxwell Boltzmann distribution. At temperatures below room temperature, the population of excited vibrational states is negligible, since the typical vibrational energy quanta are $\sim$3000~K. 
The probability to find a molecule in an initial rotational state, $J_{\rm init}$, of the ground vibrational state, $v=0$, is then
\beq
P_{\rm therm}(J_{\rm init},T) \approx \frac{(2J_{\rm init}+1) e^{-B_{v=0}J_{\rm init}(J_{\rm init}+1)/T}}{\sum\limits_{J_i} (2J_i+1) e^{-B_{v=0}J_i(J_i+1))/T}}\,.
\label{eq:Pinit_MB}
\eeq
As a result, given the temperature and energy ranges considered in this study, the molecules will populate mainly the first $\mathcal{O}(1-10)$ rotational states. 

\subsection{Kinematics}
\label{subsec:Kinematics}

As discussed above, before a DM scattering event, the gas consists of molecules with negligible kinetic energy. The molecules begin in the vibrational ground state with a small spread of low lying rotational states, $J_{\rm init}$. A DM scattering interaction that excites the molecule to some higher vibrational and rotational mode with $(v^\prime,J^\prime)$, transfers momentum $\vec{q}$, angular momentum, and energy to the nuclei of the molecule. The energy transfer is converted to vibrational and rotational excitation energy, $\DEvJ \equiv E_{v^\prime J^\prime} - E_{0J_{\rm init}}$, and to recoil energy of the entire molecule.

Conservation of energy for a molecule initially at rest requires that 
\beq
\DEvJ = \vec{v}_\chi \cdot \vec{q} - \frac{q^2}{2\mu_{\chi {\rm m}}}\,,
\label{eq:vqctheta}
\eeq
where $\vec{v}_\chi$ is the DM velocity and $\mu_{\chi {\rm m}}$ is the reduced mass of the system comprised of the DM ($m_\chi$) and molecule ($m_{\rm m}$). Strictly speaking, Eq.~\eqref{eq:vqctheta} is the only constraint on the kinematics of the process and all other information should be captured by the wavefunction overlap of the initial and final states. This is given by the function $\FmolvJsqAvg$, the averaged target form factor, which is calculated by
\beq
\FmolvJsqAvg \equiv \left \langle \left \vert \int d^{3}r \, \mathcal{O}(\mathbf{q}\cdot\mathbf{r}) \Psi_{v' J'}^{*}(\mathbf{r})\Psi_{0 J_{\rm init}}(\mathbf{r}) \right \vert^2 \right \rangle\,,
\label{eq:FmolAvg}
\eeq
with
\bea
\mathcal{O}(\mathbf{q}\cdot\mathbf{r}) & = & f_{PN}^{(1)} e^{i\frac{\mu_{12}}{m_1}\mathbf{\mathbf{q}\cdot r}} + f_{PN}^{(2)} e^{-i\frac{\mu_{12}}{m_2}\mathbf{\mathbf{q}\cdot r}} \nonumber \\
f_{PN}^{(i)} & = & \begin{cases} 
      f_{P,\text{SI}}^{(i)} Z^{(i)} + f_{N,\text{SI}}^{(i)} (A^{(i)}-Z^{(i)}) & {\rm SI} \\
      2\times[f_{P,\text{SD}}^{(i)} \langle S_P^{(i)} \rangle + f_{N,\text{SD}}^{(i)}\langle S_N^{(i)} \rangle] & {\rm SD} \nonumber \,
\end{cases} ,\
\label{eq:SDSI_Operators}
\eea
Here $\Psi_{0 J_{\rm init}}(\mathbf{r})$ and $\Psi_{v' J'}^{*}(\mathbf{r})$ are the initial and final wavefunctions respectively and $\mathbf{r}$ is the distance between the two atoms. The operator $\mathcal{O}(\mathbf{q}\cdot\mathbf{r})$ controls the interactions with each atom. The $f_{PN}^{(i)}$ control the spin dependence and are scalars for the spin independent (SI) case and non-trivial angular momentum operators for the spin dependent (SD) case. The index in $f_{PN}^{(i)}$ and $m_i$ corresponds to the two atoms within the molecule, $Z^{(i)}(A^{(i)})$ is the atomic (mass) number of atom $i$ and $\langle S_P^{(i)} \rangle$ ($\langle S_N^{(i)} \rangle$) is the zero-momentum spin structure of the protons (neutrons) within nucleus $i$~\cite{Aprile:2019dbj}. The spin dependent result is a good approximation as long as the inverse momentum transfer is larger than the typical size of the nucleus, as is always the case for the signal considered in this study. Finally, if the DM couples to electric charge (as for example in the case of a dark photon mediator), then $f_N^{(i)}=0$ and $Z^{(i)}\rightarrow Z_{\rm eff}^{(i)} = F_A(q) Z^{(i)}$ with 
\beq
F_A(q) = \frac{a^2q^2}{1+a^2q^2}\,,
\label{eq:Thomas_Fermi}
\eeq
where $a$ is the Thomas-Fermi radius~\cite{ashcroft2010solid}. This takes into account screening of the nuclear charge by the electron cloud surrounding the nucleus. The average in Eq.~\eqref{eq:FmolAvg} is taken over the Maxwell Boltzmann distribution of the initial $J_{\rm init}$ states, Eq.~\eqref{eq:Pinit_MB}, and the result is therefore temperature-dependent. A full evaluation of Eq.~(\ref{eq:FmolAvg}) is given in Appendix~\ref{app:FF}.

In order to gain intuition as to what is the typical energy transfer in the scattering process, one can consider the impact approximation. In this approximation, the interaction can be thought of as a two-step process: (1) the DM particle interacts with a single nucleus with mass $m_i$, transferring momentum $\vec{q}$ and energy $E_q$, after which (2) this momentum and energy are transferred to excite the vibrational and rotational modes and also transferred to the molecule's center of mass recoil momentum and energy. Thus, energy conservation can be calculated with initial conditions corresponding to the moment after step (1) and before step (2). Additionally, there is some typical momentum spread associated with the ground state of the molecule, which is approximately given by 
\beq
\Delta p_e \approx \sqrt{\frac{1}{2} \mu_{12} \omega_e} \approx \frac{1}{\sqrt{2}} \left(\frac{\mu_{12}}{m_{\rm e}}\right)^{1/4} \alpha_{\rm EM} m_{\rm e}\,,
\label{eq:Delta_p0}
\eeq
where $\omega_e$ has been approximated as in Sec.~\ref{subsec:Eigenstates}. This means that the momentum of $m_i$ directly after step (1) is given by $\vec{q} + \Delta\vec{p}_e$, where the magnitude of $\Delta\vec{p}_e$ is given by Eq.~\eqref{eq:Delta_p0}. Under these assumptions, there exists the additional energy conservation constraint 
\beq
\DEvJ \approx \frac{\mu_{12}}{m_i}\frac{q^2}{2m_i} + \frac{\vec{q} \cdot \Delta\vec{p}_e}{m_i}\,.
\label{eq:Eq_Impact_Approx}
\eeq
Finally, one should also consider the effects of rotation. In terms of the rotational quantum number of the final molecular state, $J^\prime$, conservation of angular momentum implies,
\beq
|J^\prime - J_{\rm init}| \lesssim q \cdot (\alpha_{\rm EM} m_{\rm e})^{-1} \lesssim J^\prime + J_{\rm init}\,.
\label{eq:Jvals}
\eeq
This is a good approximation, since the Bohr radius is approximately the impact parameter of the incoming DM particle.

If the DM scattering cross-section does not have any additional momentum dependence (we will describe the full cross section in Sec.~\ref{subsec:Cross_Sec}), Eqs.~\eqref{eq:vqctheta}--\eqref{eq:Jvals} can be used to estimate how energy is extracted from the incoming DM particle, as well as the range of momentum transfer that maximizes the cross section for given final $v^\prime$ and $J^\prime$. Specifically, since the maximal momentum transfer is approximately $q \approx 2 m_\chi v_\chi$ for DM masses much smaller than the target mass, Eq.~\eqref{eq:Eq_Impact_Approx} can be used to provide an estimate of the typical energy that can be extracted from the DM,
\beq
{\DEvJ}_{\rm typ} \approx 4 \left(\frac{\mu_{12} m_\chi}{m_i^2} + \frac{\Delta p_e}{m_i v_\chi}\right) E_\chi\,,
\label{eq:EvJ_max_Impact}
\eeq
where $E_\chi$ is the total kinetic energy of the DM particle. The first term within the parenthesis is the usual energy of the recoiling target when the DM scatters elastically off a much heavier target and decreases rapidly for decreasing $m_\chi$. On the other hand, the second term is $m_\chi$ independent and dominates for low DM masses, allowing for a much larger fraction of energy to be extracted from the DM compared to the elastic case. It is important to note that while $\Delta p_e$ from Eq.~\eqref{eq:Delta_p0} is the typical momentum spread in the initial state, larger momentum spreads are possible with increasingly small probability, allowing for larger momentum transfers and thus larger $\DEvJ$ for a given $m_\chi$.

One can also use Eq.~\eqref{eq:Eq_Impact_Approx} to estimate the ``typical'' DM mass, i.e., the DM mass that is most likely to excite the state $(v^\prime,J^\prime)$ with energy $\DEvJ$. Substituting $q\sim m_\chi v$, and taking $\vec{q}\cdot\Delta \vec{p}_e\approx0$ (averaged over angles), the solution for $m_\chi$ for a given $\DEvJ$ is 
\beq
m_{\chi{\rm typ}} \approx \frac{m_i}{v_\chi \mu_{12}} \sqrt{2 \DEvJ \mu_{12}}\,.
\label{eq:mchi_typ}
\eeq

\begin{figure}
\centering
\includegraphics[width=1\columnwidth,keepaspectratio]{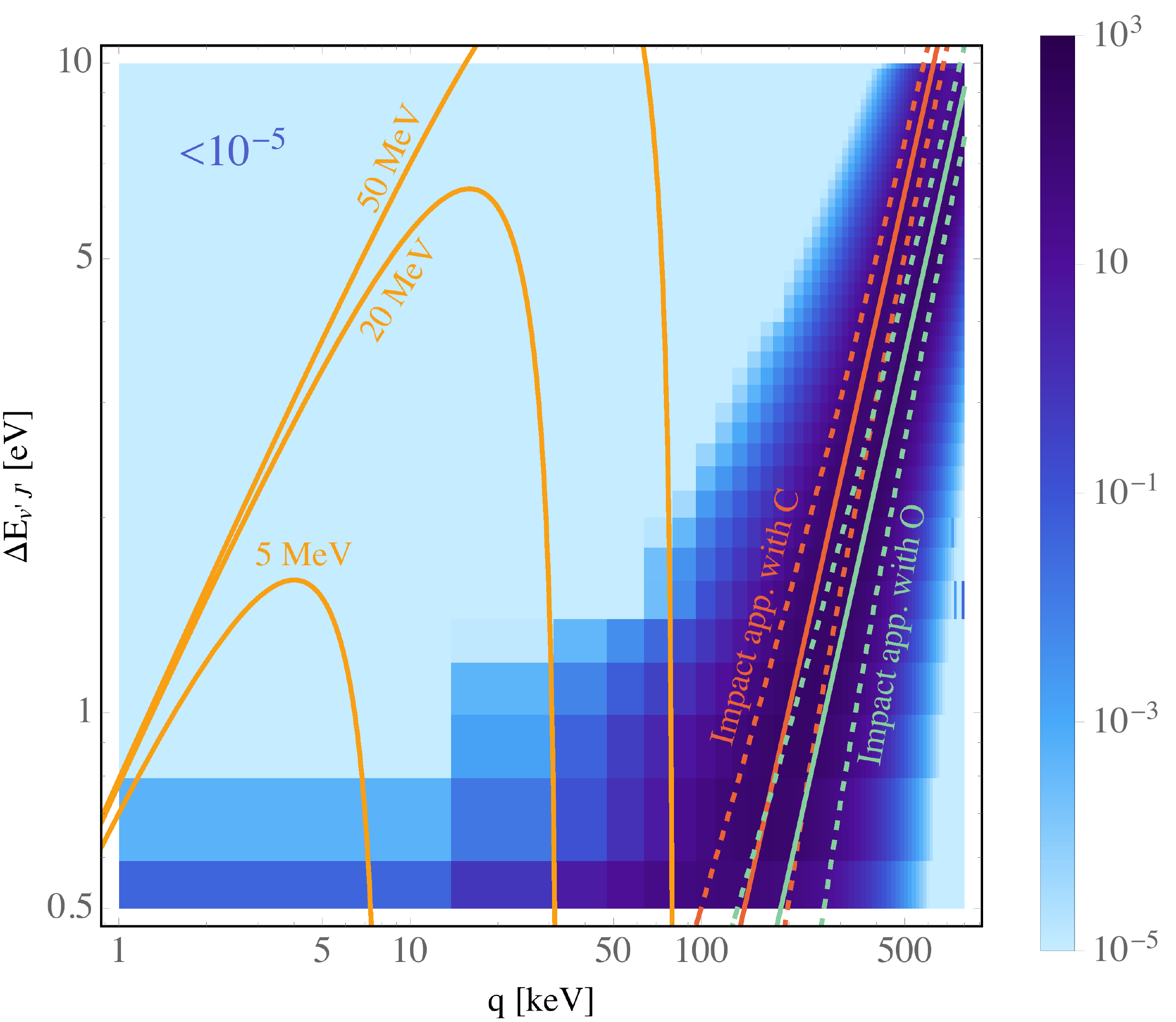}
\caption{The averaged, spin independent, molecular form factor, $\FmolvJsqAvg$, with $f_{P,{\rm SI}}^{(i)} = f_{N,{\rm SI}}^{(i)} = 1$, for a gas of CO molecules at 55 K, plotted as a function of vibrational-rotational energy, $\DEvJ$, and momentum transfer, $q$. In every energy-momentum bin, the color corresponds to the sum over all form factor elements within that bin. The red and green curves correspond to the relation expected from energy conservation under the impact approximation, Eq.~\eqref{eq:Eq_Impact_Approx}, for interaction with the C or O atom respectively. For each, the solid curves follow the first term in the equation while the dashed curves correspond to the minimal and maximal values allowed by the approximation. The orange curves correspond to Eq.~\eqref{eq:vqctheta}, with a DM velocity of $v_\chi = 240$ km/sec, and for three DM masses of $m_\chi = 5$, $20$ and $50$~MeV.}
\label{fig:FF}
\end{figure}

In Fig.~\ref{fig:FF}, the averaged, spin independent, molecular form factor with $f_{P,{\rm SI}}^{(i)} = f_{N,{\rm SI}}^{(i)} = 1$ is shown as a function of the vibrational-rotational energy, $\DEvJ$, and the momentum transfer $q$, for a DM particle scattering off a gas of CO molecules at 55 K. The solid orange lines are contours of Eq.~\eqref{eq:vqctheta} (which is the only condition that the kinematics need to satisfy), evaluated at $v_\chi = 240$~km/sec for three distinct DM masses, namely 5, 20 and 50~MeV. Above these contours there is sizable suppression of the velocity-averaged cross section from the DM's velocity distribution. The red and green curves correspond to the approximation Eq.~\eqref{eq:Eq_Impact_Approx} for impact with a C or O atom respectively. For each set of curves, the solid is the first term of the equation while the dashed curves correspond to the positive or negative contributions of the second term. Evidently, the typical energy transfer estimated with the impact approximation and given by Eq.~\eqref{eq:Eq_Impact_Approx} is an excellent measure of the energy transfer for which the form factor is maximized. 

The experimental concept presented in this paper will be sensitive to a range of $(v,J)$ modes and, therefore, a range of DM masses.  
Taking characteristic values for the vibrational energies of a CO molecule, namely ${\DEvJ}\approx0.25-2.5$~eV, which are the approximate values for the first ten vibrationally excited states of CO, the ``typical'' DM mass that can be probed is $m_{\chi, {\rm typ}} \approx \mathcal{O}(10-100\text{ MeV})$.  For smaller DM masses, the DM would prefer to excite lower ${\DEvJ}_{\rm typ}$ than those available in CO, and the rate is therefore suppressed; nevertheless, there is some probability that almost the \textit{entire} DM kinetic energy is transferred to $\DEvJ$, implying that even DM as light as $\mathcal{O}$(100~keV) can excite a vibrational mode.  For DM masses much larger than the ``typical'' mass, there is a phase space suppression to excite a particular $\DEvJ$, since a heavier DM particle prefers to transfer more energy than $\DEvJ$; in particular, it prefers to dissociate the molecule completely.  Since our detection concept is sensitive only to the photons coming from the de-excitations of higher vibrational states, and not to dissociation, the reduced available phase space at high DM masses will force the scattering rate to decrease faster than $m_\chi^{-1}$, as most of the rate at high DM masses goes into dissociating the molecule.  Finally, note that $m_{\chi{\rm typ}}$ can be reduced by using molecules that contain a low-mass nucleus, such as hydrogen.  Although Eq.~\eqref{eq:mchi_typ} shows that this reduces $m_{\chi{\rm typ}}$ only by the square root of the molecule's reduced mass, it is worth exploring hydrogen halides and a metal hydride for this reason.  Another reason to explore these molecules is that they will also provide sensitivity to spin-dependent proton couplings.

\subsection{Cross Section and Rates}
\label{subsec:Cross_Sec}

The velocity-averaged cross section for exciting a final state with vibrational-rotational quantum numbers $(v',J')$ is 
\bea
\left\langle \sigma v_\chi\right\rangle_{v^\prime J^\prime} & = & \sigmap \int \frac{q dq}{2 \mu_{\chi n}^2} \FDMsq \nonumber \\
& & \times \FmolvJsqAvg \eta[v_{\rm min}(q)]\,,
\label{eq:sigmav}
\eea
where $\mu_{\chi n}$ is the DM-nucleus reduced mass, $\eta[v_{\rm min}(q)]$ is the integrated velocity distribution defined as in~\cite{Essig:2015cda}, and $v_{\rm min}$ is the $q$-dependent minimal DM velocity allowed by Eq.~\eqref{eq:vqctheta}. We evaluate rates assuming the Earth's velocity to be $v_{\rm Earth} = 240$~km/sec, a DM velocity dispersion of $v_0 = 230$~km/sec, and a DM escape velocity from the galaxy to be $v_{\rm esc} = 600$~km/sec.

The function $\FDM$ is the DM form factor. Its form, together with a reference cross section $\sigmap$, are defined as 
\begin{eqnarray}
\FDMsq & \equiv & \frac{|\mathcal{M}_{2\to2}(\mathbf{q})|^{2}}{|\mathcal{M}_{2\to2}(\mathbf{q}^{2} = q_0^2)|^{2}} \label{eq:DD_FF} \,, \\
\sigmap & \equiv & \frac{|\mathcal{M}_{2\to2}(\mathbf{q}^{2} = q_0^2)|^{2}}{16\pi(m_{\chi}+m_{n})^{2}}\,.
\label{eq:sigma_p}
\end{eqnarray}
The form factor encodes information about the DM-target interaction such that $\sigmap \cdot \FDMsq$ is the DM-nucleon interaction cross section.  With this definition, $\sigmap$ is the DM-nucleon cross section for $f_P^{(i)} = f_N^{(i)} = 1$, at some reference momentum-transfer value, $q_0$. We take this value to be the typical momentum transfer, which is approximately the momentum spread in Eq.~\eqref{eq:Delta_p0},~$q_0 = \Delta p_e$, relevant for the process of molecular excitations. Moreover, $\FDMsq = 1$ ($\FDMsq = q_0^4/q^4$) corresponds to a ``heavy'' (``light'') particle mediating the interaction between the DM and nucleus, where heavy (light) refers to a mediator mass that is larger (smaller) than the momentum transfer.

Two points should be noted regarding the discussion from Sec.~\ref{subsec:Kinematics} of the DM masses that maximize the scattering rate. First, since much of phase space is unavailable because scattering results in excited states only (and not dissociated states) for our signal, the integral over $\int q dq$  in Eq.~\eqref{eq:sigmav} does not have the usual scaling $\propto \mu_{\chi n}^{2}$.  This can, in principle, reduce $m_{\chi\text{typ}}$ to smaller values. Moreover, even for large DM masses, the cross section is no longer constant (and the rate no longer decreases proportional to $m_\chi^{-1}$) when compared to elastic nuclear scattering. Second, the arguments regarding $m_{\chi\text{typ}}$ given in the previous section are valid only for DM interactions that are mediated by a heavy mediator. In that case, the microscopic scattering process is momentum-independent, and any momentum dependence of the process is captured entirely by the the molecular form factor described above and shown in Fig.~\ref{fig:FF}. However, if the DM interactions are mediated by a light mediator, then the momentum dependence of the final scattering rate is given by the molecular form factor combined with the $1/q^4$ behavior of Eq.~\eqref{eq:DD_FF}. In this case, the scattering rate is enhanced at low momentum transfers compared to the heavy-mediator case. If the couplings, $f_{PN}^{(i)}$, are proportional to the masses of each atom $m_i$ (as is the case for spin-independent scattering with $f_{P,\text{SI}}^{(i)} = f_{N,\text{SI}}^{(i)}$), then the low momentum behavior of the molecular form factor scales as $q^4$, or higher powers (see e.g.~\cite{Cox:2019cod}). This is not typically the case for spin-dependent scattering, since in that case the couplings are proportional to the nuclear spin structures and not their masses. Finally, we find that for the molecules considered in this study and for $\FDMsq \sim 1/q^4$, $m_{\chi\text{typ}}$ is reduced to a value close to the mass threshold and thus the experiment becomes extremely sensitive to very low DM masses.

\begin{figure*}[t]
\centering
\includegraphics[width=0.49\textwidth]{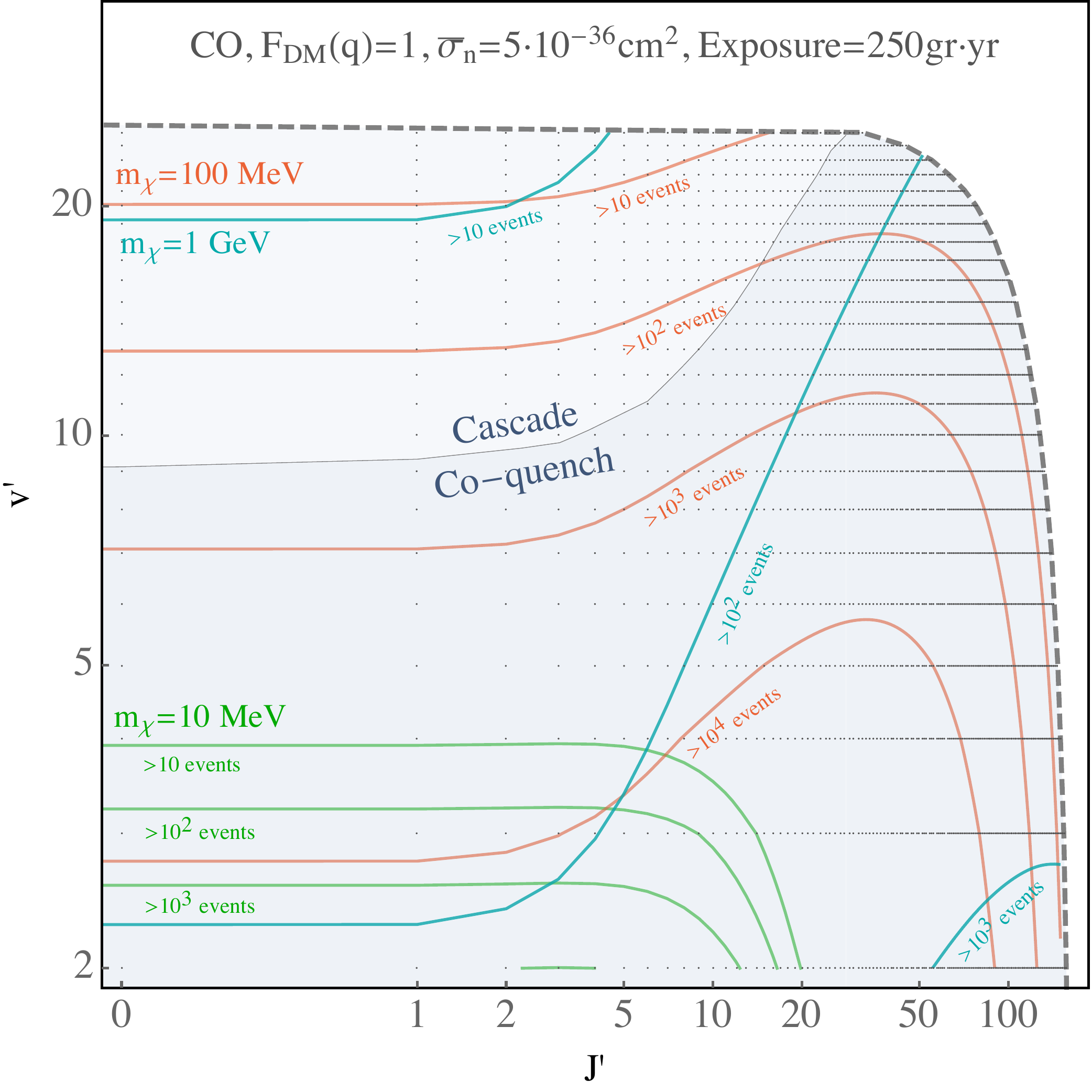}
\includegraphics[width=0.49\textwidth]{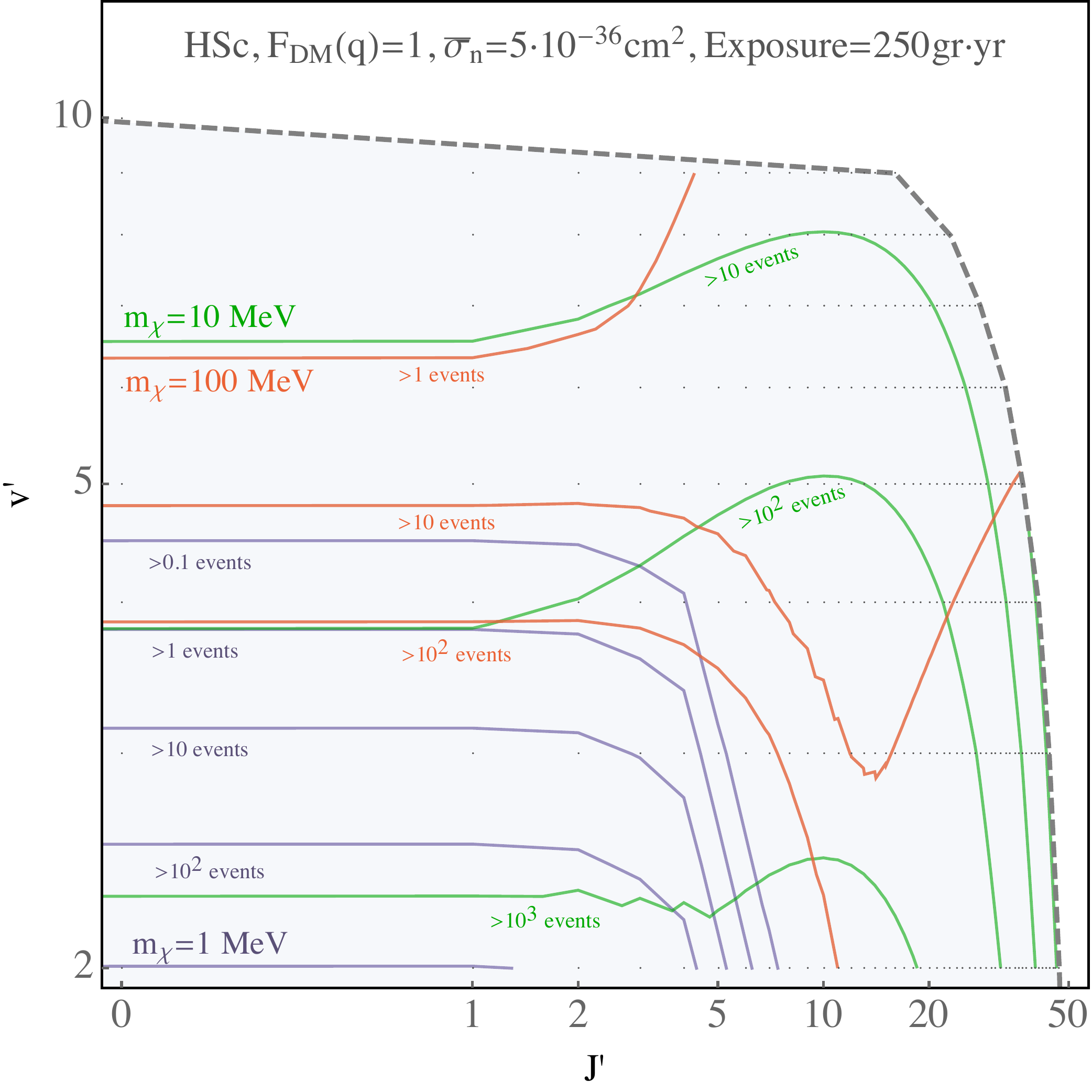}
\caption{Contours of $\rho_\chi/m_\chi \cdot \left\langle \sigma v_\chi\right\rangle_{v^\prime J^\prime}$ as a function of $(v^\prime,J^\prime)$, i.e., the number of events to scatter into a particular $(v^\prime,J^\prime)$ state, for a CO molecule ({\bf left}) and an HSc molecule ({\bf right}).  Results take the reference cross section per nucleon to be $\sigmap = 5\times 10^{-36}$ cm$^2$, an exposure of $250$~gr-yr, and a DM form factor of $\FDM=1$ (corresponding to a heavy particle mediating DM interactions with molecules). For CO, the dashed curve corresponds to the range of $(v^\prime,J^\prime)$ for which the states are unable to mix with the first excited electronic state, which could provide additional non-radiative energy decay processes (see Sec.~\ref{subsec:nearby-e-states} for further details). For HSc, this curve has been chosen arbitrarily at values of $(v^\prime,J^\prime)$ that have energies equal or below that of $E_{v^\prime=10,J^\prime=0}$, see Sec.~\ref{subsec:nearby-e-states} for details. Shaded regions in the left panel for CO indicate values of $(v^\prime,J^\prime)$ that produce either ``cascade" or ``co-quench" signals.  In the right panel for HSc, the photon production is less well-understood and could involve couplings to electronic states (see Sec.~\ref{subsec:nearby-e-states}). The different colored contours represent different DM masses, and modes below each contour are excited at a rate equal to or larger than the label on that contour.}
\label{fig:RvJ}
\end{figure*}

While Eq.~\eqref{eq:sigmav} gives the cross section to scatter into some excited state, the signal of interest in this paper is the photons that are emitted after the excited state relaxes to the ground state. For each set of $v^\prime,J^\prime$ values, there is some branching ratio that accounts for the probability to decay while emitting $\ngemit$ coincident photons, BR($v^\prime,J^\prime,\ngemit$). These branching ratios depend on the properties of the molecules and the gas. Furthermore, if $\ngemit$ photons are emitted, there is some efficiency, $\epsilon_{\rm col}(\ngemit,\ngcol)$, to collect $\ngcol$ of them with a photon detector, as well as some efficiency, $\epsilon_{\rm det}(\ngcol)$, to detect them as signal photons. These efficiencies depend on properties of the experiment, such as its geometry, the reflectivity of mirrors, and photodetector acceptances.  The efficiencies could also depend on energy, although we ignore this below.  A detailed discussion of the branching ratios and efficiencies will be the topic of subsequent sections. Currently, we focus only on their impact on the signal rate.

Putting everything together, the total signal rate per target molecule is given by,
\bea
R & = & \frac{\rho_\chi}{m_\chi} \sum \text{BR}(v^\prime,J^\prime,\ngemit) \epsilon_{\rm col}(\ngemit,\ngcol) \nonumber \\
&& \times \epsilon_{\rm det}(\ngcol) \cdot \left\langle \sigma v_\chi\right\rangle_{v^\prime J^\prime} \,,
\label{eq:RoverNtar}
\eea
where $\rho_\chi/m_\chi$ is the local DM number density, and we take $\rho_\chi=0.4$~GeV/cm$^3$~\cite{Sivertsson:2017rkp}. The sum is over $v^\prime$, $J^\prime$, $\ngemit$, and $\ngcol$, which depend on each other in a non-trivial fashion. For example, $\ngemit$ cannot be larger than $v^\prime$, since each photon emission corresponds to a decrease of at least one vibrational mode. The sum over the number of collected photons should include all values of $\ngcol$ that are defined to be part of our signal; in our case, we require two or more photons. 

Fig.~\ref{fig:RvJ} shows $\rho_\chi/m_\chi \cdot \left\langle \sigma v_\chi\right\rangle_{v^\prime J^\prime}$ as a function of $(v^\prime,J^\prime)$ for the candidate molecules CO (left) and HSc (right) for $\FDMsq=1$ and for spin independent scattering. This is the rate to scatter into all relevant $(v^\prime,J^\prime)$ states (not including any branching ratios or efficiencies for photon emission and detection). The results are given for $\sigmap = 5\times 10^{-36}$ cm$^2$ and an exposure of $250$~gr-yr. For CO, the dashed line corresponds to the values of $(v^\prime,J^\prime)$ above which the states are able to mix with the first excited electronic state, which could provide additional non-radiative energy decay processes (see Sec.~\ref{subsec:nearby-e-states} for further details). 
For HSc, this curve has been chosen arbitrarily at values of $(v^\prime,J^\prime)$ that have energies equal or below that of $E_{v^\prime=10,J^\prime=0}$; we discuss the reason for this in Sec.~\ref{subsec:nearby-e-states}. The shaded regions indicate values of $(v^\prime,J^\prime)$ that correspond to either ``cascade" or ``co-quench" signals. One notices that for CO in Fig.~\ref{fig:RvJ} (left), cascade photons require excitation to vibrational states with larger values of $v^\prime$ than those for co-quench photons. For HSc, the photon production is less well-understood and the presence of excited electronic states could play a substantial role.  The different colored contours represent the number of excited events for different DM masses (drawn as continuous lines, although of course only integer values of $v^\prime, J^\prime$ can be excited). Modes below each contour are excited at a rate equal to or larger than the label on that contour. The general behavior of Eq.~\eqref{eq:Eq_Impact_Approx} can be seen. The DM mass that maximizes the excitation of the $\DEvJ$ values in CO is $\sim$100~MeV, so that the $100$~MeV contours show the largest number of events.  Event numbers for DM masses below and above this value are suppressed as such DM particles preferentially transfer either too little or too much momentum and energy to the molecule. The preferred $J^\prime$ values given in Eq.~\eqref{eq:Jvals} can also be seen. Larger DM masses prefer to excite larger values of $J^\prime$. 
For HSc shown in Fig.~\ref{fig:RvJ} (right), the available $\DEvJ$ are smaller, and the small mass of the hydrogen nucleus implies that the number of events are largest for lower DM masses, $\sim$10~MeV. 

\section{Energy Transfer Mechanisms in Molecules}
\label{sec:Exp_Considerations}

The experimental concept proposed in this study requires understanding the different relaxation mechanisms that follow the vibrational excitation of a molecule. Of these, the spontaneous emission of photons produces the ``cascade" photon signal and is responsible for the de-excitation of the $v=1$ to $v=0$ states in the ``co-quench" photon signal. Additionally, there are four non-radiative collisional energy transfer processes (also known as collisional quenching processes), which are often referred to as ``$V-V$'', ``$V-T$'', ``$V-R$'' and ``$V-E$'' transfer, where $V$, $T$, $R$, and $E$ correspond to vibrational, translational, rotational, and ``electronic'' degrees of freedom, respectively. These four processes are both temperature and pressure dependent and compete with spontaneous emission.  Finally, if a molecule is excited to a high vibrational state of the electronic ground state that is close in energy to a vibrational state of an excited electronic state, then the two states can mix. The interplay between spontaneous emission, the four collisional processes, and the mixing between excited vibrational and electronic  states, as well as their dependence on temperature and pressure, determine the possible photon signal. For example, these processes set the branching ratios for spontaneous emission. Additionally, the choice of optimal pressure and temperature depend critically on the competing rates of spontaneous emission and collisional energy transfer. 

Of the four collisional energy transfer processes, the $V-V$, $V-R$, and $V-T$ transfers are relevant for excited vibrational states whose energy is below the first excited electronic state.  Among these three, $V-V$ transfer provides an alternative pathway to produce photons, namely the ``co-quench" photons; $V-R$ and $V-T$ processes provide pathways for the excited molecule to lose its entire energy without producing any observable photons. On the other hand, $V-E$ transfer and the mixing between excited vibrational states with excited electronic states are only relevant for higher vibrational and rotational modes whose energy is above that of some electronic state, i.e., the electronic state is energetically available. This mechanism could, in principle, provide an additional pathway to produce a photon due the transition of the excited ro-vibrational state to an excited electronic states (via mixing or $V-E$ transfer) and the subsequent decay to the electronic ground state. Since we are primarily interested in probing low-mass DM that excites the lower-lying ro-vibrational states, rather than the higher-energy electronic states, the $V-E$ energy transfer process is usually not of particular importance. This is, for example, the case for a target of CO molecules. Nevertheless, it could play an important role in other molecules. 

In the following subsections, we discuss the spontaneous emission process in Sec.~\ref{subsec:spontaneous}, the $V-V$, $V-R$, and $V-T$ transfer processes in Sec.~\ref{subsec:De-Ex_Proc}, and then describe the resulting branching ratios in Sec.~\ref{subsec:BR}. We defer a discussion of the $V-E$ transfer process as well as vibration-electronic state mixing to Sec.~\ref{subsec:nearby-e-states}. 

\subsection{Spontaneous Emission}\label{subsec:spontaneous}
The spontaneous emission rate of a molecule from state $(v^\prime,J^\prime)$ to state $(v,J)$ is set by its Einstein coefficients $A_{v^\prime J^\prime;vJ}$; henceforth, we will suppress the $J$ dependence in the notation and use the abbreviation $A_{v^\prime v}$. Spontaneous emission is a fast decay mechanism in molecules, when the transition is between different electronic states. However, spontaneous emission can also occur between two vibrational states within the same electronic state, albeit at a slower rate. In particular, for transitions within the same electronic state, a particular $(v^\prime,J^\prime)$ molecular mode will decay preferentially to a ($v'-1, J' \pm 1$) state.  While the final rotational number is determined by selection rules, the final vibrational quantum number $v$ is determined by the Franck-Condon overlap of the wavefunctions describing the $v$ and $v'$ vibrational states, as well as the transition dipole moment of the particular electronic state (in our case, the electronic ground state).  
Since $\Delta v=1$ transitions dominate over transitions with larger $\Delta v$, a cascade of transitions typically occurs, and multiple photons are emitted to produce the \textbf{cascade photon} signal.

In this study, we focus on polar molecules, which have a permanent dipole moment. In addition, we focus on molecules in which 
spontaneous emission rates are faster than the collisional energy transfer rates, so that photons will be produced efficiently (see Appendix~\ref{sectMol}). In what follows, we describe the various energy transfer mechanisms in detail. It will be shown that the total number of observed photons and their energies depend on the interplay between spontaneous emission rates and those of the various collisional energy transfer processes. Details of calculations and estimates for Einstein coefficients can be found in Appendices~\ref{COappend1} and~\ref{HXappend1}.

\subsection{$V-V$, $V-T$, and $V-R$ Transfer}
\label{subsec:De-Ex_Proc}

In this subsection, we focus on collisional energy transfer processes that do not involve electronic modes.
The collisional energy transfer mechanisms are processes whereby a molecule, denoted AB, in an excited ro-vibrational state $(v^\prime,J^\prime)$, decays into a different ro-vibrational state by colliding with another molecule of the same species, a molecule of a different species, or an atom. The precise types of processes depend strongly on the properties of the interaction potential between AB and the other object, and are classified as follows:

\begin{itemize}[leftmargin=*]
\item {\it V-V transfer} is a process whereby two molecules interact and energy is transferred primarily from the vibrational energy of one molecule to the vibrational energy of the second molecule.  This process occurs in particular when the energy gap required to excite one molecule is approximately equal to the energy gap required to de-excite the other. This is the case for the energy gaps between two nearby vibrational states for low $v'$, which are nearly equal due to the approximate harmonic nature of the potential. For a molecule initially in some excited $v'$ state interacting with another molecule of the same species, this is generally written as 
\begin{align}
 &\text{AB}(v',J') +\text{AB}(v,J) \rightarrow \nonumber \\ &\text{AB}(v'-1,J'-\Delta J') +\text{AB}(v+1,J+\Delta J) + \Delta E
\end{align}
The approximate matching of the energy levels results in very small $\Delta E$, and hence this process is also called quasi-resonant vibrational quenching.  The change in $J$ is empirically restricted, and one finds that $|\Delta J|$ and $|\Delta J'|$ are limited to $\lesssim 6$~\cite{Levon1974}. For our detection concept, we will predominantly have a molecule in an excited state $v'$ scatter with a molecule in the ground state, $v=0$. The rate for this process can be computed by thermally averaging over the population of $\text{AB}(0,J_{\rm init})$ and depends on both the excited vibrational and rotational quantum numbers, $(v^\prime,J^\prime)$. This rate is denoted by $\Gamma_{\rm VV}(v^\prime,J^\prime)$ and depends on the \textit{partial} pressure of the AB molecules in the gas. 

Since the rate is maximized when the energies between the up-scattered and down-scattered molecules are matched, this is the dominant non-radiative decay mechanism in molecules with a nearly harmonic intra-molecular potential such as CO~\cite{Wittig1972,Smith1973,Levon1974,Powell1975,Gower1975,Delenon1986,Flament1992,Billing2003}. For higher $v'$, the harmonic potential approximation breaks down, and increasing $v'$ results in decreasing  $\Gamma_{\rm VV}(v',J')$. For increasing values of $J'$, calculations suggest a slight increase in the quenching rates.

Importantly, unlike for the $V-T$ and $V-R$ transfers discussed below, the total vibrational number is conserved in intra-species $V-V$ transfers and an initial state $v'$ can still produce $v'$ photons. 
In particular, for the case of CO, the excited CO molecule can scatter off a ground-state CO molecule, which is excited to the $v=1$ state. This $v=1$ state can de-excite via spontaneous emission to produce a co-quench photon.  Multiple $V-V$ scatters can produce multiple CO molecules in the $v=1$ state and therefore multiple co-quench photons, up to a maximum of $v'$. 

\item {\it V-T transfer} is a process whereby a molecule in an excited vibrational state, $v^\prime$, decays via vibrational-kinetic energy transfer, i.e., the energy lost by the molecule goes into the translational degrees of freedom of the colliding partner, namely,
\beq
{\rm AB}(v^\prime,J^\prime) + {\rm M} \rightarrow {\rm AB}(\text{any state}) + {\rm M} + \Delta E.
\eeq
Here AB(any state) is any ro-vibrational state allowed by this process, M is another molecule or atom, and $\Delta E$ is the energy that goes into the translational degrees of freedom of the colliding partner (in our case, AB is our target molecule, and M will be a buffer gas). The collisional quenching rate, $\Gamma_{\rm VT}(v^\prime,J^\prime)$, depends on the pressure of M. Compared to $V-V$ and $V-R$ transfer, the $V-T$ transfer mechanism dominates in molecule-atom collisions~\cite{Krems2001,Krems2002} and also in some molecule-molecule collisions when the AB molecule is in a highly excited vibrational state~\cite{Park1994,Dayou2004,Hancock2009}.

\item {\it V-R transfer} is a process whereby a molecule in an excited vibrational state $v^\prime$ decays to a lower vibrational state by transferring its vibrational energy into the rotational degrees of freedom of the molecule, namely, 
\begin{align}
& \text{AB}(v',J') +\text{AB}(v,J)  \rightarrow \nonumber \\ 
&\text{AB}(v'-1,J'+\Delta J') +\text{AB}(v,J+\Delta J) + \Delta E,
\end{align}
where the energy of the AB($v',J'$) and AB$(v'-1,J'+\Delta J')$  molecules are similar, $E_{v' J'}\approx E_{v'-1 J'+\Delta J'}$. 
This process depends on the anisotropy of the molecule-molecule interaction to effectively couple different rotational states of the molecule.  The process has a large rate for molecules with anharmonic potentials and large rotational constant $B_v$ (see Eq.~\eqref{eq:E_rot}), such as the hydrogen halides H-X~\cite{Bott1971,Shin1971,Hancock1972,Zittle1973,Wilkins1979,Dash1980,Jurisch1981,Shin1983}. The rate for this process is denoted as $\Gamma_{\rm VR}(v',J')$ and depends on the \textit{partial} pressure of the AB molecules in the gas.   

\end{itemize}

\subsection{Branching Ratios}\label{subsec:BR}
The spontaneous emission rate, together with $V-V$, $V-T$, and $V-R$ transfer rates, determine the branching ratios of photon emission originating from an excited molecule decaying into lower-energy states. In particular, we are interested in calculating branching ratios to obtain $\ngemit$ photons from an initial $(v^\prime,J^\prime)$ state. These values enter $R$, the signal rate per target molecule, see Eq.~\eqref{eq:RoverNtar}. The branching ratio to $\ngemit$ photons, $\text{BR}(v^\prime,J^\prime,\ngemit)$, is the product of branching ratios for emitting single photons, denoted $\text{BR}(v',J')$, that are given by 
\begin{equation}
\text{BR}(v',J')=\frac{A_{v^\prime v^\prime-1}}{A_{v^\prime v^\prime-1} + \Gamma(v^\prime,J^\prime)}\,,
\label{eq:BR}
\end{equation}
where,
\begin{equation}
 \Gamma(v^\prime,J^\prime)=\Gamma_{\rm VT}(v^\prime,J^\prime)+\Gamma_{\rm VV}(v^\prime,J^\prime)+ \Gamma_{\rm VR}(v^\prime,J^\prime)\,.
\end{equation}

\begin{figure*}[t]
\centering
\includegraphics[width=0.48\textwidth]{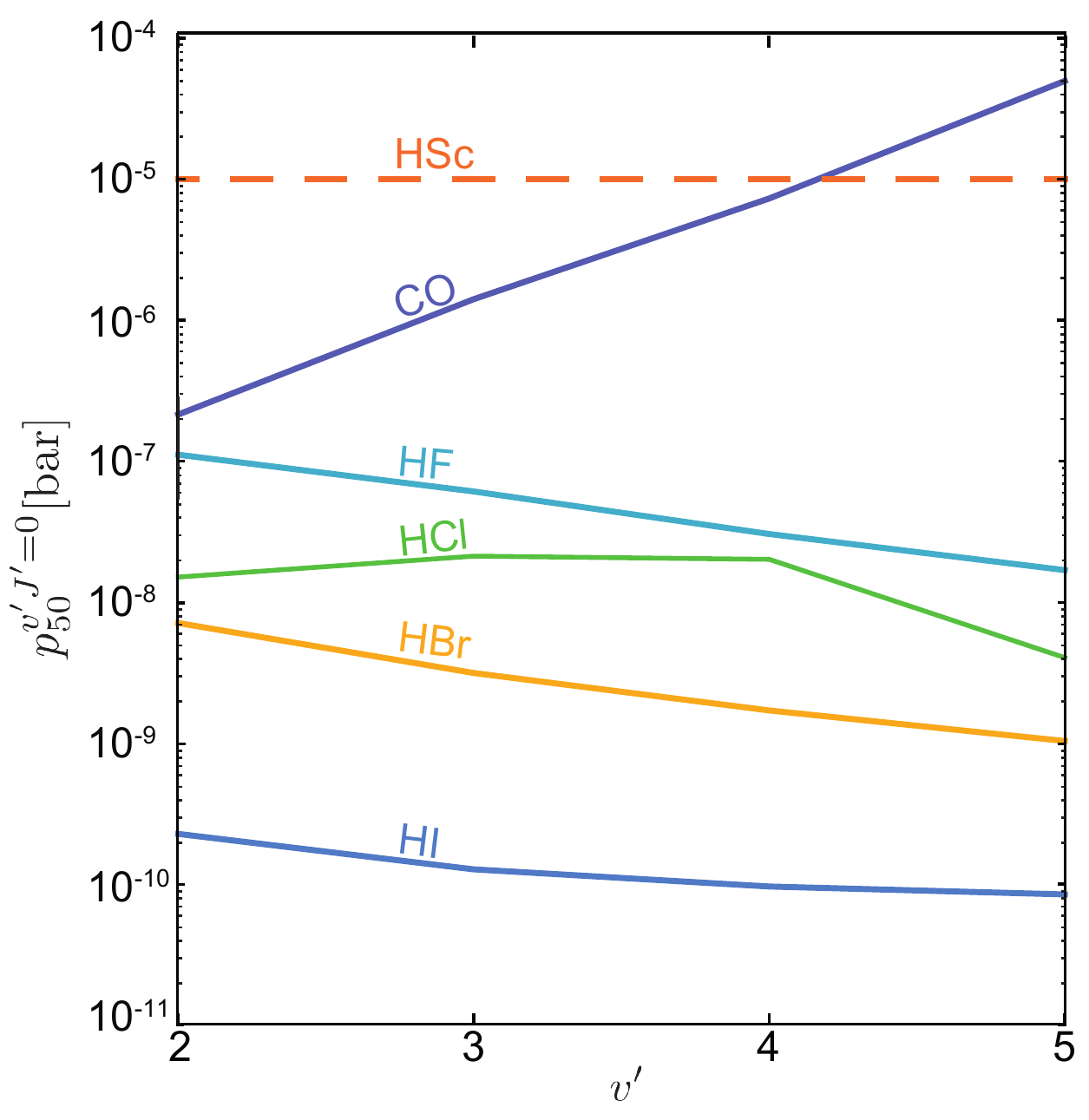}
\includegraphics[width=0.48\textwidth]{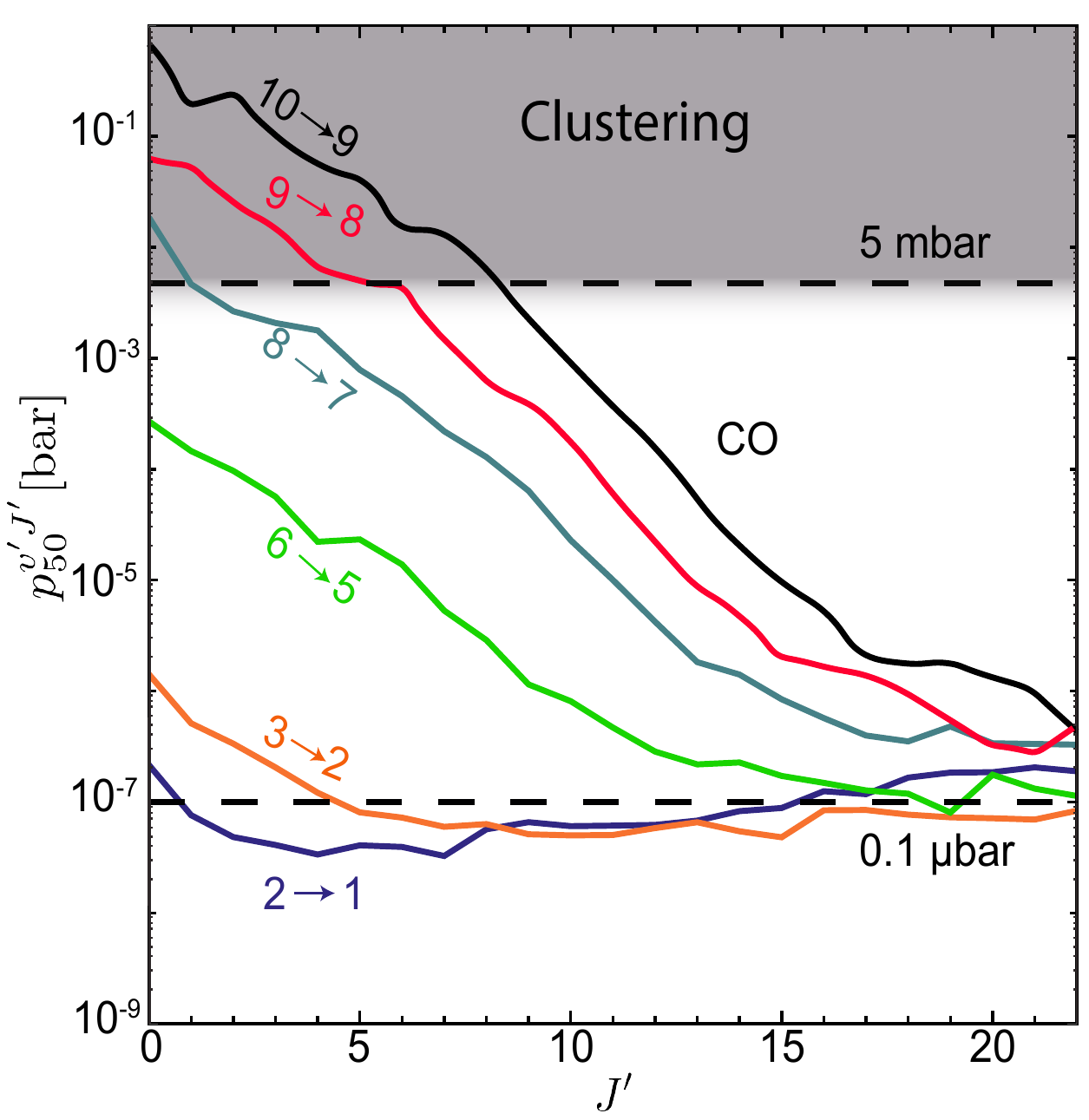}
\caption{{\bf Left:} For various molecular gases, we show the pressure $p^{v',J'=0}_{50}$ as a function of $v'$, for which the branching ratio of the ($v',J'=0)$ state to photons, $\text{Br}(v',J'=0)$, is 50\%.  Below each contour, $\text{BR}(v',J'=0)$ is larger than 50\% for that particular molecule. For halides, $p^{v',J'=0}_{50}$ decreases monotonically as a function of $v'$, while for CO, $p^{v',J'=0}_{50}$ increases monotonically.  For HSc, we show a flat dashed contour (shown in dashed to illustrate that this curve is speculative), valid under the assumption that $V-E$ transfer is extremely efficient. The temperature is chosen to be $T_{\rm BBR}= 55$~K, \{44~K, 72~K, 115~K, 83~K, and 60~K\} for CO \{HSc, HBr, HF, HCl, and HI\}. See text for details.  {\bf Right:} The pressure $p^{v',J'}_{50}$ as a function of $J'$ for various $v'\rightarrow v'-1$ transitions in CO.  The notation on each curve indicates the transitions with $v^\prime \rightarrow v^\prime -1$. The value of $p^{v',J'}_{50}$ should be smaller than the pressure above which the CO molecules begin to cluster, which for CO at $55$~K occurs for pressures above $\sim5$~mbar (gray region).} 
\label{fig:rates}%
\end{figure*}

The product of the signal rate per target molecule and the number of target molecules should be maximized in order to extract the largest signal for a particular DM mass. Roughly, this requires maximizing the individual $\text{BR}(v',J')$. If $\Gamma(v',J')$ dominates over $A_{v^\prime v^\prime-1}$, then the branching ratios in Eq.~\eqref{eq:BR} increase approximately linearly as the pressure decreases, since $\Gamma(v',J')$ increases linearly with pressure. However, decreasing the pressure also decreases the number density of target molecules, which decreases the total DM scattering rate for a given volume.  The maximum sensitivity to DM is achieved by maximizing the product ``pressure $\times$ branching ratio'', which occurs when the branching ratios are $\mathcal{O}(1)$ numbers. 

Importantly, the pressure and temperature of the gas are also coupled. A maximum temperature, denoted $T_{\rm BBR}$, can be chosen by requiring that the number of coincident blackbody radiation photons mimicking a DM signal consisting of two-or-more photons is negligible. We find $T_{\rm BBR}= 55$~K, \{44~K, 72~K, 115~K, 83~K, and 60~K\} for CO \{HSc, HF, HCl, HBr, and HI\}, respectively (see Sec.~\ref{sec:bkg} for details).  For the remainder of this section, we present results for $T=T_{\rm BBR}$.  Furthermore, there is a temperature-dependent maximum pressure, $p_{\rm max}$, above which the target molecules begin to cluster. While it may be possible to produce observable photons from clustered molecules, in this study we consider only a gaseous state. For our candidate molecules, the value of $p_{\rm max}$ is a strictly monotonically increasing function of temperature in our range of interest. Therefore, the pressure that maximizes the number of target molecules for a given volume is obtained by choosing the maximum temperature allowed before being limited by blackbody radiation backgrounds. However, as will be discussed below in Sec.~\ref{sec:optimization}, operating at pressures below $p_{\rm max}$ can provide additional sensitivity to low DM masses, even though the high-mass reach is decreased. For CO, the reason for this increased sensitivity is that low DM masses predominantly produce co-quench photons, which have a longer mean free path for lower pressures. 

Fig.~\ref{fig:rates} presents the pressure for which $\text{BR}(v',J') = 50\%$, denoted as $p^{v',J'}_{50}$. In Fig.~\ref{fig:rates} (left), we show $p^{v',J'=0}_{50}$ as a function of $v'$ for CO, HF, HCl, HBr, HI, and HSc.  Fig.~\ref{fig:rates} (right) shows $p^{v',J'}_{50}$ as a function of  $J'$ for CO for several $v'\to v'-1$ transitions. In both panels, the 
$\text{BR}(v',J')$ for a given molecule is \textit{larger} than 50\% in the region \textit{below} the corresponding line. In other words, for a given pressure, only those $v',J'$ states for which the pressure is smaller than $p^{v',J'}_{50}$ will have a sizable branching ratio to photons. In Fig.~\ref{fig:rates} (right), the gray region has a pressure $\gtrsim$0.5~mbar, for which the CO gas clusters at $T_{\rm BBR}=55$~K.

As can be seen in the figure, resonant quenching rates for CO for small $J^\prime$ increase with decreasing $v^\prime$ (i.e., $p^{v',J'}_{50}$ decreases as $v^\prime$ decreases). This general behavior mostly persists also for larger values of $J^\prime$ and $v^\prime$. The underlying reason for this is an increasing mismatch of energy between the excited state transition to $v^\prime - 1$ and the transition of the ground state to $v=1$. 

Evidently, in Fig.~\ref{fig:rates} (left) a qualitatively different behavior of $p^{v',J'=0}_{50}$ is observed for CO molecules compared to the halides (HF, HCl, HBr, and HI): while $p^{v',J'=0}_{50}$ \textit{increases} for larger $v'$ in CO, $p^{v',J'=0}_{50}$ \textit{decreases} for larger $v'$ in the halides. Moreover, the maximum operating pressure at which one can obtain a sizable photon signal, is much larger for CO than for halides, at least for larger $v'$. For example, HF, HCl, and HBr require pressures of less than about $0.01~\mu$bar to $0.1~\mu$bar, while the maximum pressure in CO can be as large as 0.5~mbar and is set by the need to avoid clustering. The different behavior of $p^{v',J'=0}_{50}$ is due to the different behavior of  $\Gamma(v',J')$ for CO compared to the halides. In particular, as mentioned in Sec.~\ref{subsec:De-Ex_Proc}, the halides studied here suffer from an efficient $V-R$ transfer, which not only efficiently quenches excited states even at moderate pressures, but also efficiently quenches higher $v'$ states at a higher rate than low $v'$ states. On the other hand, for CO, the non-radiative collisional energy transfer rates are smaller; moreover, the quenching rate is dominated by $V-V$ transfer and hence suppressed for higher $v'$ and lower $J'$.  
A detailed discussion of these rates is provided in Appendices~\ref{COappend1} and~\ref{HXappend1}. Since $p^{v',J'=0}_{50}$ is largest for HF among the halides, we will mostly ignore discussing the other halides below.  

For HSc, we have simply drawn a flat (dashed) line; we will discuss the reason for this in Sec.~\ref{subsec:nearby-e-states}. 
We expect that $p^{v',J'=0}_{50}$ is larger for HSc than for the halides and for CO with low $v'$.  

\subsection{Mixing With Nearby Electronic States and $V-E$ Transfer} 
\label{subsec:nearby-e-states}

The presence of an energetically accessible excited electronic state can allow the wavefunctions to mix between an excited vibrational state of the electronic ground state and an excited electronic state, or allow for $V-E$ collisional energy transfer (a vibrational state quenching to a nearby excited electronic state).  For example, this could occur in CO for any $(v,J)$ states with energies larger than that of $(v=26,J=0)$ (the dashed curve in the left panel of Fig.~\ref{fig:RvJ}), and for HSc for any states with energies larger than that of  $(v=2,J=0)$.\footnote{For the halides, the electronic state is above the dissociation threshold.}  While such mixing and $V-E$ transfer can have important consequences for $v^\prime$ above these values, these phenomena are often poorly understood. 

For CO, the lowest-lying excited electronic state, $a^3\Pi$, has an energy of $\sim6$~eV above the ro-vibrational and electronic ground state, $X^1\Sigma^{+}$. The lifetime of the $a^3\Pi$ state that decays to the $X^1\Sigma^{+}$ state is 2.63~ms~\cite{MetastableCO}. It is possible that the DM-induced excitation of a high ro-vibrational states of the electronic ground state $X^1\Sigma^{+}$ can therefore create a single photon of energy $\mathcal{O}$(6~eV) by mixing with the excited electronic state $a^3\Pi$ or by $V-E$ transfer. Moreover, it may be possible to create a photon of energy $\mathcal{O}$(6~eV) together with a number of co-quench photons: the excited molecule might transition to an excited electronic state with a non-zero vibrational quantum number, which decays via photon emission to the electronic ground state with a non-zero vibrational quantum number, which subsequently produces additional co-quench photons. However, these possibilities, in particular the mixing with the nearby electronic state and the $V-E$ transfer rate are only poorly understood~\cite{Flament1992,Schulz1995}. We therefore only consider ro-vibrational states with energies below that of $a^3\Pi$, which is in any case conservative, since the inclusion of higher states would only increase the total expected signal rate.  Nevertheless, we emphasize that additional work to understand these processes is highly motivated, since this could be a way to produce photons of relatively high energies, which would be easier to detect than the infrared photons.

For HSc, the first excited electronic state has an energy of only $\sim0.74$~eV above the ro-vibrational and electronic ground state.  
There is also a $B^1\Pi$ state with an energy of $\sim1$~eV~\cite{RAM1997263,Ram1996}. This $B^1\Pi$ state is believed to have a lifetime of $\mathcal{O}$($10^{-5}$~s), which has been calculated using state-of-the-art {\it ab initio} potential energy curves including spin-orbit couplings in~\cite{Lodi2015,Tennyson_2016}. While not verified empirically or theoretically, it is conceivable that an excited vibrational state with $v>2$ can transition to the first excited electronic state via an efficient $V-E$ transfer.  The subsequent electronic decay will be extremely efficient and produce a photon with an energy of $\sim1$~eV.  
If empirically verified, HSc would be an excellect candidate for DM detection, as it provides a favorable mix of both target mass (hydrogen), a sizable rate to produce observable photons, and photons of higher energy that are easier to detect than the co-quench and cascade photons.

However, HSc is not a thoroughly studied molecule and further information regarding its Einstein coefficients and quenching rates is essential in verifying this (see Appendix~\ref{HXappend1}). While the required quantum dynamics calculations are beyond the scope of this paper, for illustration purposes we assume that the timescale for photon decay from different $v^\prime$ states in HSc are given by the radiative decay rate of the B$^1\Pi$ state. We also model $V-V$, $V-T$, and $V-R$ for this molecule as shown in Appendix~\ref{HXappend1}, finding that the maximum pressure $p_{50}^{v',J'} \sim 10~\mu\text{bar}$, as shown in Fig.~\ref{fig:rates} (left), is far higher than any of the halides. 
Fig.~\ref{fig:rates} (left) shows an HSc curve that is flat as a function of $v'$, since we have assumed that all $v'>2$ states have the same pressure-dependent quenching rates, under the assumption that they would all transition to the same electronic state and decay with the same lifetime.

\section{Signal Generation}
\label{sec:Sig_Gen}

In this section, we discuss the generation of the photon signal in more detail.  We consider processes that produce at least two coincident infrared photons following a molecular excitation. As mentioned above, this is possible in one of two ways. Either the molecule up-scatters to a high $v^\prime$ state and then cascades down emitting multiple cascade photons, or the up-scattered molecule undergoes several $V-V$ transfers to produce multiple $v=1$ states that each decay, producing multiple co-quench photons. 
Co-quench photons are produced only if the $v=1$ state has subdominant $V-T$ and $V-R$ transfer rates compared to its spontaneous  emission rate. This is the case for CO, even when choosing the pressure to be the maximum allowed before the molecules cluster at the maximum temperature allowed before blackbody radiation dominates. 

\begin{figure}[t]
\centering
\includegraphics[width=0.52\textwidth]{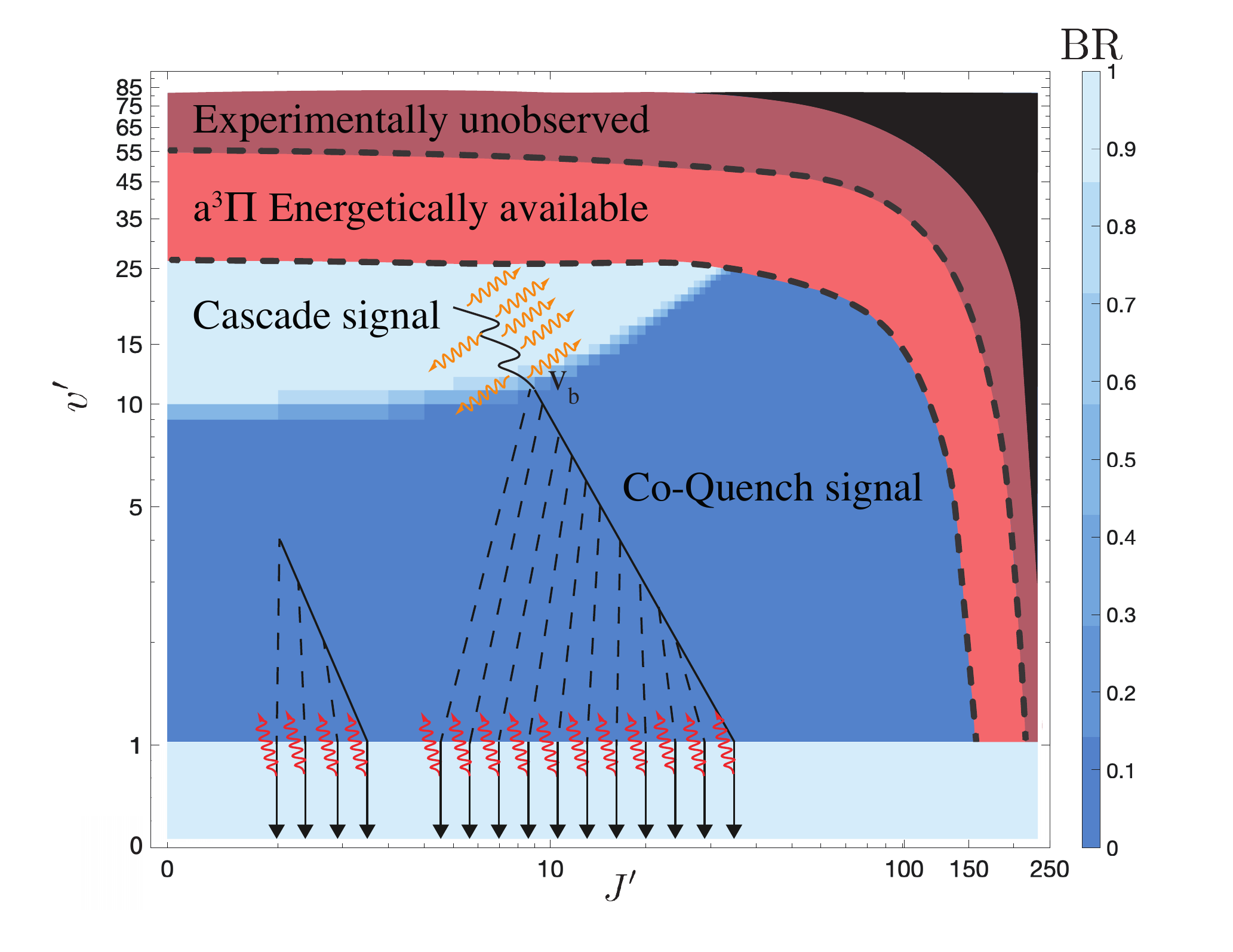}
\caption{
The single-photon branching ratios for a ($v'$, $J'$) state in CO, BR$(v^\prime,J^\prime)$ (see Eq.~\eqref{eq:BR}), and the type of expected signal (cascade or co-quench photons).  We assume a temperature equal to $T_{\rm BBR}= 55$~K and a pressure $p=5$~mbar. The blue shading corresponds to the value of BR$(v^\prime,J^\prime)$ for each $(v^\prime,J^\prime)$ pair. The light blue region corresponds to values of ($v'$,$J'$) for which a cascade signal is possible. The darker blue region corresponds to values for which only a co-quench signal is possible. The red shaded regions are values of ($v'$, $J'$) for which there is not enough information currently available to ensure that a photon signal is produced.  Two example events are shown: excitation to (1) the state $(v^\prime,J^\prime)=(18,7)$ within the cascade region and (2) the state $(v^\prime,J^\prime)=(4,1)$ within the co-quench region. In event (1), the molecule cascades down several steps of size $\Delta v=1$ until reaching $v_{\rm b}$, emitting $v'-v_b$ low-energy photons that are transparent to the medium.  At $v_b$, the molecule resonantly quenches and the remaining vibrational energy of the state is converted into $v_b$ excited $(1,J)$ states, which each emit a photon with a very small mean free path within the medium. Event (2) immediately gets co-quenched to $v'$ excited $(1,J)$ states, which each emit a single photon with small mean free path.}
\label{fig:v_j_Plane}
\end{figure}

To gain intuition, Fig.~\ref{fig:v_j_Plane} shows the single-photon branching ratios for a ($v'$, $J'$) state, BR$(v^\prime,J^\prime)$ in Eq.~\eqref{eq:BR}, as well as the type of expected signal (cascade versus co-quench photons) for a CO target.  We assume a temperature equal to $T_{\rm BBR}= 55$~K and a pressure $p=5$~mbar. The blue shading corresponds to the value of BR$(v^\prime,J^\prime)$ for each $(v^\prime,J^\prime)$ pair. Evidently there is a sharp transition from BR$(v^\prime,J^\prime)\approx1$ to BR$(v^\prime,J^\prime)\approx0$ corresponding to the transition from cascade to co-quench signals.
The $J$-dependent value of $v$ at this transition is denoted $v_{\rm b}$. Importantly, as discussed below, this transition curve depends very sensitively on pressure. The red-shaded regions above the dashed curves correspond to states that either energetically overlap with the excited electronic states or have not been observed experimentally.  We conservatively do not include the DM-induced scattering rates to these 
$(v^\prime,J^\prime)$ states in our DM sensitivity estimates. 

The co-quench signal is absent for our candidate halides and HSc, since $V-T$ and $V-R$ transfers dominate over $V-V$ transfer for all pressures and $v'$, as is further explained in Appendices~\ref{sectMol}, \ref{COappend1}, and \ref{HXappend1}.  
The $V-T$ and $V-R$ transfer processes quench high $v'$ states to lower $v$, until the transfer rates are slower than the spontaneous emission rate, at which point cascade photons are produced.  The decay chains for the halides and HSc are therefore rather simple, and we focus our discussion on CO for the remainder of this section. 

\subsection{Cascade Signal}

To produce two or more photons of this type in CO, the molecule must be excited to a state with $v^\prime \geq 3$ and then cascade decay in multiple $\Delta v=1$ steps via spontaneous emission.\footnote{If the photodetectors have zero, or very low, dark counts, one could also consider excitations to the $v'=2$ state that produces a single photon.  However, in this study, we do not consider this case for a CO target.} This proceeds efficiently as long as the pressure is low enough such that all collisional quenching mechanisms are subdominant, i.e, 
\bea
\Gamma(v') & \lesssim & A_{v' v'-1}\,, \nonumber \\ 
& \vdots & \nonumber \\
\Gamma(v_{\rm b}+1) & \lesssim & A_{v_{\rm b}+1 v_{\rm b}},
\label{eq:res_col_constraints}
\eea
where the $J$ indices in $\Gamma(v, J)$ have been suppressed for brevity. 

For a large range of $J^\prime$ and above some value of $v^\prime$ (see Fig.~\ref{fig:v_j_Plane}), the excited molecule will spontaneously emit photons until it reaches $v_{\rm b}$, producing $(v^\prime  - v_{\rm b})$ cascade photons as follows: 
\begin{align}
&{\rm AB}(v') \rightarrow {\rm AB}(v'-1) + \gamma_{v' \rightarrow v'-1} \nonumber \\ &\rightarrow {\rm AB}(v'-2) + \gamma_{v' \rightarrow v'-1} + \gamma_{v'-1\rightarrow v'-2} \rightarrow ... \nonumber \\
& \rightarrow {\rm AB}(v_{\rm b}) +(v' -v_{\rm b}) \cdot \gamma\,.
\end{align}
Each photon, $\gamma_{v,v-1}$, has a slightly different energy, depending on the various values of $(v,J)$ in the cascade chain 
(the $J$-dependence on the $\gamma$ indices are suppressed). 
The medium is almost completely transparent to these photons, since the photon energies are smaller than that of any possible 
vibrational excitation from the ground state. 
The exact decay chain, i.e., the exact values of $(v,J)$ along the chain, depends strongly on the BR($v,J$) for spontaneous emission from each $(v,J)$ state in the chain to $(v-1,J\pm1)$. 

As the pressure decreases, the values of $v_{\rm b}$ (which determine the boundary between the cascade and co-quench regions) decrease. One can therefore create a cascade signal with a lower energy threshold by decreasing the pressure.  This provides enhanced sensitivity to lower DM masses at the expense of a smaller number of targets (for a given volume). In order for the lowest possible DM mass to produce at least two cascade photons in CO, $v^\prime = 3$ must be excited, which requires a pressure of $\sim 50$~nbar.

As an example of the behavior of the cascade signal for a different molecule, for HF, the signal always consists of cascade photons, but the pressures required to avoid collisional quenching for all $v'$ states are much smaller than for the case of CO.  For example, to produce the largest number of cascade photons for $v' \leq 5$ from DM scattering, the pressure is 100~nbar for HF. We discuss this further in Sec.~\ref{sec:optimization}. 

\subsection{Co-Quench Signal}

Co-quench photons are produced if (i) the $V-V$ transfer rate of the excited state is larger than the spontaneous emission rate and the other collisional energy transfer rates, and (ii) if the spontaneous emission rate of the $v=1$ state is larger than the collisional energy transfer rates,  $\Gamma_{\rm VT}(1)+\Gamma_{\rm VR}(1) \lesssim A_{10}$. For CO at 55~K, this condition is trivially satisfied for pressures below the clustering pressure. 

For CO, the co-quench region is shown in the dark blue region of Fig.~\ref{fig:v_j_Plane}, corresponding to $(v', J')$ for which $V-V$ transfer is extremely efficient.  
This occurs for low $v'$ due to the almost harmonic nature of the nuclear potential. 

A CO molecule can be excited directly to $v^\prime<v_{\rm b}$ or can be excited to a higher value of $v'$ and cascade down to $v_b$. Denoting the first vibrational mode that produces a co-quench photon by $v_{\rm min,b} \equiv \text{min}(v^\prime,v_{\rm b})$, the process can be written as
\begin{align}
&\text{AB}(v_{\rm min,b}) + v_{\rm min,b} \ \text{ AB}(0) \rightarrow \text{AB}(v_{\rm min,b}-1) \nonumber \\
&+ \text{AB}(1) + (v_{\rm min,b}-1) \ AB(0) + E_{k,i} \rightarrow \nonumber \\
&... \rightarrow v_{\rm min,b} \ \text{AB}(1,J_{{\rm fin},i}) + E_k^{\rm tot} \nonumber \\
&\rightarrow v_{\rm min,b} \ \gamma + v_{\rm min,b} \ AB(0) + E_k^{\rm tot}\,.
\label{eq:Coll_Quenching_Process}
\end{align}
Here, $J_{{\rm fin},i}$ is the final rotational quantum number of each molecule $i$ that was excited during a collision.  The AB(0) molecules are vibrational ground state molecules in the gas, with rotational values corresponding to the Maxwell-Boltzmann distribution.  For notational simplicity, we do not write the $J$-values during the intermediate steps, and have also suppressed the $J_{\rm init}$ symbol. $E_k^{\rm tot}$ is the small amount of residual kinetic energy released, and accounts for the excitation energies of the AB(0) molecules not being precisely equal to the de-excitation energies of the excited AB molecule.  The $v_{\rm min,b}$ molecules in the state ${\rm AB}(1)$ have a timescale of $\sim$ms to decay to the ground state, releasing $v_{\rm min,b}$ coincident photons with equal energies $\Delta E_{1,0}$.

The co-quench photons typically have a very small mean free path because their energies are close to those needed to excite the ubiquitous ground state molecules in the gas. 
Since they provide the best sensitivity to DM with the lowest masses in CO, we will present a detailed discussion of techniques to maximize both their mean free path and the photon collection efficiency in Sec.~\ref{sec:photon-collect}.  

\section{Photon Collection and Detection}
\label{sec:photon-collect}

Once the signal photons are emitted, they must be efficiently collected and detected. However, the photons could be absorbed by the gas or walls of the tank, or fail to be detected due to the photodetector's efficiency, $\epsilon_{\rm det}(\ngcol)$, being below unity. This section discusses how one can minimize some of these losses and detect the emitted photons. Furthermore, the section presents the final ingredients required to calculate the observed signal rate per unit molecule in Eq.~\eqref{eq:RoverNtar}, namely $\epsilon_{\rm col}(\ngemit,\ngcol)$ and $\epsilon_{\rm det}(\ngcol)$.  

\subsection{Mean Free Path}
\label{subsec:MFP}

The probability of absorbing a co-quench or cascade photon can be quantified in terms of the mean free path of the photons in the gas.  The mean free path depends on the molecules in the gas that could absorb a photon. 
For the molecules considered in this study, at a temperature of $T_{\rm BBR}$, 
the molecules are dominantly in a state with $v=0$ and $J_{\rm init}\lesssim10$, see Eq.~\eqref{eq:Pinit_MB}. 
The absorption cross section is given by 
\begin{equation}
\sigma_{\text{abs}}(\omega)= \omega B_{0 J_{\rm init} \rightarrow v_jJ_j} g(\omega)\,,
\end{equation}
where $\omega$ is the photon frequency, $B_{0 J_{\rm init} \rightarrow v_jJ_j}$ is the Einstein absorption coefficient from the $(v=0, J=J_{\rm init})$ to $(v_j,J_j)$ state, and $g(\omega)$  is the line profile of the absorption process. 
The line profile receives contributions from both collisional broadening and Doppler broadening, which are well described by a Lorentzian profile (with scale parameter $\gamma_{\rm el}$) and a Gaussian profile (with standard deviation $\sigma_{\rm Dop}$), respectively.  Therefore, the line shape takes the form of a Voigt profile, $g(\omega-\omega_{\rm p})$, which peaks at the resonant frequency $\omega_{\rm p}$ and is given by 
\beq
g(\omega-\omega_{\rm p})=\int\frac{\gamma_{\text{col}}}{\pi [\gamma_{\rm col}^2+(\omega-\omega_{\rm p}-\omega')^2 ]}\frac{e^{-\omega'^2/(2\Delta_{\rm Dop}^2)}}{\sqrt{2\pi}\Delta_{\rm Dop}} d\omega'\,.
\label{eq:voigt}
\eeq
Here, $\gamma_{\rm col}$ and $\Delta_{\rm Dop}$ are given by 
\begin{eqnarray}
\gamma_{\rm col} & = & n_{\rm col} \left< \sigma_{\rm col} v \right>\,, \nonumber \\
\Delta_{\rm Dop} & = & \sqrt{\frac{T}{M_{\rm mol}}} \omega_{\rm p}\,,
\label{eq:Col_Dop_Width}
\end{eqnarray}
where $M_{\rm mol}$ is the molecular mass, $n_{\rm col}$ is the number density of particles that contribute to elastic collisional broadening, and $\left< \sigma_{\rm col} v \right>$ is the velocity averaged pressure broadening cross section.\footnote{If multiple species contribute to $n_{\rm col}$, their number densities should be summed.} The mean free path is then 
\beq
\lambda_{\rm MFP}(\omega) = \frac{1}{n_{{\rm tar},J_{\rm init}}\cdot\sigma_{\text{abs}}(\omega)}\,,
\label{eq:MFP}
\eeq
where $n_{{\rm tar},J_{\rm init}}$ is the number density of molecules in the state that can absorb the photon.  

For cascade photons, the photons are not on resonance with any molecule in the gas. Moreover since $\omega_p-\omega \gg \gamma_{\rm col}  , \Delta_{\rm Dop}$, only the tail of the line-shape contributes, leading to extremely large $\lambda_{\rm MFP}$.
We find that, for cascade photons in CO, $\lambda_{\rm MFP}$ can essentially be taken as infinite, and instead the efficiency of the reflective coating in a large-volume tank will be the limiting factor determining how many cascade photons are observed. 

For co-quench photons in CO, however, the photons are on resonance with the vibrational ground state.  This leads to a small $\lambda_{\rm MFP}$. For typical values of the pressure and temperature, and for a gas comprised of $^{12}$C$^{16}$O molecules only, $\lambda_{\rm MFP}\ll1$~cm. Therefore, if these photons are to be collected in an experiment, the mean free path must be increased. This can be done through collisional broadening with the addition of a buffer gas such as He. 

\begin{figure*}[t]
\centering
\includegraphics[width=0.49\textwidth]{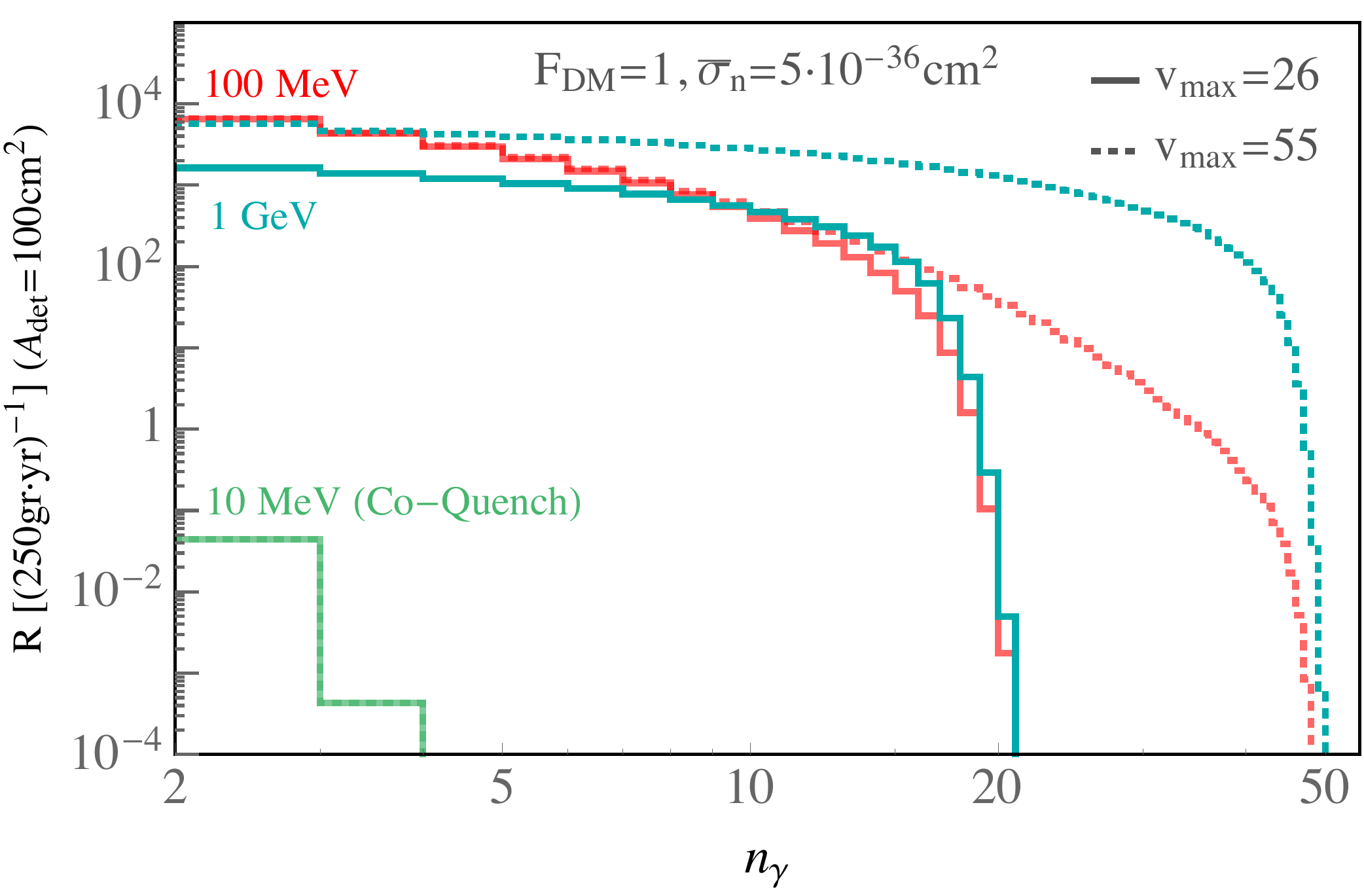}
\includegraphics[width=0.49\textwidth]{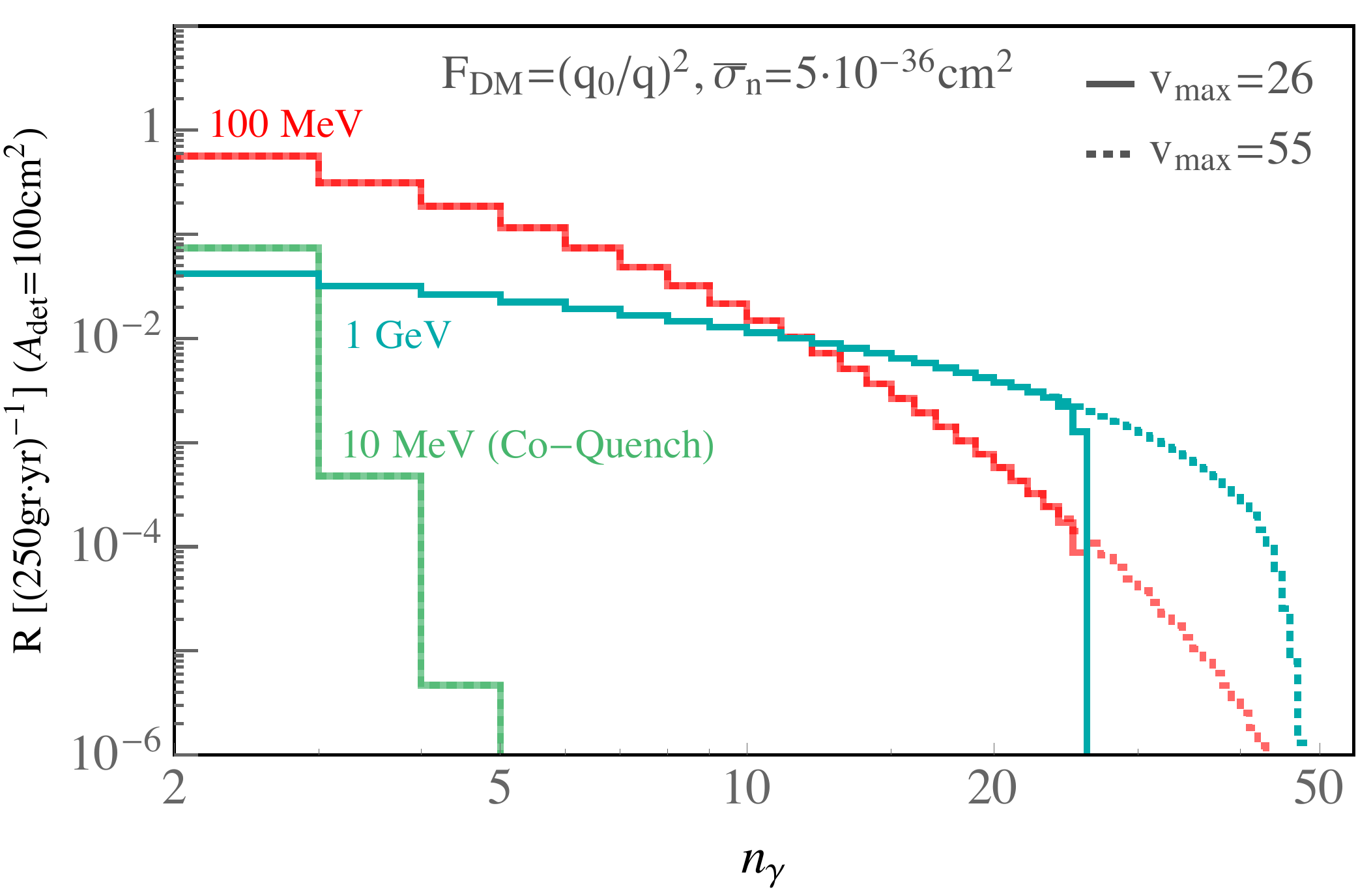}
\caption{Cascade photon spectrum for $m_\chi = 100$ MeV and $1$ GeV and co-quench spectrum for $m_\chi = 10$ MeV. The rate is given for CO at $55$ K with $\sigmap = 5\times 10^{-36}$ cm$^2$, for $p=5$~mbar, $\epsilon_{\rm abs} = 10^{-5}$, and $\epsilon_{\rm det}(\ngcol)=1$. The tank has volume $(2\text{m})^3$ and a photo-detector surface area of $A_{\rm det}=100$ cm$^2$. The volume, temperature and pressure correspond to a total CO mass of $250$~gr. The \textbf{left} panel corresponds to $\FDM=1$ and the \textbf{right} panel to $\FDM=(q_0/q)^2$ with $q_0=\Delta p_e \approx 28$ keV from Eq.~\eqref{eq:Delta_p0}. The cascade signal is always much larger than the co-quench signal since co-quench photons are not transparent to the medium. On the other hand, co-quench signals are able to access much lower DM masses since they correspond to a lower energy threshold. The co-quech result for $m_\chi=10$ MeV corresponds to a setup with an internal reflective cylinder based on the photo-detector with length twice the mean free path, within the main tank. Solid curves in the figure correspond to values of $(v,J)$ with energies equal to or below that of $(v,J)=(26,0)$, i.e. below electronic energy levels. For illustration, if higher modes are found to be experimentally accessible, the dotted curves correspond to maximal values of $(v,J)$ equal or below the energy of $(55,0)$. In this case, many more photons are observed.}
\label{fig:Spectrum}
\end{figure*}

An He buffer gas at high enough pressure can dominate the collisional broadening scale parameter such that $n_{\rm col}$ from Eq.~(\ref{eq:Col_Dop_Width}) becomes the He number density. In this case, the Voigt profile simplifies to a Lorentzian, and for 
$\omega \sim \omega_{\rm p}$ is given by,
\begin{equation}
g(\omega-\omega_{\rm p}=0)=\frac{1}{\pi \gamma_{\rm col}}\,.
\end{equation}
Thus, Eq.~\eqref{eq:MFP} simplifies to,
\begin{equation}
\lambda_{\rm MFP}=\frac{n_{\rm He}}{n_{\rm CO}} \frac{\left<\sigma_{col} ^{\text{CO-He}} v\right>}{\pi \omega B_{ij}}  \propto \frac{p_{\rm He}}{p_{\rm CO}}.
\label{eq:lammfp}
\end{equation}
Clearly, $\lambda_{\rm MFP}$ increases with increasing $p_{\rm He}$ due to collisional broadening, and decreases with increasing $p_{\rm CO}$ due to the larger number of absorbers.

There exists a maximal He pressure above which the $v=1$ state of CO quenches too fast via $V-T$ transfer. This pressure is $p_{\rm He}^{\rm max}\approx4$~bar at 55~K. Taking this maximum pressure for the He buffer gas and the maximal allowed CO partial pressure of $p_{\rm CO}=5 \text{ mbar}$, the mean free path for the $v=1\to0$ photons would be $\lambda_{\rm MFP}\approx0.8$~cm if all the CO molecules act as absorbers. However, different $J$ levels are mutually transparent and, at $55$~K, using Eq.~\eqref{eq:Pinit_MB}, not more than $18\%$ of CO molecules are in any particular $J_{\rm init}$ level. Additionally, one may introduce different isotope combinations of ($^{12}$C,$^{13}$C) and ($^{16}$O, $^{17}$O, $^{18}$O), which are also mutually transparent. This further decreases the number density of absorbers of any particular photon. With these modifications, one can achieve a mean free path of $\lambda_{\rm MFP} \approx 27~\text{cm}$.

\subsection{Collection Efficiency}
\label{subsec:Det_Geometry}

In what follows, we explicitly calculate the collection efficiency, $\epsilon_{\rm col}(\ngemit,\ngcol)$ from Eq.~\eqref{eq:RoverNtar}, i.e., the efficiency with which the emitted photons are collected within the photodetectors.

\subsubsection{Cascade Signal}

Due to their large mean free path, cascade photons cannot reach the photodetector only if they get absorbed by the non-instrumented, mirrored walls of the tank. For simplicity, we model the gas tank as a cube with sides $L$, as depicted in Fig~\ref{fig:Setups}. Photons  emitted within the tank reflect off the mirrors until they reach the photodetector which has a surface area $A_{\rm det}$. During each reflection there is some probability for the photon to get absorbed, $\epsilon_{\text{abs}}$.\footnote{In a realistic setup, this absorption probability can depend on many factors such as photon wavelength and incidence angle. Here we take the probability to be constant. This can be thought of as an absorption efficiency averaged over all relevant parameters.} For a single photon, the probability of hitting the detector area after exactly $n$ bounces is 
\begin{equation}
P_n = (1-\epsilon)^n \left(1-\frac{A_{\text{det}}}{6L^2}\right)^n \frac{A_{\text{det}}}{6L^2}\,.
\end{equation}
Summing over all $n$ from zero to infinity, we get 
\begin{equation}\label{eq:cas}
P_{\rm col,cas} = \frac{A_{\text{det}}}{A_{\text{det}}+6L^2 \epsilon}\,.
\end{equation}

The efficiency for $\ngcol$ cascade photons to be collected by the photodetectors when $\ngemit$ photons are emitted is
\bea
\epsilon_{\rm col,cas}(\ngemit,\ngcol)  & = & ^{\ngemit}C_{\ngcol} (P_{\rm col,cas})^{\ngcol} \times \nonumber \\
& & (1-P_{\rm col,cas})^{\ngemit-\ngcol} \,,
\label{eq:pgivennspec}
\eea
where $^n C_r$ is the combinatorial factor. The efficiency for two or more photons to make it to the photodetector is 
\bea
\epsilon_{\rm col,cas}(\ngemit,\ngcol\geq2) & = & 1-(1-P_{\rm col,cas})^{\ngemit} \nonumber \\
&  & -\ngemit (1-P_{\rm col,cas})^{\ngemit} P_{\rm col,cas}. \nonumber \\
\label{eq:pgivenn}
\eea
The efficiency from the above equation should be used for calculating the total cascade signal rate from Eq.~\eqref{eq:RoverNtar}, since all events with two or more collected photons are included in the signal.

On the other hand, if one is interested in calculating the photon spectrum (the rate of events with $n_\gamma$ photons) for some DM mass, the efficiency of Eq.~\eqref{eq:pgivennspec} must be used within Eq.~\eqref{eq:RoverNtar}. Fig.~\ref{fig:Spectrum} presents the expected cascade photon spectrum for CO, for $m_\chi = 100$~MeV and $1$~GeV and for two different DM form factors corresponding to a heavy (left panel) and a light (right panel) mediator. The temperature is taken to be $T_{\rm BBR}=55$~K, the reference cross section $\sigmap = 5\times 10^{-36}$~cm$^2$, a pressure of 5~mbar, a mirror absorption efficiency of $\epsilon_{\rm abs} = 10^{-5}$, a detector efficiency of $\epsilon_{\rm det}(\ngcol)=1$, a tank volume of $(2 \text{m})^3$ and a photodetector surface area of $A_{\rm det}=100$~cm$^2$. The chosen parameters correspond to a total CO mass of $250$~gr. For the solid curves, only those $(v,J)$ states with energies less-than-or-equal-to that of $(v,J)=(26,0)$ are included, i.e., below the first excited electronic state.  For illustration, the dotted curves show the expected photon spectrum from all $(v,J)$ states with energies less-than-or-equal-to that of $(v,J)=(55,0)$; however, as discussed in Sec.~\ref{subsec:nearby-e-states}, the presence of an electronic state may drastically reduce the cascade photon signal from the higher-lying states. 

We note that only very few co-quench photons make it to the detector and therefore have not been included for $m_\chi =100$~MeV or $m_\chi =1$~GeV. In contrast, the photon spectrum for $m_\chi = 10$~MeV in Fig.~\ref{fig:Spectrum} consists entirely of co-quench photons, which will be discussed next.

\subsubsection{Co-quench Signal} 
\label{sec:photoncollect}

When searching for cascade photons from DM, a larger volume at fixed pressure and temperature will usually produce a higher observed DM signal rate (we will discuss varying experimental parameters in Sec.~\ref{sec:optimization}). In contrast, only those co-quench photons produced close to the photodetector of area $A_{\rm det}$ have a chance of being observed. Therefore, for a fixed set of experimental parameters, increasing the tank volume beyond some value no longer increases the co-quench signal. Parametrically, the ideal tank dimensions that maximize the observed number of co-quench photons for a particular DM candidate is a cylinder with a base area set by $A_{\rm det}$ and a length set by $\lambda_{\rm MFP}$ (this statement will be refined below).

For a single co-quench photon, produced within such a cylinder at some vertical distance $z$ from the photodetector, the probability of detection (averaging over all emission angles) is given by 
\beq
P_{\rm col,co}(z)=\frac{1}{2}\, f\left (\frac{z}{\lambda_{\rm MFP}}\right)\,,
\eeq
where $f(x)=e^{-x}- x\ \Gamma(0, x)$ and $\Gamma$ is the incomplete gamma function. The probability of collecting two or more co-quench photons, $\epsilon_{{\rm col,co},z}(\ngemit,\ngcol\geq2,z)$, when $\ngemit$ photons are emitted from some point within the cylinder, is then simply Eq.~\eqref{eq:pgivenn} with $P_{\rm col,cas} \rightarrow P_{\rm col,co}(z)$ (the result for any value of $n_{\gamma,{\rm col}}$ is given by making the same replacement in Eq.~\eqref{eq:pgivennspec}). The result is $z$-dependent and must be averaged over the length of the cylinder.  
The number of photons that make it to the detector drops rapidly as $z$ increases. Numerically, we find that even if $\ngemit$ is as large as 50, essentially no photons make it to the photodetectors from distances greater than $\sim$2$\lambda_{\text{MFP}}$. 
The ``ideal'' cylinder volume is then simply 
\begin{equation}
V_{\rm co}= 2 A_{\rm det} \lambda_{\rm MFP}\,. 
\label{eq:vco}
\end{equation}
For this volume, and taking all the signal photons to originate from the same point, $z$, in the cylinder,\footnote{This is a good assumption, since the 
molecules do not diffuse far before they decay to produce the co-quench signal.} the collection efficiency is
\begin{eqnarray}
&  \epsilon_{\rm col,co}(\ngemit,\ngcol\geq2)  = \frac{1}{2\lambda_{\rm MFP}}  \times  \nonumber \\
& \int_0^ {2\lambda_{\rm MFP}} [ 1-(1-P_{\rm col,co}(z))^{\ngemit} \nonumber \\
& \hspace{2.3cm} \ngemit (1-P_{\rm col,co}(z))^{\ngemit} P_{\rm col,co}(z) ] dz \nonumber \\. 
\end{eqnarray}
We show the expected co-quench photon spectrum in Fig.~\ref{fig:Spectrum} for $m_\chi=10$~MeV for a cylinder with a base area 
of $A_{\rm det}=(10\text{ cm})^2$ and a cylinder height equal to $2\lambda_{\rm MFP}\simeq 54~\text{cm}$. 

We note that the tank volume that maximizes the co-quench signal in Eq.~\eqref{eq:vco} is limited by $A_{\rm det}$ and thus by the challenge of constructing a large photodetector array.  In contrast, it is comparatively easy to construct a larger tank volume (keeping $A_{\rm det}$ fixed), which usually increases the observed cascade signal. Ideally, however, one would build a single experiment that is optimized to search for both cascade and co-quench signals. One way to do this is to construct a large tank as depicted in Fig.~\ref{fig:Setups} and additionally place a half-open cylinder within the tank, with the dimensions discussed above and thus a volume given by Eq.~\eqref{eq:vco}. The inside of the closed base of the cylinder is instrumented with the photodetector array, and all other cylinder walls (inner and outer) are covered in reflective coating. 

\begin{figure*}[ht]
\centering
\includegraphics[width=0.49\textwidth]{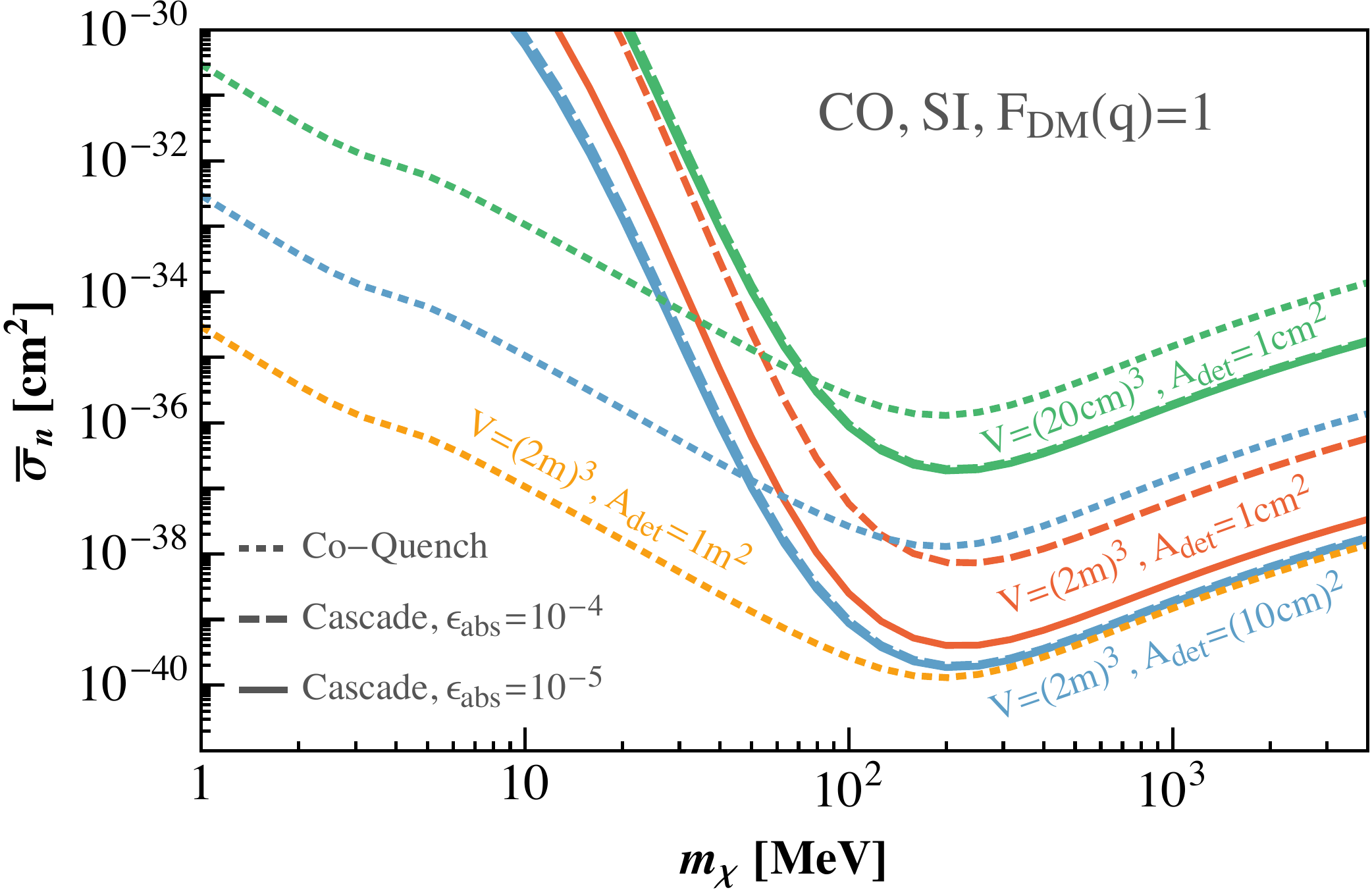}
\hspace{0.0cm}
\includegraphics[width=0.49\textwidth]{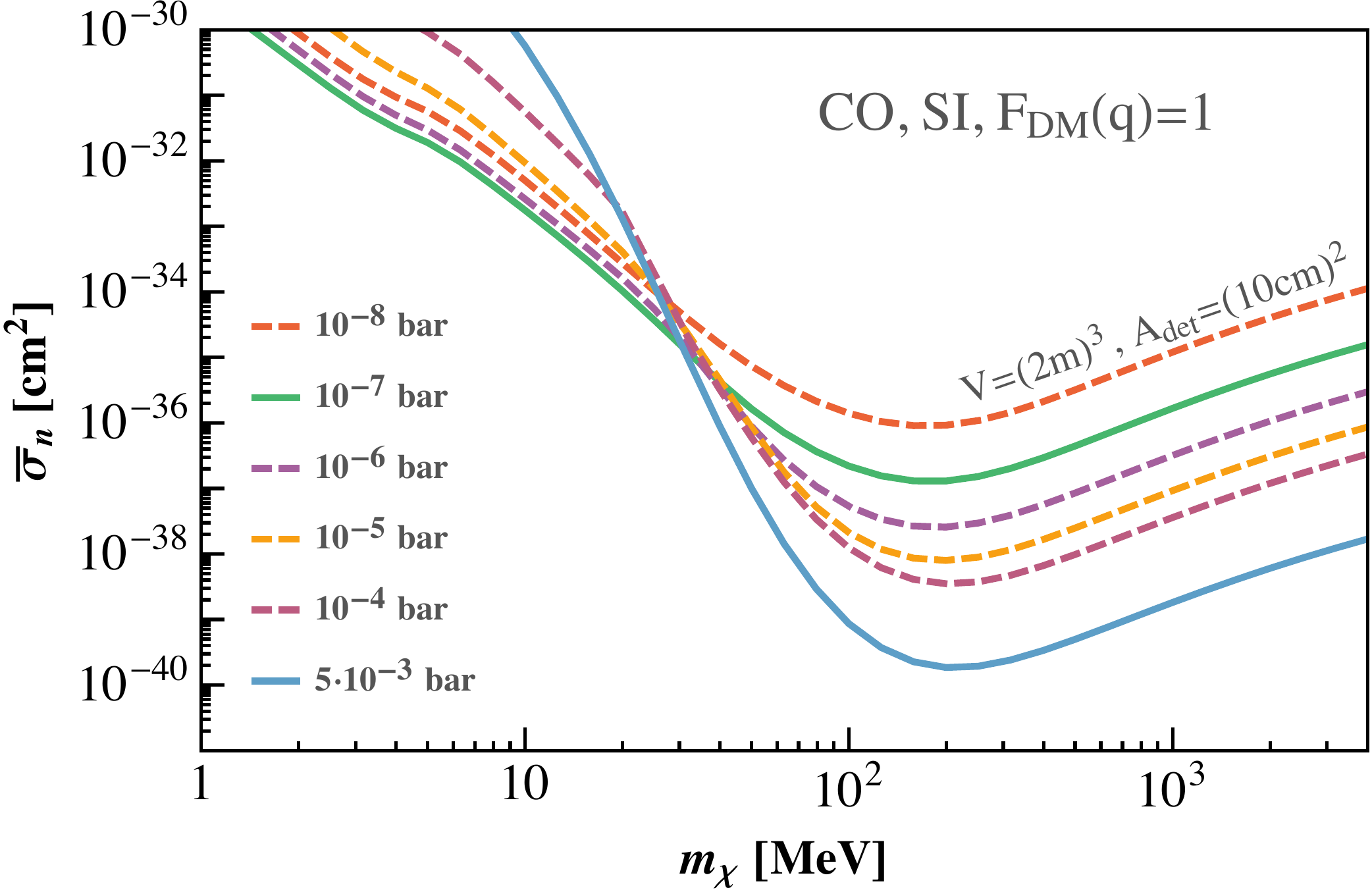}
\caption{{\bf Left:}  The sensitivity to DM spin-independent nuclear scattering off CO molecules for different target volumes, $V$, different photodetector areas, $\text{A}_{\rm det}$, and different photon absorption coefficients for the mirrors, $\epsilon_{\rm abs}$.  The sensitivities to signals of cascade photons are shown with dashed and solid curves for a fixed pressure of $5$~mbar, while the sensitivities to signals of co-quench photons are shown in dotted curves and are independent of the partial pressure of CO. {\bf Right:}  The sensitivity to DM spin-independent nuclear scattering off CO molecules for various values for the pressure and for signals of cascade photons only.}
\label{fig:pandabsdep}%
\end{figure*}

\subsection{Photon Detection}
The ideal sensor with which to detect the collected photons should have a zero (or very small) intrinsic dark count, a high detection efficiency, and be able to cover a large area. The last few years have seen rapid progress in several technologies that may be suitable for detecting these infrared photons. The technologies include the Superconducting Nanowire Single Photon Detector (SNSPD)~\cite{SEMENOV2001349,Goltsman2001,Natarajan2012}, the Microwave Kinetic Inductance Detector (MKID)~\cite{Mazin:12}, and the transition edge sensor (TES)~\cite{Irwin:2005,Pyle:2015pya}.  All three devices are expected to have intrinsically low dark counts and a high efficiency of detecting the photon, although more work is required to quantify the ultimate limits of these devices. Additional work is also required to increase their detection areas, either by scaling up single devices in size or by constructing large arrays of small devices.  Since these devices currently operate at very low temperatures in comparison to the gas temperature envisioned in this study, one challenge will be coupling these devices to the gas tank. This could be done, for example, by developing detectors that work at higher temperatures, or by connecting optical fibers to the tank that transport the light from the tank to photodetectors within a colder environment~\cite{private-Berggren-Nam-Shaw}.

We will derive the DM projections below assuming a single-photon detection efficiency of $\epsilon_{\rm det}(\ngcol)=1$ (see Eq.~\eqref{eq:RoverNtar}).  Moreover, we will present results for three different photodetector areas, $A_{\rm det}=1$~cm$^2$, $A_{\rm det}=(10~{\rm cm})^2$, and $A_{\rm det}=1~{\rm m}^2$.  
Based on preliminary discussions regarding the use of an array of SNSPDs as photodetectors~\cite{private-Berggren-Nam-Shaw}, this spans the range between ``definitely achievable'' to ``ambitious, but possible''.

\section{Experimental parameters}
\label{sec:optimization}
In this section, we discuss how the sensitivity to different DM masses changes as one varies the volume of the tank, the pressure of the molecular gas and the absorption efficiency of the mirrors. These quantities all enter the observed signal rate through Eq.~\eqref{eq:RoverNtar}. The cross section $\left\langle \sigma v_\chi\right\rangle_{v^\prime J^\prime}$ depends on the DM mass and, as was shown in Sec.~\ref{sec:Exp_Considerations}, larger DM masses preferentially excite higher $v^\prime$ and $J^\prime$ states.  The pressure dictates $\text{BR}(v^\prime,J^\prime,\ngemit)$, while pressure, $p$, and tank volume, $V$, together determine the number of targets, $N_T$, through 
\begin{equation}
N_T=  \frac{p}{1 \text{ bar}}\frac{273 \text{ K}}{T}\frac{V}{22~\text{ liter}} \text{ mol}\,.
\end{equation}
Pressure, volume, and mirror reflectivity will determine $\epsilon_{\rm col}(\ngemit,\ngcol)$. Below, we also discuss the effects of varying the target molecule.

\subsection{Cascade Signal}

While the target volume increases as a function of the length, $L$, of the side of a cubic tank, Eq.~\eqref{eq:cas} shows that the probability of detection decreases with increasing $L$.  As a result, these effects compete with each other and there exists some optimal value of $L$ which maximizes the overall signal and depends on $\epsilon_{\text{abs}}$ and $\ngemit$. To illustrate this, we consider a few concrete setups for CO at~55 K and 5~mbar. 

First, we consider a prototype setup consisting of a tank with volume $V=(20 \text{ cm})^3$ and detector area $A_{\rm det}=1~{\rm cm}^2$, and assume $\epsilon_{\text{abs}}=10^{-4}$.  With high probability, all emitted photons will be detected in this prototype. The prototype's sensitivity is shown for the cascade signal by the dashed-green curve in Fig.~\ref{fig:pandabsdep} (left) for $\FDM=1$, assuming $p=5~\text{mbar}$ (we will discuss the co-quench signal below). We find a nearly identical DM sensitivity for $\epsilon_{\text{abs}}=10^{-5}$ (solid-green curve), illustrating that $\epsilon_{\text{abs}}=10^{-4}$ is sufficient for this prototype tank.

For a tank with $V=(2\text{ m})^3$, two or more photons will only be observed if a large number of photons are emitted. Despite having a volume that is a factor of $10^3$ times larger than the prototype detector volume, the increase in sensitivity is far less significant, as can be seen in Fig.~\ref{fig:pandabsdep} (left) comparing the dashed-green and dashed-red curves.  However, better mirrors with, e.g., $\epsilon_{\rm abs}=10^{-5}$ (solid-red), and/or a larger detector area of $A_{\rm det} = (10~{\rm cm})^2$, improve the sensitivity dramatically (solid- and dashed-blue). Note that, for this large tank volume and for $\epsilon_{\rm abs}=10^{-5}$, there is little improvement in sensitivity when $A_{\rm det}$ is increased from $1\text{ cm}^2$ to $(10\text{ cm})^2$ since the collection efficiency is already $\mathcal{O}(1)$ for both these setups. Any additional increase in $A_{\rm det}$ will result in essentially no further improvement in reach unless $\epsilon_{\rm abs}$ also increases. Finally, we note that for $\FDM\propto 1/q^2$, as well as for the halides and HSc, we expect the DM sensitivity to show slightly greater dependence on $\epsilon_{\rm abs}$ and $A_{\rm det}$.  The reason for this is that fewer photons are typically produced in a DM scattering event. 

In Fig.~\ref{fig:pandabsdep} (right), we show how the DM sensitivity varies with pressure when using only cascade photons (for $\FDM=1$).  For $p=5~\text{mbar}$ (the maximum pressure before clustering), only excitations to states with $v' \ge 10$ result in a cascade signal, while lower $v'$ produce co-quench photons.  However, as the pressure decreases, a cascade signal can be produced also for lower $v'$.  This increases sensitivity to lower DM masses, albeit at the expense of reducing the overall number of molecular targets and hence reducing the experiment's sensitivity for larger masses. The low-mass DM sensitivity is saturated at a pressure of $10^{-7}~\text{bar}$, since the $v'=3$ state already produces an appreciable cascade signal; decreasing the pressure further only reduces the DM sensitivity for all masses.

\begin{figure}[t]
\centering
\includegraphics[width=0.49\textwidth]{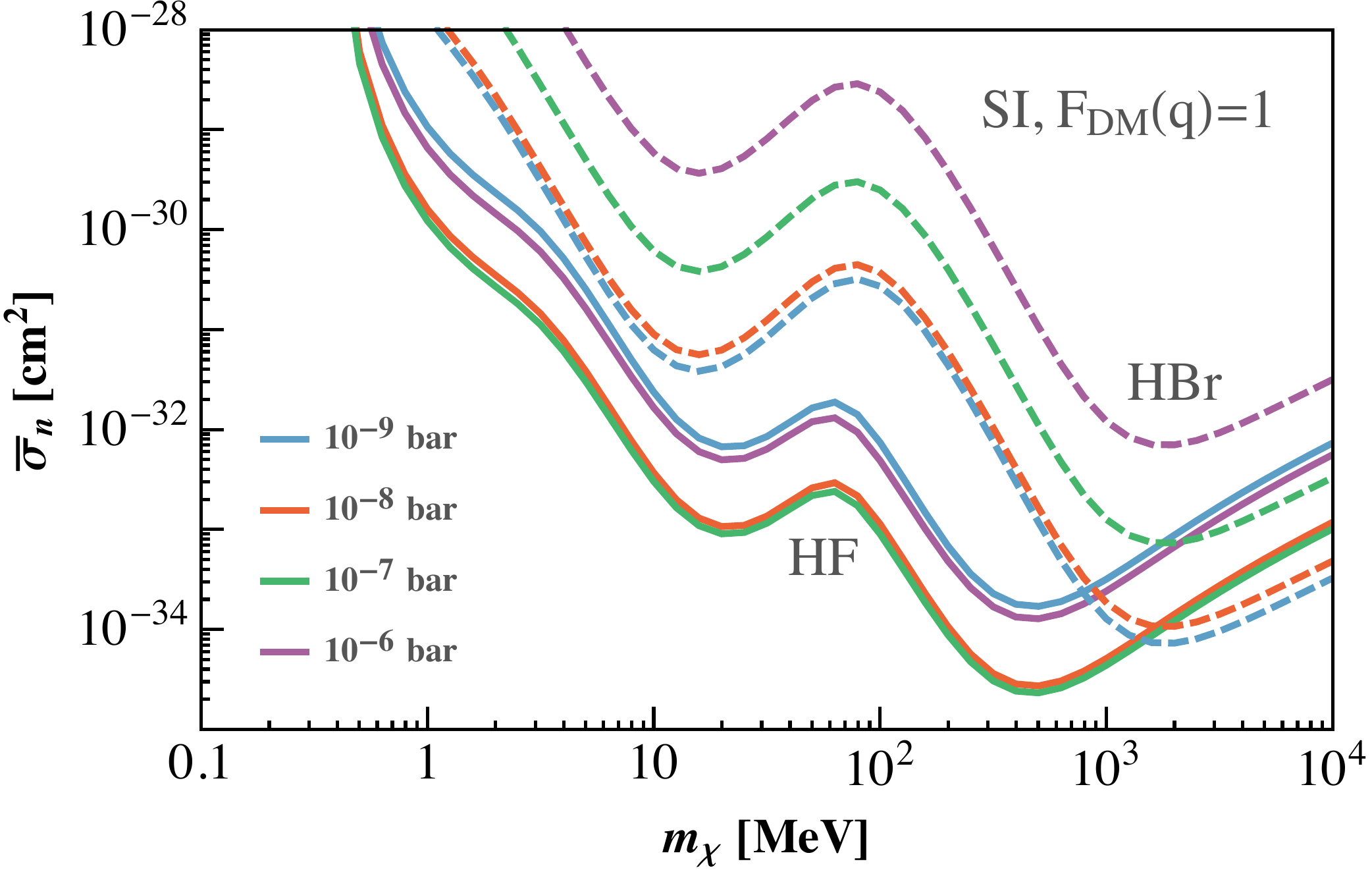}
\caption{The sensitivity to DM spin-independent nuclear scattering off HF and off HBr molecules for various values for the pressure. For these targets, essentially only cascade photons are produced.}
\label{fig:Halides}%
\end{figure}

Fig.~\ref{fig:Halides} presents the sensitivity to spin-independent DM scattering with HF for various values of the pressure (only cascade photons exist for this target). For comparison, we also show results for an HBr target. For both these targets, quenching increases with increasing $v'$. The pressure that maximizes the sensitivity to DM for HF, $p=100~\text{nbar}$, is two orders of magnitude higher than the optimal pressure for HBr, $p=1~\text{nbar}$, which is due to the much larger Einstein coefficients for HF (we set the temperature to be the respective $T_{\rm BBR}$).  The reason there are two peak sensitivities, one at tens of MeV and one near 1~GeV, is due to DM scattering either off H or off F/Br, which have very different masses.  
In the case of HSc (not shown), even the $v'=3$ state could be accessible at $p=50~\mu \text{bar}$ if efficient $V-E$ transfer is experimentally confirmed.

\subsection{Co-Quench signal}
\label{co-quench nt}
The co-quench photons are collected with only an $\mathcal{O}(1)$ number of reflections, and as a result $\epsilon_{\rm abs}$ is not a critical parameter for optimizing the setup. However, since the relevant target volume scales as $V_{\rm co}\sim A_{\rm det} \lambda_{\rm MFP}$ (Eq.~\eqref{eq:vco}), there is a linear increase in reach as a function of $A_{\rm det}$. This is seen in Fig.~\ref{fig:pandabsdep} (left) where there is a 100-fold increase in reach from $A_{\rm det} = 1~\text{cm}^2$ (dotted-green), to $A_{\rm det} = (10~\text{cm})^2$ (dotted-blue), and again a 100-fold increase to $A_{\rm det} = 1~\text{m}^2$  (dotted-yellow).

As the pressure of CO is increased, the number of relevant target molecules, proportional to the partial pressure of CO and the accessible volume, is 
\begin{eqnarray}
N_T \sim p_{\rm CO}\  V_{\text{co}} & \sim & p_{\rm CO}\ A_{\rm det}\ \lambda_{\rm MFP} \sim p_{\rm CO} \ A_{\rm det} \ \frac{p_{\rm He}}{p_{\rm CO}} \nonumber \\ 
\sim A_{\rm det}\ p_{\rm He}\,,
\end{eqnarray} 
where we have used Eqs.~\eqref{eq:lammfp} and~\eqref{eq:vco}. We see that the number of target CO molecules, and hence the DM sensitivity, is independent of the partial pressure of CO.

\begin{figure*}[t]
\centering
\includegraphics[width=0.49\textwidth]{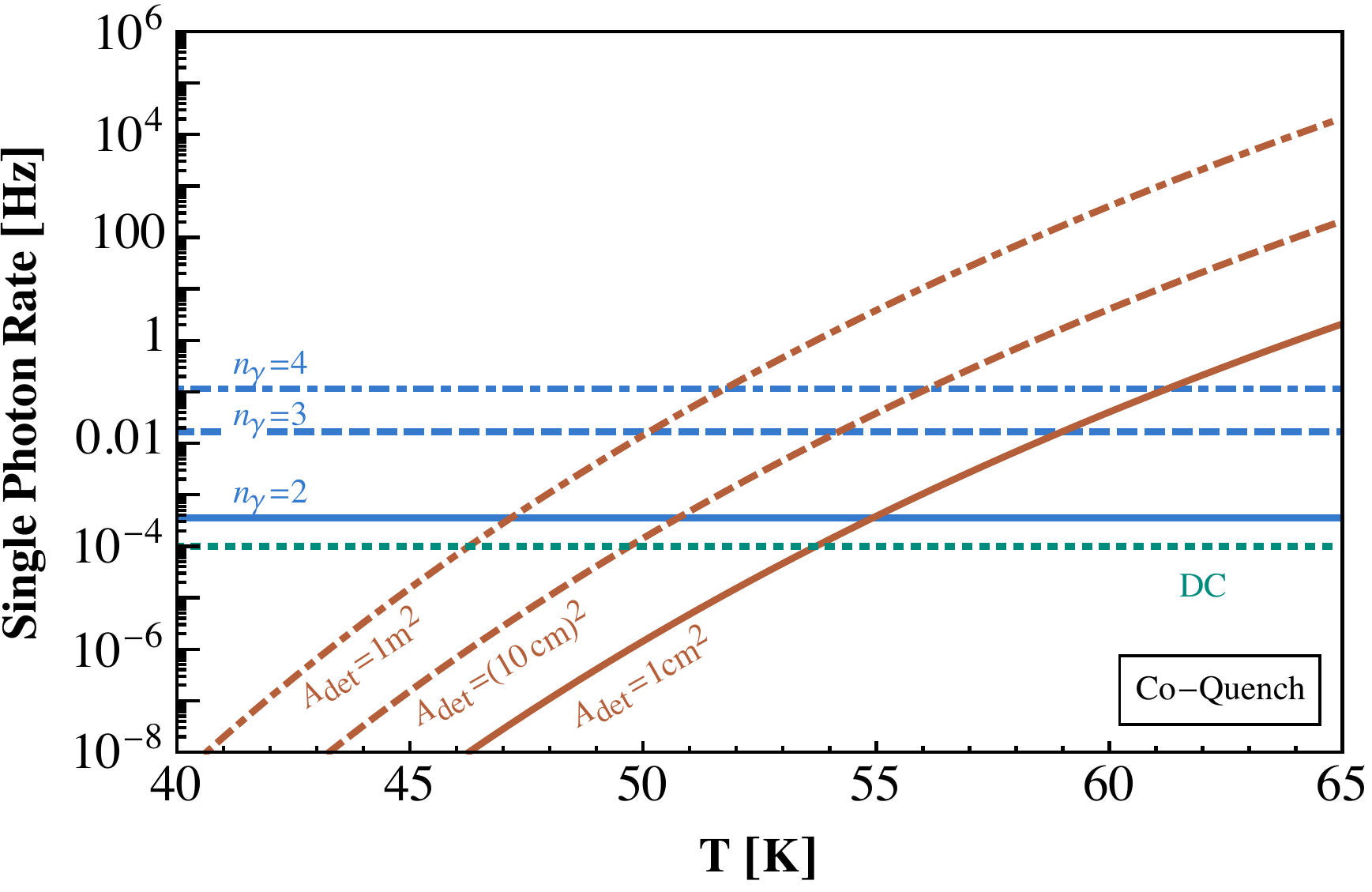}
\includegraphics[width=0.49\textwidth]{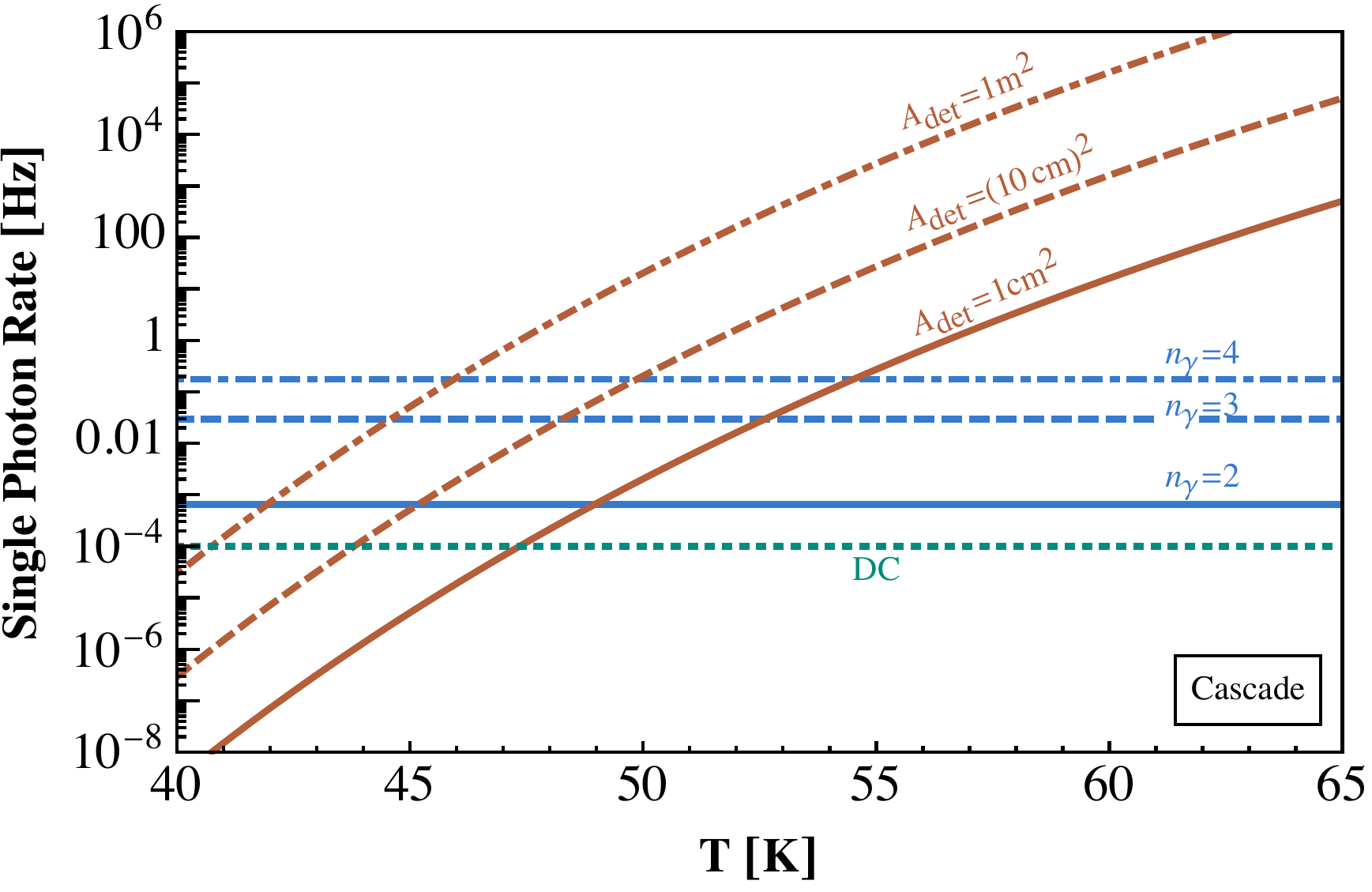}
\caption{\textbf{Left:} The maximum allowed single-photon rate (blue curves) versus temperature for achieving an $n_\gamma$-photon coincidence rate of $R_{n\gamma} < 0.1/{\rm year}$ for CO for $n_\gamma =4$ (dot-dashed), $n_\gamma =3$ (dashed), and $n_\gamma =2$ (solid) co-quench photons ($v=1\to 0$ transitions). Red curves show the single-photon rates from blackbody radiation for different detector areas, $A_{\rm det}=1$~cm$^2$ (solid), $A_{\rm det}=100$~cm$^2$ (dashed), and $A_{\rm det}=1$~m$^2$ (dot-dashed). The dotted-green curve shows a dark count rate of $10^{-4}$~Hz, as demonstrated for SNSPDs in~\cite{Wollman:17}. 
\textbf{Right:} Same as left panel, but for cascade photons. Here the blue curves show the results for $n_\gamma =2$, $n_\gamma =3$, and $n_\gamma =4$ cascade photons corresponding to the transitions $v'=10\to 8$, $v'=11\to 8$, and $v'=12\to 8$ in CO, respectively. The shift to lower temperatures of the red curves in the right panel compared to the left panel is a result of the lower energies of the cascade photons, leading to larger single-photon rates from blackbody radiation for a given temperature.
}
\label{fig:bbr}%
\end{figure*}

\section{Backgrounds}
\label{sec:bkg}
In this section, we briefly discuss radiogenic backgrounds, photodetector dark counts, and blackbody radiation. 

\subsection{Radioactive and Cosmogenic Backgrounds}
When designing an experiment based on this study's proposed detection concept, great care will need to be taken to understand and mitigate radioactive backgrounds, including neutrons and gamma-rays, as well as solar neutrinos scattering coherently off nuclei, all of which can mimic a DM signal. However, given that such an experiment will probe substantial new regions in DM parameter space even for relatively small exposures (the maximum exposure for our projections in Sec.~\ref{sec:Projections} is only 250~g-yr), we do not expect radioactive backgrounds to limit the DM sensitivity for first generation experiments. Solar neutrinos scattering off nuclei are known to not be an important background for small exposures either~\cite{Essig:2016crl}. We also note that the expected photon spectrum from DM is quite distinctive, which would help in distinguishing DM from backgrounds.

\subsection{Blackbody Radiation and Dark Counts} 
Background photons could come from either dark counts (which mimic a photon signal) or blackbody radiation (which are real photons). In both cases the number of observed photons is Poisson distributed in time, and a substantial reduction in the background rate is achieved simply by requiring two or more coincident photons. If the single photon rate due to blackbody radiation or dark counts is $R_{1\gamma}$, the rate to observe $n$ or more photons within a time $\Delta t_n$ is given by 
\begin{equation}
R_{n\gamma} = (R_{1\gamma})^{n-1} (\Delta t_n)^{n-2} (1-e^{-R_{1\gamma}\Delta t})\,.
\label{coinceqn}
\end{equation}
Here $\Delta t_n$ is the time needed to observe $n$ signal photons.  For co-quench photons, $\Delta t_n \sim \frac{1}{A_{10}}$ since V-V transfer occurs on much shorter timescales for our choice of parameters, while for cascade photons,
\beq
\Delta t_n \sim \sum_{v=v'-n+1}^{v'}\frac{1}{A_{v,v-1}},
\eeq
where the $A$'s are Einstein coefficients and the sum is over the relevant vibrational states. In the $R_{1\gamma} \Delta t_n \ll 1$ limit,
\begin{equation}
R_{n\gamma} =  (R_{1\gamma})^{n} (\Delta t_n)^{n-1}\,.  
\end{equation}
One can now solve for the maximum allowed single-photon rate for achieving, for example, $R_{n\gamma} < 0.1/{\rm year}$. This maximum allowed single-photon rate decreases with increasing $n$. We present the results for CO in Fig.~\ref{fig:bbr} (left) for $n_\gamma =4$, $n_\gamma =3$, and $n_\gamma =2$ co-quench photons (with $A_{10}=33.9$~Hz); and in Fig.~\ref{fig:bbr} (right) for $n_\gamma =2$, $n_\gamma =3$, and $n_\gamma =4$ cascade photons corresponding to the transitions $v'=10\to 8$, $v'=11\to 8$, and $v'=12\to 8$, respectively.

\begin{figure*}[t]
\centering
\includegraphics[width=0.49\textwidth]{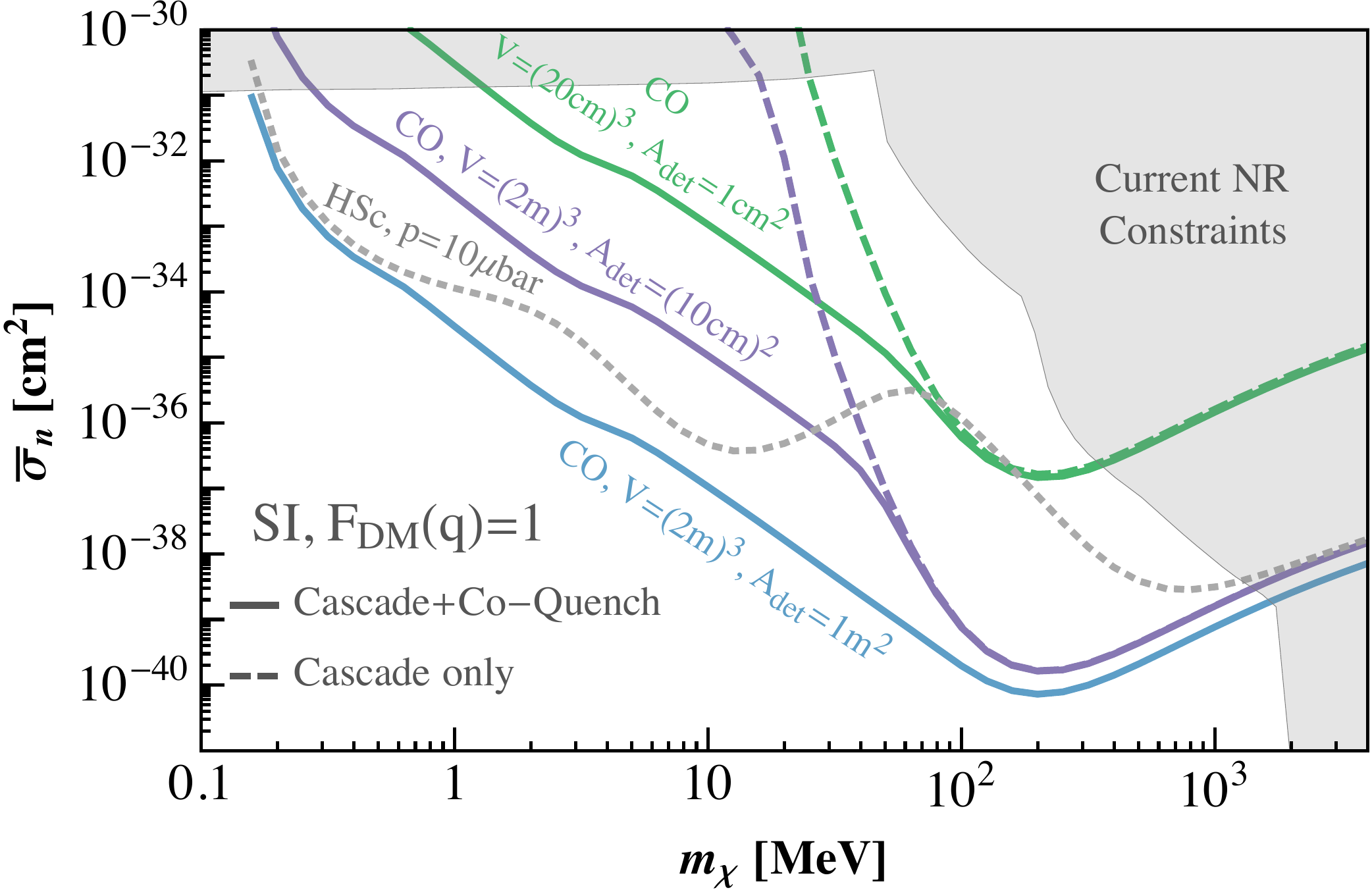}
\hspace{0.0cm}
\includegraphics[width=0.49\textwidth]{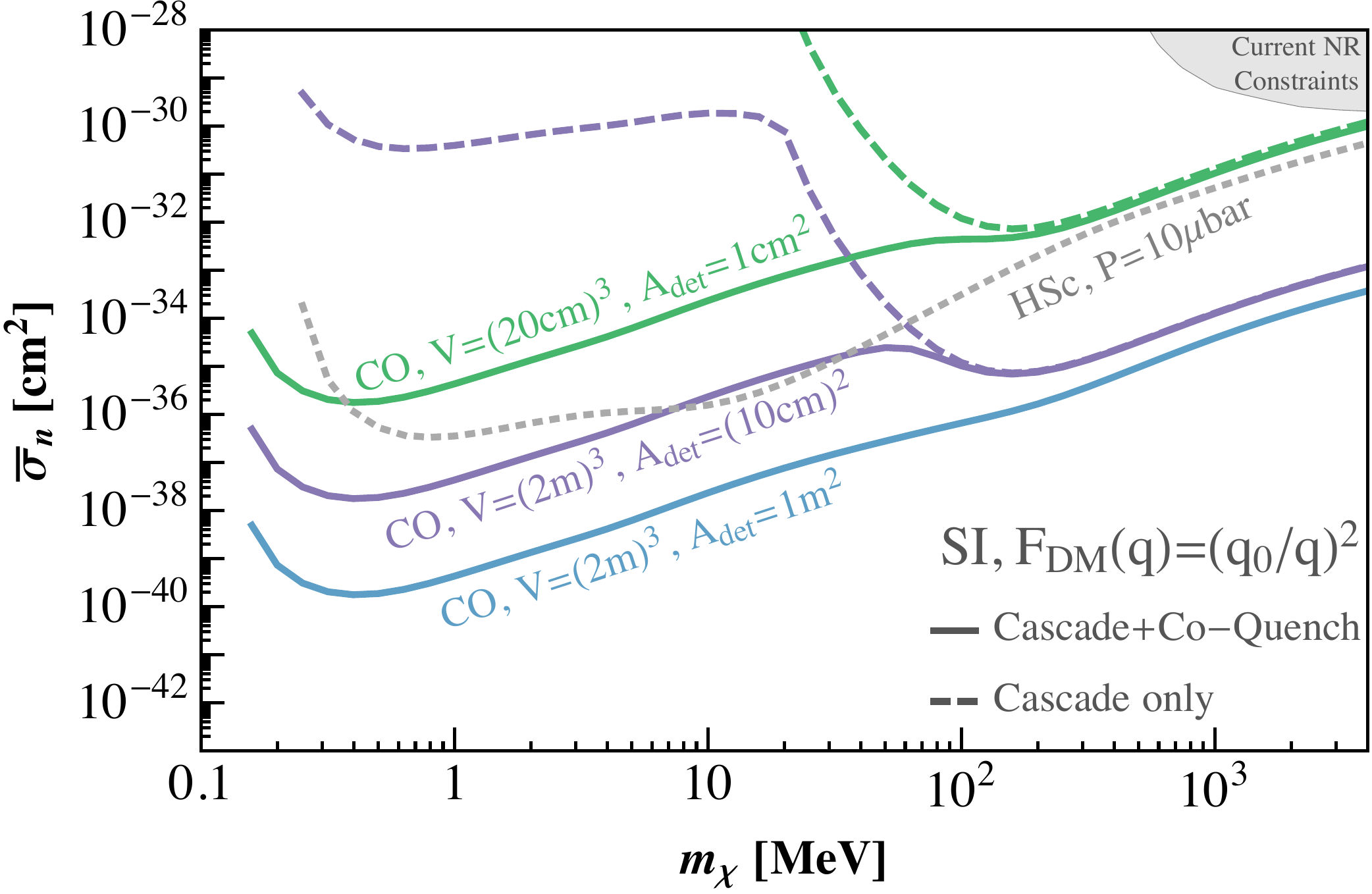}
\caption{The 95\% confidence-level sensitivity for DM spin-independent nuclear scattering off CO molecules at a partial pressure of $5$~mbar and a temperature of $55$~K, for various tank sizes and photodetector area and for a time exposure of 1~year. The signal consists of two-or-more coincident photons, and we assume zero background events. The \textbf{left} panel corresponds to a trivial DM form factor, $\FDMsq=1$, while the \textbf{right} panel corresponds to $\FDMsq=\left(q_0/q\right)^4$ (with $q_0 = (\mu_{12}/4m_e)^{1/4} \alpha_{\rm EM} m_e$). All results assume $f_{N,\text{SI}}^{(i)}=f_{P,\text{SI}}^{(i)}=1$. Colored dashed curves correspond to CO cascade photons while colored solid curves correspond to the sum of cascade and co-quench photons.  Since co-quench photons are produced at a smaller energy threshold, they reach to lower DM masses. The co-quench signal assumes the addition of an He buffer gas at a pressure of $4$~bar and an internal mirror that helps to focus photons to the detector (see text). The grey dotted curve corresponds to a cascade photon signal from an HSc gas at a pressure of $10~\mu$bar and at a temperature of $46$~K. The gray shaded regions are current constraints set by other nuclear recoil direct-detection experiments (see text).}
\label{fig:N_recoil_SensPlot}%
\end{figure*}

Blackbody radiation contributes to $R_{1\gamma}$ and depends on the detector area, $A_{\rm det}$, as well as the gas temperature, $T$.  
The single-photon blackbody radiation rate is given by 
\begin{equation}
R_{1\gamma}^{\rm BBR}\sim \frac{\Delta \omega \ \omega^2}{\pi^2} A_{\rm det} e^{- \frac{\omega}{T}}\,, 
\label{eq:R1bbr}
\end{equation}
where $\Delta\omega$ is the resolution of the photodetector.\footnote{If the detector, such as an SNSPD, simply has an energy threshold above which it detects a photon, then one would need to integrate the blackbody spectrum over all energies above that threshold; however, due to the exponential suppression of the blackbody radiation, the result is very similar to simply assuming sensitivity to a finite energy width, $\Delta \omega$.} 
Hence, there is some maximum temperature, $T_{\rm BBR}$, above which blackbody radiation becomes an irreducible background. 
To estimate $T_{\rm BBR}$, we show $R_{1\gamma}^{\rm BBR}$ from Eq.~\eqref{eq:R1bbr} versus $T$ for $A_{\rm det}=1~\text{cm}^2$,  $(10~\text{cm})^2$ and $1~\text{m}^2$ in Fig.~\ref{fig:bbr} (left) for co-quench photons with $\omega =0.265~$eV (corresponding to the $v'=1\rightarrow0$ transition) and in Fig.~\ref{fig:bbr} (right) for cascade photons with $\omega =0.23~$eV (corresponding to the $v'=12\rightarrow 11$ transition); we assume $\Delta \omega = \frac{\omega}{10}$ in both figures.

As can be seen in Fig.~\ref{fig:bbr} (left), the maximum temperature above which blackbody radiation mimics two coincident, co-quench photons in one year is $T_{\rm BBR}\simeq 55$~K for a photodetector area of $A_{\rm det}=1~{\rm cm}^2$. The temperature must be decreased to $\sim$46~K for $A_{\rm det} =1~\text{m}^2$. For all values of $A_{\rm det}$, larger temperatures are allowed when requiring a coincidence of more than two photons. Importantly, as $m_\chi$ increases, more coincident co-quench photons are expected (with a maximum of $v_{\rm b}$). The maximum energy of the cascade photons, $\omega =0.23~{\rm eV}$ is less than the energy of the co-quench photons from the $v'=1\rightarrow0$ transition, $\omega =0.265~{\rm eV}$. For this reason, the required $T_{\rm BBR}$ is shifted to lower values for cascade photons than for co-quench photons, see Fig.~\ref{fig:bbr} (right). Again, larger DM masses often produce four or more coincident photons (see Fig.~\ref{fig:Spectrum}), in which case we again obtain $T_{\rm BBR}=55$~K for $A_{\rm det}=1~{\rm cm}^2$. For larger $A_{\rm det}$, $T_{\rm BBR}$ decreases to avoid blackbody radiation mimicking cascade photons. However, for large $A_{\rm det}$, the co-quench signal increases and at some point begins to dominate. 

Motivated by these considerations, all rates for CO presented in Sec.~\ref{sec:Projections} correspond to $T_{\rm BBR}=55$~K. It should be noted that for a realistic setup, $T_{\rm BBR}$ depends on $A_{\rm det}$, $n_\gamma$, and whether one is interested in focusing on co-quench or cascade photons. A similar analysis as described above can be performed for any other candidate molecules. We find the results to be qualitatively similar as for the case of CO. In particular, we find $T_{\rm BBR}=44$~K, 115~K and 72~K for HSc, HF and HBr, respectively.

Finally, the dark count rate for single-photon photodetectors is not well understood.  For example, SNSPDs have demonstrated dark count rates as low as $10^{-4}$~Hz~\cite{Wollman:17} (shown in Fig.~\ref{fig:bbr} as a green dotted line), in which case the two-photon coincidence from dark counts is less than 0.1/year. 

\section{Sensitivity Projections}
\label{sec:Projections}

\begin{figure*}[t]
\centering
\includegraphics[width=0.49\textwidth]{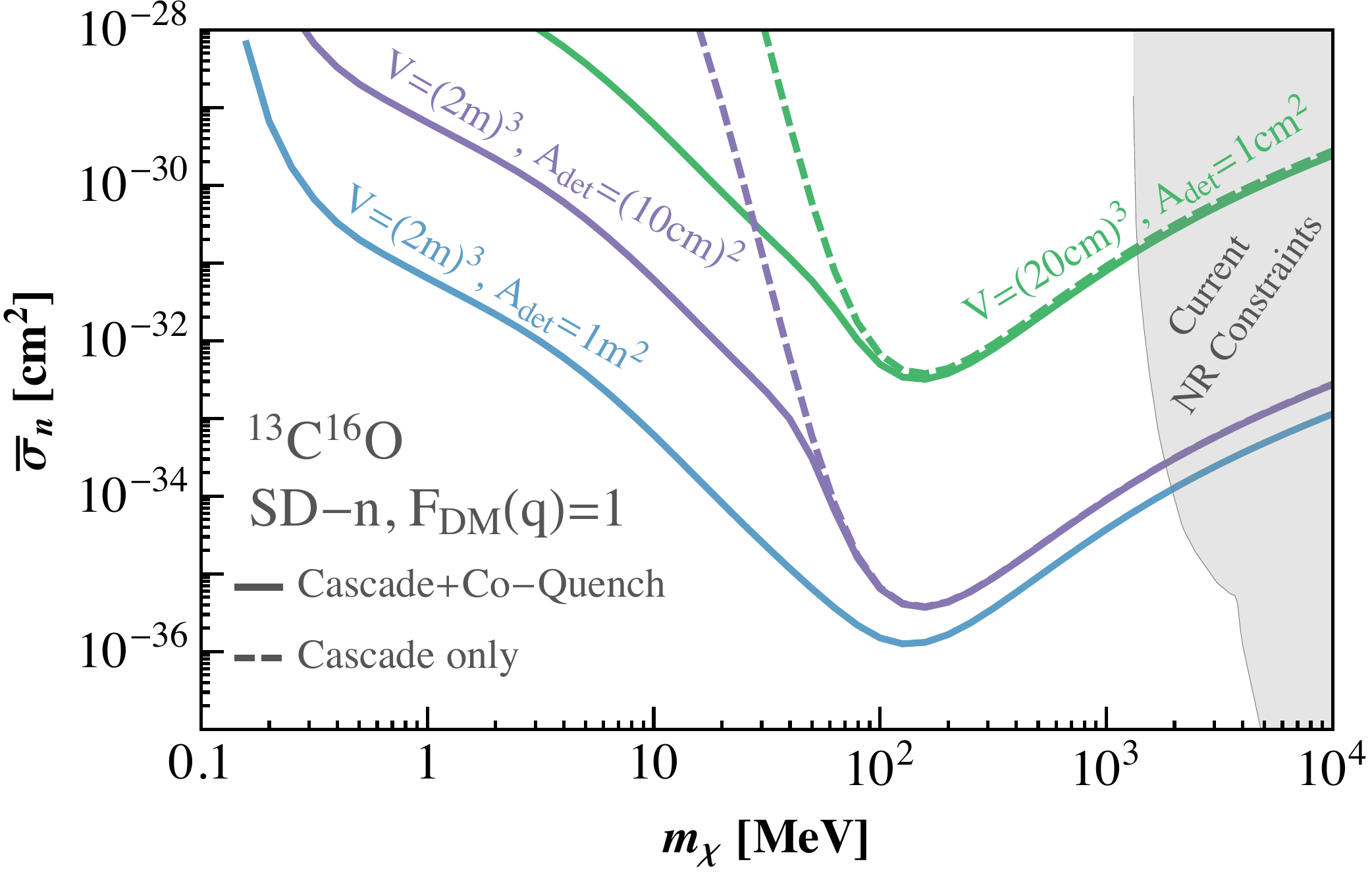}
\hspace{0.0cm}
\includegraphics[width=0.49\textwidth]{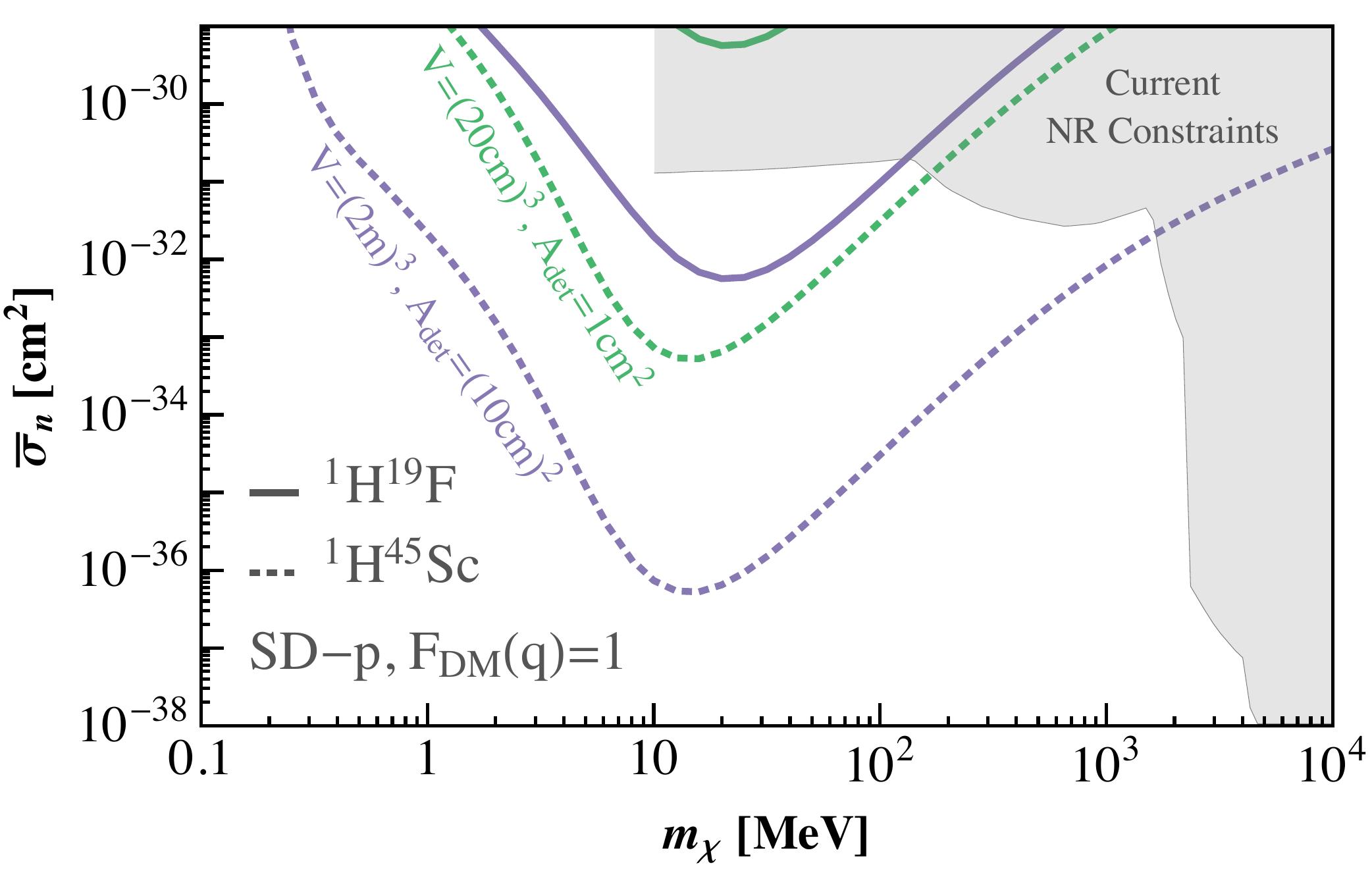}
\caption{
The 95\% confidence-level sensitivity for DM spin-dependent scattering for various tank sizes and photodetector areas and for a time exposure of 1~year.  
The signal consists of two-or-more coincident photons, and we assume zero background events. We assume a trivial DM form factor, $\FDMsq=1$. The gray shaded regions are current constraints set by other nuclear recoil direct-detection experiments (see text). 
\textbf{Left:} Sensitivity to spin-dependent neutron interactions ($f_{N,{\rm SD}}^{({\rm C})}=1$, $f_{P,{\rm SD}}^{({\rm C})}=0$) using $^{13}$C$^{16}$O molecules at a partial pressure of $5$~mbar and a temperature of $55$~K.  Colored dashed curves correspond to CO cascade photons while colored solid curves correspond to the sum of cascade and co-quench photons. 
  Since co-quench photons are produced at a smaller energy threshold, they reach to lower DM masses. The co-quench signal assumes the addition of an He buffer gas at a pressure of $4$~bar and an internal mirror that helps to focus photons to the detector (see text). 
\textbf{Right:}  Sensitivity to spin-dependent proton interactions ($f_{N,{\rm SD}}^{({\rm H})}=0$, $f_{P,{\rm SD}}^{({\rm H})}=1$ and neglecting couplings to F and Sc atoms) using $^{1}$H$^{19}$F molecules at a pressure of $0.1~\mu$bar and a temperature of $115$~K (solid curves) and $^{1}$H$^{45}$Sc molecules at a pressure of $10~\mu$bar and a temperature of $46$~K (dotted curves). 
The signal corresponds only to cascade photons with a minimal $v'=3$. 
}
\label{fig:SD_SensPlot}
\end{figure*}

 Figs.~\ref{fig:N_recoil_SensPlot} and~\ref{fig:SD_SensPlot} present projected sensitivities of the proposed experimental concept to spin-independent and spin-dependent interactions, respectively. In the appendix, we also show Fig.~\ref{fig:Sigmae_HiddenPhoton} which presents the sensitivity for the specific case of a DM particle interacting with ordinary matter through a dark photon. Each figure has several curves corresponding to different experimental parameters such as temperature, total pressure, molecular target, tank volume, and photodetector surface area. Below, we provide details of each result.

\subsection{Spin-Independent Nuclear Recoils}

Fig.~\ref{fig:N_recoil_SensPlot} shows the 95\% confidence-level (c.l.) sensitivity for spin-independent nuclear couplings with $f_{P,{\rm SI}}^{(i)} = f_{N,{\rm SI}}^{(i)} = 1$, for a signal consisting of two-or-more coincident photons with zero background events, i.e.,  $3.1$ signal events per exposure. The left panel corresponds to a trivial DM form factor, $\FDMsq=1$, while the right panel corresponds to $\FDMsq=\left(q_0/q\right)^4$ (with $q_0 = (\mu_{12}/4m_e)^{1/4} \alpha_{\rm EM} m_e$). The results are given for a gas of CO molecules at an optimal pressure of $5$~mbar at 55~K, as well as for a gas of HSc molecules at $10$~$\mu$bar at 46~K. For CO, all $(v',J')$ states with energies below that of $(v=26,J=0)$ have been included (the lowest dashed curve in Fig.~\ref{fig:v_j_Plane}). Above that energy, excited electronic states potentially become energetically available. For HSc, all states with energies below that of $(v=5,J=0)$ have been included. Finally, in each panel, gray shaded regions show currently existing bounds from various experiments. For $\FDMsq=1$, these include CDMSLite~\cite{Agnese:2017jvy}, CRESST~III and the CRESST surface run~\cite{Abdelhameed:2019hmk}, DarkSide50~\cite{Agnes:2018ves}, CDEX-1B~\cite{Liu:2019kzq}, constraints derived from an organic liquid scintillator experiment~\cite{Collar:2018ydf}, XENON1T~\cite{Aprile:2018dbl}, and an analysis of the effects of up-scattering of light DM from cosmic rays with subsequent scattering in the XENON1T experiment~\cite{Bringmann:2018cvk}. For $\FDMsq=\left(q_0/q\right)^4$, current bounds include results from LUX~\cite{Akerib:2018hck} and PandaX~II~\cite{Ren:2018gyx}.

As discussed above, since little experimental data is currently available for HSc molecules, branching ratios and an optimal working pressure are not easily calculable. Therefore, results are presented under the assumption of a constant branching ratio of unity for $p=10~\mu$bar. This should be taken as an order of magnitude estimate and requires further study for a more accurate projection. Results for CO are based on more detailed data. These projections include non-trivial photon emission branching ratios from each $(v,J)$ state and the pressure has been chosen to optimize the reach for a cascade signal for DM masses with the maximum scattering rate, namely above $\sim 50$~MeV. For lower DM mass, the pressure independent co-quench signal dominates.

In each panel of Fig.~\ref{fig:N_recoil_SensPlot}, results for CO are given for three examples of experimental setups with combinations of gas tank volumes, $V=(20~\text{cm})^3$ and $(2~\text{m})^3$ and detector surface areas $A_{\rm det}=1~\text{cm}^2$, $(10~\text{cm})^2$ and $1~\text{m}^2$. The results for HSc are given for $V=(2~\text{m})^3$ and $A_{\rm det}=(10~\text{cm})^2$. Dashed curves are sensitivities for cascade photon signals, while solid curves are sensitivities when combining cascade and co-quench photons (the dotted HSc curve corresponds to cascade-only photons). The co-quench results assume the addition of an He buffer gas as well as cylindrical mirrors with a height of $2\lambda_{\rm MFP}$ for efficient photon collection, as described in Sec.~\ref{subsec:Det_Geometry}. Results assume an absorption efficiency of $\epsilon_{\rm abs} = 10^{-5}$ and $\epsilon_{\rm det}=1$.

The behavior of the sensitivity curves for cascade and co-quench photons and for different molecular targets is evident for $\FDM=1$. First, for CO, since the two nuclei in the molecule have similar masses of $\sim$10~GeV, and since cascade signals only have sizable branching ratios above $v'\approx10$, the sensitivity reach peaks near $m_\chi \approx 100$~MeV and falls sharply for lower DM masses. For HSc, all values of $v'\geq3$ correspond to a cascade signal and the large mass difference between the nuclei cause two peaks in the rate, at $\sim$10~MeV and $\sim$1~GeV. Some of this behavior is washed out by the $q^{-4}$ weighting of the cross section for the non-trivial DM form factor. As a result, the reach improves at lower DM masses, stopping only at around $m_\chi \sim 200~\text{keV}$ below which the kinetic energy is below the threshold for molecular excitations (see discussion in Sec.~\ref{subsec:Cross_Sec}). For a given gas volume, the collection efficiency grows with detector surface area. Beyond some surface area, the efficiency saturates to unity. For the parameters chosen here, this occurs for $V=(2~\text{m})^3$ at around $A_{\rm det}\sim (10~\text{cm})^2$, and hence we do not show the DM sensitivity using cascade photons for $A_{\rm det}$ larger than $(10~\text{cm})^2$. On the other hand, the co-quench signal (relevant for CO only) is sensitive to values of $v'\geq2$. Therefore, this signal has sizable reach for much smaller DM masses. The absolute minimum around $m_\chi \approx 200$ keV corresponds to extraction of the total DM's kinetic energy. The co-quench signal scales linearly with $A_{\rm det}$, and for sufficiently large detector areas, the co-quench signals begin to dominate for all DM masses, as shown here by the blue curves. 

In Appendix~\ref{subsec:Sens_Dark_Photon} we investigate the sensitivity to DM that scatters off molecules through a heavy or ultralight dark photon mediator.  Since the dark photon couples to electric charge, there will be a suppression at low momentum transfers due to the screening of the nuclear charge. Such DM can also scatter off the electric dipole moment of the molecule, but we leave an investigation of the sensitivity to this scenario to future work.

\subsection{Spin-Dependent Nuclear Recoils}

Fig.~\ref{fig:SD_SensPlot} presents sensitivities to spin-dependent couplings to neutron spin (left) and proton spin (right), with similar assumptions as for the spin-independent case. All results are given for a trivial DM form factor, $\FDMsq=1$. Additionally, only the zero-momentum spin structure of the nuclei have been taken into account. The values of $\langle S_P \rangle$ and $\langle S_N \rangle$ have been calculated in the odd group approximation~\cite{Bednyakov:2004xq}. Current nuclear recoil constraints are shown in shaded gray. For spin-dependent neutron couplings these include CDMSLite~\cite{Agnese:2017jvy}, LUX~\cite{Akerib:2017kat} and XENON1T~\cite{Aprile:2019dbj}, while for spin-dependent proton couplings, constraints include CDMSLite~\cite{Agnese:2017jvy}, PICASSO~\cite{Behnke:2016lsk}, PICO60~\cite{Amole:2017dex}, constraints derived from an organic liquid scintillator experiment~\cite{Collar:2018ydf}, and results from up-scattering of DM by cosmic rays in the Borexino detector~\cite{Bringmann:2018cvk}. 

CO provides sensitivity to spin-dependent neutron couplings, and we show projected sensitivities for $f_{N,{\rm SD}}^{({\rm C})}=1$ and $f_{P,{\rm SD}}^{({\rm C})}=0$ for a gas of $^{13}$C$^{16}$O at $p=5$~mbar. This curve has been plotted assuming $\lambda_{\rm MFP} = 27$ cm (this is achievable if other mutually transparent atoms/molecules are present within the tank). On the other hand, hydrogen, fluorine, bromine, and scandium provide sensitivity to spin-dependent proton couplings, and we show projected sensitivities for $f_{N,{\rm SD}}^{({\rm H})}=0$ and $f_{P,{\rm SD}}^{({\rm H})}=1$ for a gas of $^{1}$H$^{19}$F at $p=0.1~\mu$bar (solid curves) and $^{1}$H$^{45}$Sc (dotted curves) at $p=10~\mu$bar. For simplicity, we present spin-dependent results under the conservative assumption of zero couplings to the F and Sc atoms. The resulting rates are expected to be approximately accurate for low DM masses and underestimated for larger DM masses. See Appendix~\ref{app:FF} for details.

\section{Conclusions and Outlook}
\label{sec:Conclusions}

We have presented a novel detection concept for sub-GeV DM interacting with nuclei, in which DM excites a ro-vibrational mode of a molecule, which subsequently decays to the ground state to produce multiple photons.  We find that CO, as well as potentially HF and HSc, are excellent candidate molecules. We also study several halides and discuss other possible candidates in the appendix.

The proposed concept has several important features.  (i) Since the DM signal consists of multiple photons that arrive in coincidence on a relatively short timescales of $\mathcal{O}$(0.1~s), dark counts of the photodetectors and background photons from blackbody radiation are not expected to be a limiting concern. (ii) The expected number of photons and their energies will be determined by the DM mass and interaction type (heavy or light mediator), allowing for an understanding of the particle nature of DM in the case of a positive signal, and helping also to distinguish signal from background.  (iii) Since the DM scattering is an inelastic process that excites internal degrees of freedom corresponding to ro-vibrational excitations of the molecule, a much larger fraction of the DM's kinetic energy can be transferred to the molecule compared to the energy transferred to a nucleus in elastic DM-nucleus scattering.  This implies that a DM particle as light as $\mathcal{O}$(100~keV) is able to excite a vibrational mode that lies $\mathcal{O}$(200~meV) above the ground state. (iv) The technological requirements for the realization of the proposed concept are expected to be available on relatively short time scales.  (v) Several technological requirements needed to realize our proposed detection concept for DM-molecular scattering---for example, the creation of  a large array of photodetectors---will also be useful for probing DM absorption by molecules or DM-electron scattering or absorption in scintillators.

Several additional calculations and experimental measurements are required to better understand the relaxation rates of an excited molecule. Some of these relaxation mechanisms compete with the emission of a photon signal and therefore could reduce the number of observed photons.  For the case of CO, which we envision operating at a temperature of 55~K, there is no experimental measurement of the CO-CO vibrational relaxation rates of states with high vibrational quantum numbers for temperatures below 75~K. However, full quantum mechanical scattering calculations for the vibrational relaxation rates of the first vibrational state for CO-H$_2$ and H$_2$-H$_2$ down to 10~K show that the relaxation rate is temperature independent below some threshold temperature, which turns out to be $T \lesssim 80$~K~\cite{Samantha2011,BalaCOH2}.  A similar behavior is predicted and observed for CO-He~\cite{Krems2002,ExpCOHe} and H$_2$-He~\cite{McGuire1975}.  Our calculation of the CO-CO vibrational relaxation rates are done using first-order time-dependent perturbation theory, which does not show the same constant temperature dependence at low temperatures. A full quantum mechanical scattering calculation, as well as measurements, will be required to fully determine the relaxation rates. This might slightly shift the transition between the cascade and co-quench signals in Fig.~\ref{fig:v_j_Plane}. Moreover, this is also needed to determine whether other collisional quenching processes, which do not produce any photons, begin to dominate for states with very high vibrational quantum numbers.  

In the case of the hydrogen halides and HSc, there is even less information on the self-quenching processes at temperatures below room temperature. However, our calculation of the vibrational relaxation rates for these molecules is very conservative.  In particular, it may be possible to operate a DM detector using a molecular halide or HSc at much higher pressures than those assumed in our study. This would dramatically increase its sensitivity to DM beyond what we has been presented in this paper.

The proposed detection concept has the potential to allow for the exploration of orders of magnitude of new DM parameter space with near-term technology.  

\acknowledgments

We thank Eden Figueroa for many useful discussions and for collaboration at the early stages of this work.  
We also thank Masha Baryakhtar, Karl Berggren, Carlos Blanco, Juan Collar, Daniel Egana-Ugrinovic, Simon Knapen, Ben Mazin, Gerard Meijer, Tongyan Lin, Sae Woo Nam, Matthew Shaw and Ken van Tilburg for useful discussions. R.E.'s work in this paper is supported by DoE Grant DE-SC0017938. R.E. also acknowledges support from the US-Israel Binational Science Foundation under Grant No.~2016153, from the Heising-Simons Foundation under Grant No.~79921, from a subaward for the DOE Grant No.~DE-SC0018952, and from Simons Investigator Award 623940. H.R. is supported in part by the DOE under contract DE-AC02-05CH11231. Some of this work was done at the Aspen Center for Physics, which is supported by NSF grant PHY-1607611 and at KITP, supported in part by the National Science Foundation under Grant No. NSF PHY-1748958.

\appendix
\section{Molecular Form Factor}
\label{app:FF}
Before averaging over the distribution of initial states, the form factor from Eq.~\eqref{eq:FmolAvg} is,
\bea
\left \vert F_{{\rm mol}, v'J'J_{\rm init}}(q) \right \vert^2 & \equiv & \frac{1}{\#_m} \sum_{m,m'} \biggr\vert \int d^{3}r \, \mathcal{O}(\mathbf{q}\cdot\mathbf{r}) \nonumber \\
& & \times \Psi_{v' J' m'}^{*}(\mathbf{r})\Psi_{0 J_{\rm init} m}(\mathbf{r}) \biggr\vert^2 \,,
\label{eq:FmolAvg1}
\eea
with $\mathcal{O}(\mathbf{q}\cdot\mathbf{r})$ is given by Eq.~\eqref{eq:SDSI_Operators}. Note that there is an implicit average over initial $m$ and sum over final $m'$ quantum numbers ($\#_m$ is the number of $m$ states). The wave-functions can be expanded in spherical harmonics,
\beq
\Psi_{vJm}(\mathbf{r}) = \phi_{vJ}(r) \mathcal{Y}_{Jm}(\Omega)\,,
\eeq
where $\phi_{vJ}(r)$ are the solutions to the radial component of the Schr\"odinger equation and $\mathcal{Y}_{Jm}(\Omega)$ are spherical harmonics. The evaluation of the form-factor is more straightforward for the spin independent case and will be considered first.

\subsection{Spin-Independent}
For the spin-independent case, $f_{PN}^{(i)}$ are just numbers (as opposed to operators). One can expand the exponent in Eq.~\eqref{eq:SDSI_Operators} using plane wave expansion,
\beq
\label{SIexp}
e^{i \frac{\mu_{12}}{m_i}\mathbf{q}\cdot \mathbf{r}} = \sqrt{4 \pi} \sum_{\ell=0}^\infty i^\ell \sqrt{2 \ell + 1} j_\ell \left(\frac{\mathrm{\mu_{12}}}{m_{i}}qr \right) \mathcal{Y}_{\ell 0}(\Omega)\,,
\eeq
where $j_\ell(\frac{\mu_{12}}{m_i}qr)$ are spherical Bessel functions. Inserting this, averaging over $m$ and summing over $m'$ gives,

\begin{eqnarray}
\left \vert F_{{\rm mol}, v'J'J_{\rm init}}(q) \right \vert^2 & = & \frac{(2J_{\rm init}+1)(2J'+1)}{2 J_{\rm min} + 1}  \nonumber \\ & &
 \sum_{\ell=|J_{\rm init}-J'|}^{J_{\rm init}+J'} (2\ell+1) \biggr\lvert \left(
\begin{array}{ccc}
J_{\rm init} & J' & \ell \\
0 & 0 & 0 \\
\end{array}
\right) \nonumber \\
& & \times \int r^2 dr \left[ f_{PN}^{(1)} j_\ell(\frac{\mu_{12}}{m_{1}}qr) + f_{PN}^{(2)} j_\ell(-\frac{\mu_{12}}{m_{2}}qr) \right] \nonumber \\ & & \phi_{v' J'}(r) \phi_{0 J_{\rm init}}(r)
 \biggr\rvert^2. \
\end{eqnarray}
Here $J_{\rm min} \equiv {\rm min}(J_{\rm init},J')$ and the SI values of $f_{PN}^{(i)}$ should be used. The sum over $\ell$ is only over values with the same parity as $J_{\rm init}+J'$ (this is enforced by the 3-$j$ symbol). In the above result, orthogonality properties of the 3-$j$ symbol have also been used.

\subsection{Spin-Dependent}
For spin-dependent interactions, both the spin quantum numbers of the individual nuclei, $S_i$, as well as the orbital angular momentum of the molecule, $J$, must be taken into account. We denote the total angular momentum quantum number as $f=J+\sum_iS_i$ (a vector sum).\footnote{Note that in nuclear physics literature, the standard notation for total angular momentum is $J$ (our $f$) and the standard notation for orbital angular momentum is $L$ (our $J$). For consistency with the rest of the text we have chosen this notation.} As a simplification in this study, we assume that DM interacts with only one of the two nuclei. This avoids sums over multiple nuclear spins and possible complications resulting from interference between the two nuclei. This is an excellent approximation for the combination of isotopes in $^{13}$C$^{16}$O, since for this molecule, only the C atoms carry unpaired nucleons (the neutrons). For the case of H-X molecules, we consider interactions with the H atom only. This is a good approximation for low DM masses since interference between the atoms is small and since the low DM mass regime is dominated by interactions with the light species, H. At large DM masses, the calculation underestimates the true rate.

For the spin dependent case, the specific expansion that corresponds to Eq.~\eqref{SIexp} is given in Appendix B of~\cite{Klos:2013rwa}. After summing over final $m'$ and averaging over initial $m$ states, the form factor is given by~\cite{Klos:2013rwa}
\begin{align}
 &{\lvert F_{{\rm mol}, v' J' f' J_{\rm init} f_{\rm init}}(q) \rvert^2}  =  \frac{4 \langle S_{P/N} \rangle^2}{(2J_{\rm init}+1)} \nonumber \\
& \times \sum_{\rho,\ell}  \left \vert \int d^{3}r \, (\mathcal{O}^{\rm SD}_{\rho,\ell}) \Psi_{v' J' f'}(\mathbf{r})\Psi_{0 J_{\rm init}f_{\rm init}}(\mathbf{r}) \right \vert^2\,,
\end{align}
where the dependence on the $f$ quantum numbers (and hence the spin) has now been written explicitly in the wavefunctions, and the sum over $\rho\in\{1,2,3\}$ corresponds to
\beq
\mathcal{O}^{\rm SD}_{1,\ell}=\mathcal{L}_{\ell}^5 \quad \mathcal{O}^{\rm SD}_{2,\ell}=\mathcal{T}_{\ell}^{el 5}\quad \mathcal{O}^{\rm SD}_{3,\ell}=\mathcal{T}_{\ell}^{mag 5},
\eeq
with
\begin{eqnarray}
\mathcal{L}_\ell^5 & = & \frac{i}{2\sqrt{2\ell+1}}[a_0+a_1 \tau_3^i \frac{m_\pi ^2}{q^2+m_\pi ^2}] \nonumber \\
&& \times [\sqrt{\ell+1} M_{\ell,\ell+1} +\sqrt{\ell} M_{\ell,\ell-1}] \nonumber \\ 
& \approx & \frac{i}{\sqrt{2\ell+1}} [\sqrt{\ell+1} M_{\ell,\ell+1} +\sqrt{\ell} M_{\ell,\ell-1}] \nonumber \\
\mathcal{T}_\ell^{el 5} & = & \frac{i}{\sqrt{2\ell+1}} \frac{1}{2} [a_0+a_1 \tau_3^i ] [-\sqrt{\ell} M_{\ell,\ell+1} +\sqrt{\ell+1} M_{\ell,\ell-1}]\nonumber \\ 
& \approx & \frac{i}{\sqrt{2\ell+1}} [-\sqrt{\ell} M_{\ell,\ell+1} +\sqrt{\ell+1} M_{\ell,\ell-1}] \nonumber \\
\mathcal{T}_\ell^{mag 5} & = & \frac{1}{2} [a_0+a_1 \tau_3^i ] M_{\ell,\ell} \approx  M_{\ell,\ell}.
\label{eq:SD_Operators}
\end{eqnarray}
In the above equations, $\tau_3^i = a_0=a_1=1$ for a nucleus with an extra proton and $\tau_3^i = -a_0=a_1=-1$ for a nucleus with an extra neutron and $m_\pi$ is the pion mass. It can be shown that the three operators in Eq.~\eqref{eq:SD_Operators} do not interfere and hence can be squared separately and then summed. The expectation value of $M_{\ell,\ell'}$, which appears in each of the three terms, is
\begin{align}
& \int d^{3}r \, M_{\ell,\ell'} \Psi_{v' J' f'}^{*}(\mathbf{r})\Psi_{0 J_{\rm init} f_{\rm init}}(\mathbf{r})  = (-1)^{\ell'} \sqrt{\frac{6}{4\pi}} \times \nonumber \\ 
& \biggr[ (2\ell'+1)(2\ell+1)(2J'+1)(2J_{\rm init}+1)(2f'+1)(2f_{\rm init}+1) \biggr]^\frac{1}{2} \nonumber \\
& \times \begin{pmatrix} J' & \ell' & J_{\rm init} \\ 0 & 0 & 0 \end{pmatrix}  \begin{Bmatrix} J' & J_{\rm init} &\ell' \\ \frac{1}{2} & \frac{1}{2} & 1 \\ f' & f_{\rm init} &\ell \end{Bmatrix} \nonumber \\
& \times \int dr \, \phi_{v' J' }(r) j_{\ell'}(\frac{\mu_{12}}{m_1} q r) \phi_{v J_{\rm init}}(r).
\end{align}
Here, $m_1$ is the mass of the particle with which the interaction occurs. Finally, the result should be summed over $f'$ and averaged over $f_{\rm init}$.

\section{Molecular Candidates}
\label{sectMol}

A DM-molecule collision can lead to a ro-vibrational excitation of the molecule within the ground electronic state. Since molecules have vibrational energy spacings of $\lesssim$0.5~eV, they are excellent targets for probing DM masses below $m_\chi\lesssim $100~MeV.  However, in order to efficiently produce observable photons after a DM scattering event and also in order to ensure that our theoretical calculations are reliable, the molecule must satisfy several properties. 

Only polar molecules possess ro-vibrational transitions within the same electronic state with sufficiently short lifetimes $\lesssim$1~ms to ensure that the molecule does not collisionally quench through the different energy exchange mechanisms described in Sec.~\ref{sec:Exp_Considerations}. Moreover, the molecules should be chemically stable and should be easily available. While polar polyatomic molecules may be viable candidates for DM detection, their density of states is large since they have a large number of degrees of freedom. This typically induces efficient intramolecular energy exchange that allow for efficient quenching~\cite{Hovis1978,Hess1980,Hager1980,Zittel1989}, but also complicates the theoretical understanding of a DM scattering event. Thus, we focus on diatomic molecules, where only a single vibrational degree of freedom is available in addition to the rotational degrees of freedom (assuming electronic transitions are negligible). We also choose to work with X~$^1\Sigma^+$ molecules due to their absence of spin-orbit coupling, which allows for an easier description of energy exchange processes (Hund case b).

We present a few candidate molecules within this group, together with their main spectroscopic properties, in Table~\ref{T3}. The table shows the harmonic frequency $\omega_e$, the anharmonicity constant $\omega_e x_e$, the rotational constants at the equilibrium distance $B_e$, the first anharmonic correction to the rotational constant $\alpha_e$ and the rotationless Einstein $A_{10}$-coefficients.  We also estimate the vibrational quenching rate $k_{10}$, i.e, the relaxation rate per unit pressure for a molecule from the $v=1$ state to the $v=0$ state (more generally, $k_{v'v}$ is the relaxation rate to relax from $v'$ to $v$). Note that Table~\ref{T3} presents values evaluated at room temperature. Details on evaluation of $k_{10}$ at the lower temperatures relevant for this study are given below.

\begin{table*}[ht]
\centering
 \caption{Molecular parameters and main relaxation properties associated with molecules relevant for DM direct detection. All parameters are defined with respect to the ground electronic state of the molecules, which is X$^1\Sigma^+$, with the exception of H$_2$, which is X$^1\Sigma_g^+$. The standard spectroscopic parameters ($\omega_e$, $\omega_e x_e$, $B_e$, and  $\alpha_e$) are taken from the National Institute of Standards and Technology (NIST), with the exception of HSc, whose constants are taken from~\cite{RAM1997263,Ram1996}.  The dissociation energy, $D_e$, is from~\cite{HerzbergDiatomics}. The rotationless Einstein A coefficients are shown in s$^{-1}$ and have been calculated as explained in the text.  Empty spaces indicate the absence of any theoretical prediction or experimental data.}
 \begin{tabularx}{12.5 cm}{c c c c c c c c} 
 \hline
 \hline
 Molecule & $D_e$(eV)&$\omega_e$(eV) & $\omega_e x_e$(eV) & $B_e$(eV) & $\alpha_e$(eV) & $A_{10}$ (s$^{-1}$) & $k_{10}$(s$^{-1}$Torr$^{-1}$) \\ [0.5ex] 
 \hline
$^{12}$C$^{16}$O &11.22 &0.269 & 1.65$\times 10^{-3}$ & 2.39$\times 10^{-4}$ &  2.17$\times 10^{-6}$ & 33.9& 2.00$\times  10^{-3}$\footnote{Experimental value at 300~K taken from~\cite{Kovacs1972}.}\\ 
 HF&6.12  &0.513 & 1.11$\times 10^{-2}$ & 2.60$\times 10^{-3}$ &  9.89$\times 10^{-5}$ & 193.7& 4.7$\times 10^{4}$\footnote{Experimental rate at 300~K taken from~\cite{Hancock1975}.}\\ 
 HCl &4.62& 0.371 & 6.55$\times 10^{-3}$ & 1.31$\times 10^{-3}$ &  3.81$\times 10^{-5}$ & 42.7& 780\footnote{Experimental value at 298~K taken from~\cite{Zittel1973}.}\\
 HBr &3.92& 0.328 & 5.61$\times 10^{-3}$ & 1.05$\times 10^{-3}$ &  2.89$\times 10^{-5}$ & 8.09& 605\footnote{Experimental value at 296~K taken from~\cite{Losert1988}.}\\
 HI &3.20&  0.286& 4.91$\times 10^{-3}$ & 7.96$\times 10^{-4}$ &  2.09$\times 10^{-5}$ &0.06 & 375\footnote{Average experimental value at 295~K taken from~\cite{Chen1968}.}\\
 HSc  & & 0.192& 3.00$\times 10^{-3}$& 6.72$\times 10^{-4}$ &  1.54$\times 10^{-5}$ & 72.86~\footnote{See~\cite{Lodi2015}.}& \\
  AlF & 6.94 & 0.100 &  5.91$\times 10^{-4}$ & 6.85$\times 10^{-5}$  &6.18$\times 10^{-7} $ & & \\
 H$_2$&4.75 &  0.546 & 1.50$\times 10^{-2}$ & 7.54$\times 10^{-3}$ &  3.80$\times 10^{-4}$ &8.54$\times 10^{-7}$\footnote{See~\cite{Wolniewicz1998}.}  & 0.45\footnote{Experimental value at~300 K taken from~\cite{Kovacs1972}.}\\ 
 \hline
 \hline
\end{tabularx}
\label{T3}
\end{table*}

In order to evaluate whether a molecule is a good candidate for a DM search and in order to characterize the observed photon signal, one should consider three main properties: 
\begin{itemize}
\item \textbf{Blackbody radiation.} Molecules with larger vibrational spacing, $\omega_e$, will have a lower blackbody radiation rate at a given temperature. Note however that larger $\omega_e$ implies less sensitivity to lower DM masses.
\item \textbf{Vibrational quenching.}  In order to produce co-quench photons, the process $\text{AB}(1)+\text{AB}(0)\xrightarrow{k_{10}}\text{AB}(0)+\text{AB}(0)$, in which an excited molecule collisionally decays without emitting a photon, must be smaller than the spontaneous emission rate, i.e., $k_{10}\ \times p\gtrsim A_{10}$. This vibrational quenching rate is dominated by $V$-$T$ and $V$-$R$ energy transfer mechanisms. For a given temperature, the allowed maximum pressure will be $A_{10}/k_{10}$, which is directly related to the number of available target molecules. We see in Table~\ref{T3} that CO is an excellent candidate, since its ratio $A_{10}/k_{10}$ is five orders of magnitude larger than for HX molecules, where X is either a metal or a halogen atom.
\item \textbf{Collisional quenching.} The process $\text{AB}(v')+\text{AB}(0)\xrightarrow{k_{v'v}}\text{AB}(v)+\text{AB}(\text{any state})$ plays a major role in determining whether a molecular target produces photons or is instead non-radiatively quenched, and hence whether the photons that are produced will be cascade or co-quench photons. In CO, this process is governed by $V-V$ transfer, which is slow and conserves the total vibrational quantum number, hence allowing for the production of multiple photons.  Moreover, for high vibrational states ($v\gtrsim 10$, depending on pressure and $J$), spontaneous emission dominates over $V-V$ transfer, and hence cascade photons are produced, while for low vibrational states with $1\lesssim v \lesssim 10$, $V-V$ transfer is more efficient than spontaneous emission and co-quench photons are produced. In contrast, for the HX molecules, the $V-T$ and $V-R$ transfers are more efficient than the $V-V$ transfer~\cite{Jurisch1981,Shin1983,Ahl1973}. Hence, for the HX systems, the pressure must be extremely low and only cascade photons are produced.

\end{itemize}

We conclude that CO is an excellent candidate, since it allows for a large number of target molecules at a given temperature and pressure. Moreover, in order to increase the mean free path of co-quench photons, one must add a buffer gas, such as He, in order to broaden absorption lines. Fortunately, the available data and theoretical results of CO-He collisions~\cite{Krems2002} suggest that the rate for the process CO(1)+He$\rightarrow$ CO(0) + He, which would eliminate a co-quench signal, is the lowest rate among the diatomics considered for T$\lesssim$300~K in that study. In addition, CO has one of the most energetic molecular bonds in chemistry.

Finally, we emphasize that aluminium fluoride (AlF) seems to be a great alternative to CO, since it has similar electronic properties and a large dissociation energy.  Indeed, the interest in this molecule is growing due to its unique properties for laser cooling~\cite{Wells2011} and its importance in astrophysics~\cite{Yousefi2018}.

\section{Relevant Molecular Processes in CO}
\label{COappend1}

\subsection{Spontaneous Emission Rate}

The state-to-state Einstein $A_{v'J';vJ}$ coefficients for the decay processes $(v',J')\rightarrow (v,J)$ are given by~\cite{Hansson2005}
\begin{equation}
\label{A}
A_{v'J';vJ}=\frac{64\pi^4\nu_{v'J';vJ}^3}{(4\pi \epsilon_0)3h} \frac{S_{J'J}}{2J'+1} |\langle \phi_{v'J'=0}(r) |d(r)|\phi_{vJ=0}(r) \rangle|^2,
\end{equation}
where $h$ is Planck's constant, $\epsilon_0$ is the electric permittivity in vacuum (for clarity, we will no longer use natural units from here and below), $d(r)$ is the dipole moment function of the X$^1\Sigma^+$ electronic state of CO and $\nu_{v'J';vJ}=(E_{v'J'}-E_{vJ})/hc$ is the frequency of the transition, where $c$ is the speed of light in vacuum. In Eq.~(\ref{A}), $S_{J'J}$ denotes the so-called H\"onl-London factor, which for transitions $^1\Sigma^{\pm} \rightarrow ^1\Sigma^{\pm}$,  is given by $S_{J'J'+1}=J'+1$ and $S_{J'J'-1}=J'$.

\begin{table}[t]
\centering
    \caption{Comparison of calculated and experimental dipole matrix elements, $\langle \phi_{v',J'=0}(r) |d(r)|\phi_{v,J=0}(r) \rangle$, for CO(X$^1\Sigma^+$) in units of Debyes. The experimental data are taken from~\cite{Roux1972,Wisbach1973,Roux1974}. }
 \begin{tabular}{c c c c} 
 \hline
 $v'$ & $v$ & Our Calculation & Measured \\ [0.5ex] 
 \hline
 1 & 0 & 1.05$\times 10^{-1}$ &1.05$\times 10^{-1}$ \\ 
 2 & 0 & -6.00$\times 10^{-3}$& -6.43$\times 10^{-3}$ \\
 3 &1 &  -1.07$\times 10^{-3}$& -1.27$\times 10^{-2}$ \\
  3& 1 & -1.08$\times 10^{-2}$ & -1.07$\times 10^{-2}$ \\
 5& 3 & -2.00$\times 10^{-2}$& -2.31$\times 10^{-2}$\\
 6& 4 & -2.50$\times 10^{-2}$ & -2.66$\times 10^{-2}$\\ [1ex] 
 \hline
\end{tabular}
\label{T1}
\end{table}

The vibrational wavefunctions have been obtained by solving numerically the pertinent Schr\"odinger equation by means of the Numerov method between 1.7~$a_0$ and 15~$a_0$ with a step size of 0.021~$a_0$ ($a_0$ is the Bohr radius), reaching a convergence better than one part per million. The dipole matrix elements have been obtained by means of the dipole moment function of Young and Eachus~\cite{Young1966} through numerical integration of  $\langle \phi_{v',J'=0}(r) |d(r)|\phi_{v,J=0}(r) \rangle$. Some of its values in comparison with the available experimental data are shown in Table~\ref{T1}, where we note that the present theoretical results agree with an accuracy better than $10\%$ for most of the considered transitions.

It is possible to introduce the rotationless Einstein A coefficient as
\begin{equation}
A_{v'v}=\frac{64\pi^4\nu_{v'v}^3}{(4\pi \epsilon_0) 3h}|\langle \phi_{v',J'=0}(r) |d(r)|\phi_{v,J=0}(r) \rangle|^2,
\end{equation}
where $\nu_{v'v}=\nu_{v'0;v0}$. This $A$ coefficient represents a characteristic vibrational spontaneous emission rate, whose inverse is very close to the realistic lifetime of the state. These coefficients for different initial states $v'$ and different final states $v=v'-\Delta v$ are shown in Table~\ref{T2}. The lifetime of the $v=1$ state in CO is about 30~ms. 

\begin{table}[t]
\centering
 \caption{Calculated rotationless Einstein $A_{v'v'-\Delta v}$ coefficients for CO(X$^1\Sigma^+$) in units s$^{-1}$.   }
 \begin{tabular}{c c c c c} 
 \hline
 & \multicolumn{3}{c}{$\Delta v$} \\
 $v'$ & 1 & 2 & 3 \\ 
 \hline
 1 & 33.90 & &  \\ 
 2 & 65.28 & 0.91 &  \\
 3 & 94.20 & 2.70 & 0.01 \\
 4 & 120.70 &5.37& 0.04  \\
 5&  144.83& 8.88 & 0.10 \\
  10&231.97 & 37.98 & 1.37\\
 15& 265.18 & 72.68 & 4.42\\
  20& 266.67& 122.13& 11.19\\
  23 &247.36 & 145.39& 13.78 \\
 26& 214.33 & 154.82 & 11.25\\ [1ex] 
 \hline
\end{tabular}
 \label{T2}
\end{table}

In order to check how the rotationless Einstein $A$ coefficients change for different dipole moment functions, we have calculated them also with the dipole moment function of~\cite{CHACKERIAN1983431}, finding a deviation up to 15$\%$ in the $A$ coefficients. This has a negligible impact on the projected DM sensitivity.

\subsection{Elastic Collisions}
\label{xsel}
\begin{figure}[t]
\centering
\includegraphics[width=1\columnwidth,keepaspectratio]{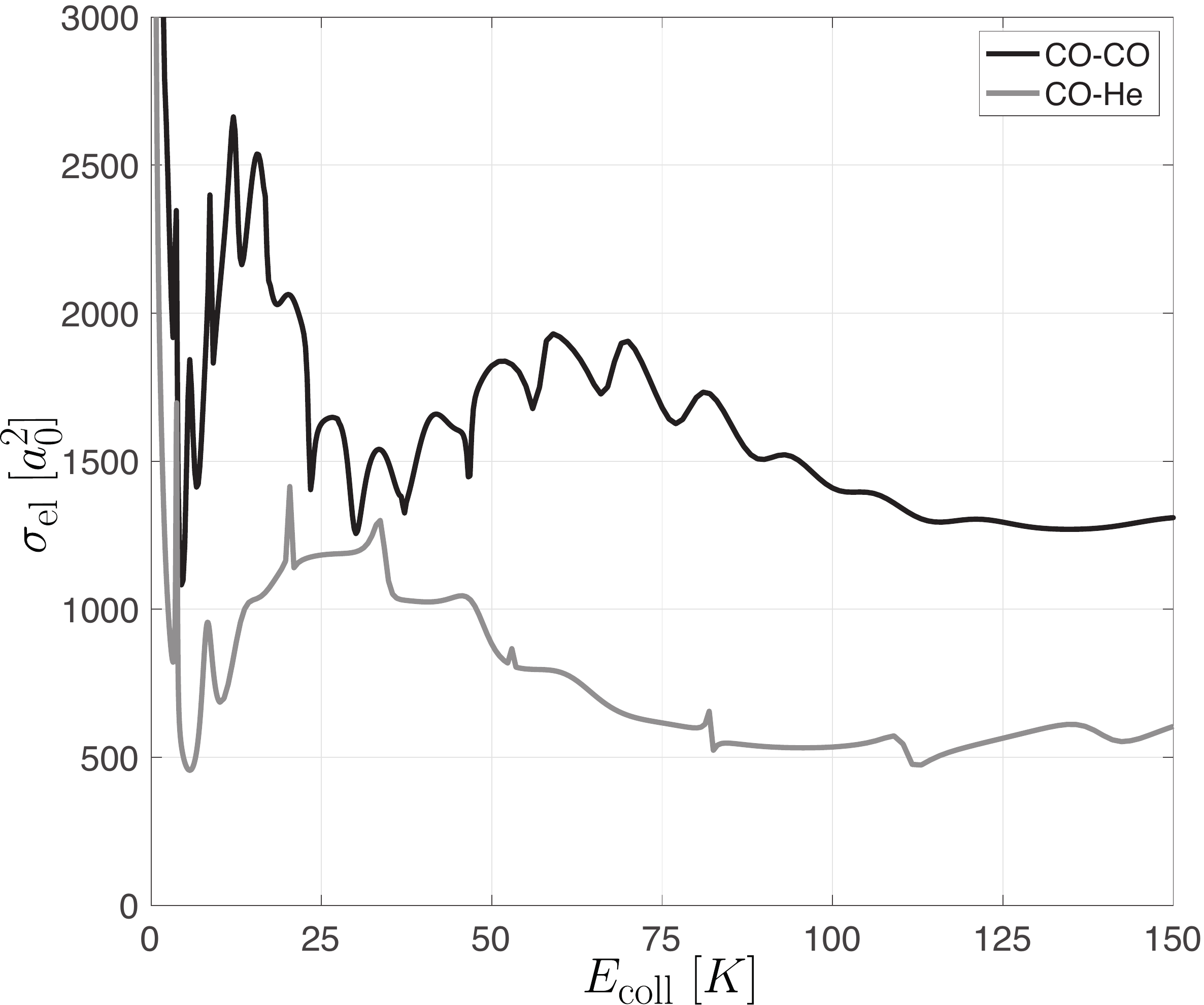}
\caption{Elastic cross section in atomic units for CO($v$=0) + He $\rightarrow$ CO($v$=0) + He and CO($v$=0) + CO($v$=0) $\rightarrow$ CO($v$=0) + CO($v$=0) as a function of the collision energy (in K). The cross sections do not show dependence on the rotational quantum number $J$, since we assume an isotropic interaction and an interaction potential that is spherically averaged. 
}
\label{figelast}
\end{figure}

Elastic collisions between molecules and between molecules and He atoms play an important role in determining the observable photon signal; they determine the collisional frequency (together with the gas temperature and density) and are crucial in broadening the absorption lines. We have calculated the relevant elastic collisions for atom-molecule and molecule-molecule collisions using a single-channel scattering approach and assuming that the collision takes place along the spherical component of the potential energy surface. 
The calculations follow Numerov's method; we propagate the single-channel wave functions from 3$a_0$ to 50$a_0$ with a uniform radial step size of 0.01$a_0$ (reaching a convergence of better than 0.01$\%$), and at larger radii match the wave functions with the regular and irregular Bessel functions.  This gives the $S$-matrix and, hence, the cross section. 
We now describe the details and results for CO-He and CO-CO collisions: 

\begin{itemize}

\item $\bf{CO-He}$. 
This scattering process plays a key role for the co-quench signal, since it determines the broadening of the 4.5 $\mu$m absorption line in CO, corresponding to the CO(1)$\rightarrow$CO(0) radiative decay process. We have solved the single channel scattering problem employing the Lennard-Jones potential of~\cite{Keil1979}. Our calculations include up to 100 partial waves and the results for the elastic cross section, $\sigma_{\rm el}$, are presented in Fig.~\ref{figelast} as the dashed-gray curve. We see that for a range of energies relevant for this work, the average elastic atom-molecule cross section is about 750$a_0^2$. Also, we observe several shape resonances, which originate from the large number of partial waves included in the simulations. 

\item $\bf{CO-CO}$. We have employed the potential energy surface of~\cite{Barreto2017} to obtain the spherical component of the CO-CO interaction. We use this potential in Numerov's propagation method, and include 120 partial waves. The results for the elastic cross section are shown as a black solid curve in Fig.~\ref{figelast}; we find that the molecule-molecule elastic cross section is approximately 1500 a$_0^2$. 
\end{itemize}

In Fig.~\ref{figelast}, independently of the collision partners under consideration, the elastic cross section shows a large number of shape resonances on top of an overall oscillatory behavior that are known as Glory undulations~\cite{Ford1959,Bernstein1962,Bernstein1973,Aquilanti1998,PEREZRIOS201228}. 

\subsection{CO($v=1,J$)+M Inelastic Collisions}

CO($v=1$)+M collisions are dominated by the $V$-$T$ transfer mechanism and are extremely inefficient, leading to the small relaxation rates shown in Table~\ref{T3}. This process plays a major role in constraining the maximum pressure allowed for CO at a given temperature before the spontaneous emission from the $v=1$ state is collisionally quenched, see Sec.~\ref{sec:Exp_Considerations}. In particular, for CO there are two relevant processes of this type:
\begin{itemize}
\item  $\bf{CO-He}$. The vibrational quenching of CO($v$=1) molecules in a buffer gas of helium has been theoretically studied by Krems~\cite{Krems2002} following a full quantum mechanical scattering approach, finding good agreement between theory and the available experimental data. The values reported by Krems have been employed in the present work. 
\item $\bf{CO-CO}$. To the best of our knowledge, there is no experimental data, nor theoretical study of the process ${\rm CO}(v=1) + {\rm CO}(v=0)\to {\rm CO}(v=0) + {\rm CO}(v=0)$ below 100~K. However, at 55~K there is a large amount of data for H$_2$+H$_2$ self-quenching reactions.  At 300~K the self-quenching rates for H$_2$ are two orders of magnitude larger than for CO (as can be seen in Table~\ref{T3}). However, we expect the description for both interactions to be similar at lower temperatures, due to the low anisotropy in both the CO and H$_2$ self-quenching interactions. As a conservative estimate, we therefore assume that the self-quenching rate of H$_2$ is the same as CO at 55~K. In principle, one would expect the CO rate to be lower than that for H$_2$, since CO is more harmonic and has a smaller rotational constant and therefore $V-T$ transfer is the only energy transfer mechanism of relevance. On the other hand, it is known that the vibrational relaxation rate becomes constant for sufficiently low temperatures (which in the case of H$_2$-H$_2$ occurs at approximately 80~K~\cite{Samantha2011}). However, it is not known at which temperature this occurs in CO-CO interactions. The elucidation of this point will be the matter of future work. 
\end{itemize}

\subsection{CO($v$)+CO($0$) Collisions}

\begin{figure}[t]
\centering
\includegraphics[width=1\columnwidth,keepaspectratio]{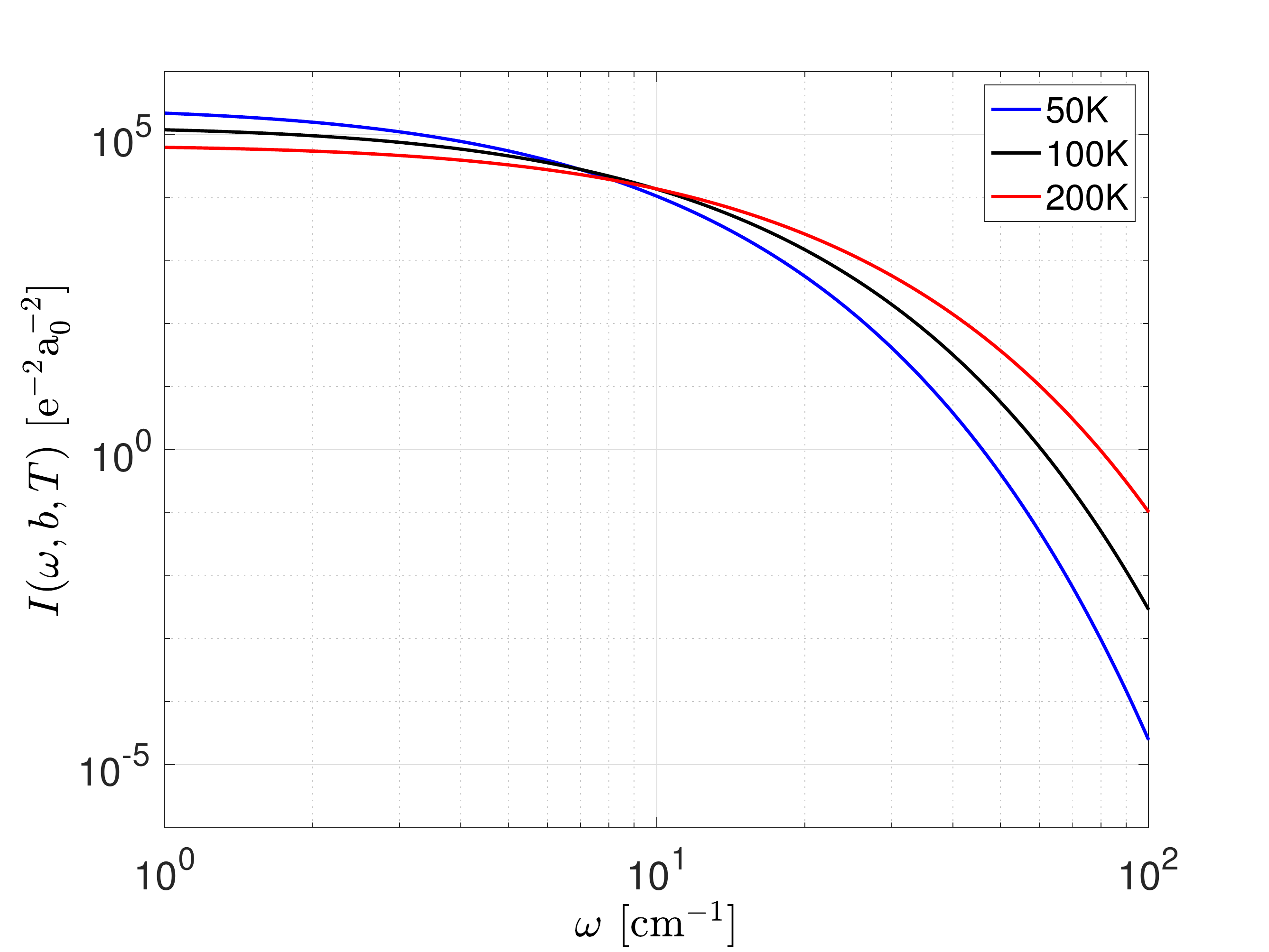}
\caption{The function $I(\omega,b,T)$ from Eq.~\eqref{eq12} in atomic units for $b=7$a$_0$, which is needed to determine the probability of resonant quenching through the process CO(0)+CO($v$)$\rightarrow$CO(1)+CO($v-1$), see Eq.~\eqref{eqvib}.}
\label{figprob}%
\end{figure}

This scattering process is dominated by $V-V$ transfer, which becomes efficient when the excitation energy of one of the colliding partners is very close to the de-excitation energy of the other partner. This process is also known as vibrational resonant quenching. For some molecules with permanent dipole moment (which are infrared active), this $V-V$ energy exchange is due to the long-range dipole-dipole interaction between the colliding partners. Thus, applying first-order time dependent perturbation theory within the impact parameter approximation, it was shown in~\cite{Levon1974,Levon1973} that the probability of resonant quenching for the process AB(0,$J_1$)+AB($v,J_2$)$\rightarrow$AB($1,J_1+\Delta J_1$)+AB($v-1,J_2+\Delta J_2$), thermally averaged over $J_1$ (which is just $J_{\rm init}$ for the specific case described in the body of this study) is given by
\begin{eqnarray}
\label{eqvib}
P_{v,v-1,J_2}(T)&=&C_{v,v-1}^2\sum_{J_1}\sum_{\Delta J_2,\Delta J_1}I(\omega(v,J_1,J_2,\Delta J_1,\Delta J_2),b,T) \nonumber \\  
&&\times \xi(\Delta J_1)\xi(\Delta J_2)P_{\rm therm}(J_1,T)\,,
\end{eqnarray}
where
\begin{equation}
C_{v,v-1}=\left(\frac{2}{3} \right)^{1/2}\frac{J+1}{\sqrt{(2J+1)(2J+3)}}d_{0,1}d_{v,v-1} 
\end{equation}
is the root-mean-squared of the dipole-dipole interaction energy within Margenau's approach~\cite{Mahan1966}, where $J$=max$(J_1,J_2)$ and $d_{v',v}=\langle \phi_{v'J'=0}(r)|d(r)| \phi_{vJ=0}(r)\rangle$ is the dipole matrix element of a given vibrational transition. The quantity $\omega(v,J_1,J_2,\Delta J_1,\Delta J_2)$ is the energy difference between the internal energy of the initial and final scattering states, also known as the energy defect, and
\begin{eqnarray}
\label{eq12}
I(\omega,b,T)&=&8\left(\frac{\omega \mu}{2bk_BT}\right)^2  \\
& & \times \int_{0}^{\infty} \frac{dv}{v} \ \big [ K_0^2(\omega b/v)+2K_1^2(\omega b/v) \big ] 
 e^{-\frac{\mu v^2}{2k_BT}} \nonumber
\end{eqnarray}
is proportional to the thermal averaged cross section divided by the hard-sphere cross section $\pi b^2$, with $b=7$a$_0$; $K_{\eta}(x)$ denotes the modified Bessel function of order $\eta$. The dependence of $I(\omega,b,T)$ on the temperature $T$ and energy difference between the initial and final scattering state $\omega$ is shown in Fig.~\ref{figprob}.  We see that the resonant quenching probability depends only weakly on the temperature, but depends strongly on the energy difference. In particular, for moderate 
$\omega\sim50$~cm$^{-1}$, there is a suppression between three to five orders of magnitude with respect to $\omega\sim10$~cm$^{-1}$. Since the molecular potential is more anharmonic for higher vibrational states, $\omega$ increases with the vibrational quantum number $v$, and hence highly vibrational states will have a lower $V-V$ transfer rate, which is experimentally confirmed~\cite{Gower1975,Smith1973,Wittig1972,Powell1975}. However, at even larger vibrational states, it is expected that the $V$-$T$ transfer will begin dominating the quenching process, as has been observed in NO and O$_2$~\cite{Yang1992,Hancock2009,Park1994}.

\begin{figure}[t]
\centering
\includegraphics[width=1\columnwidth,keepaspectratio]{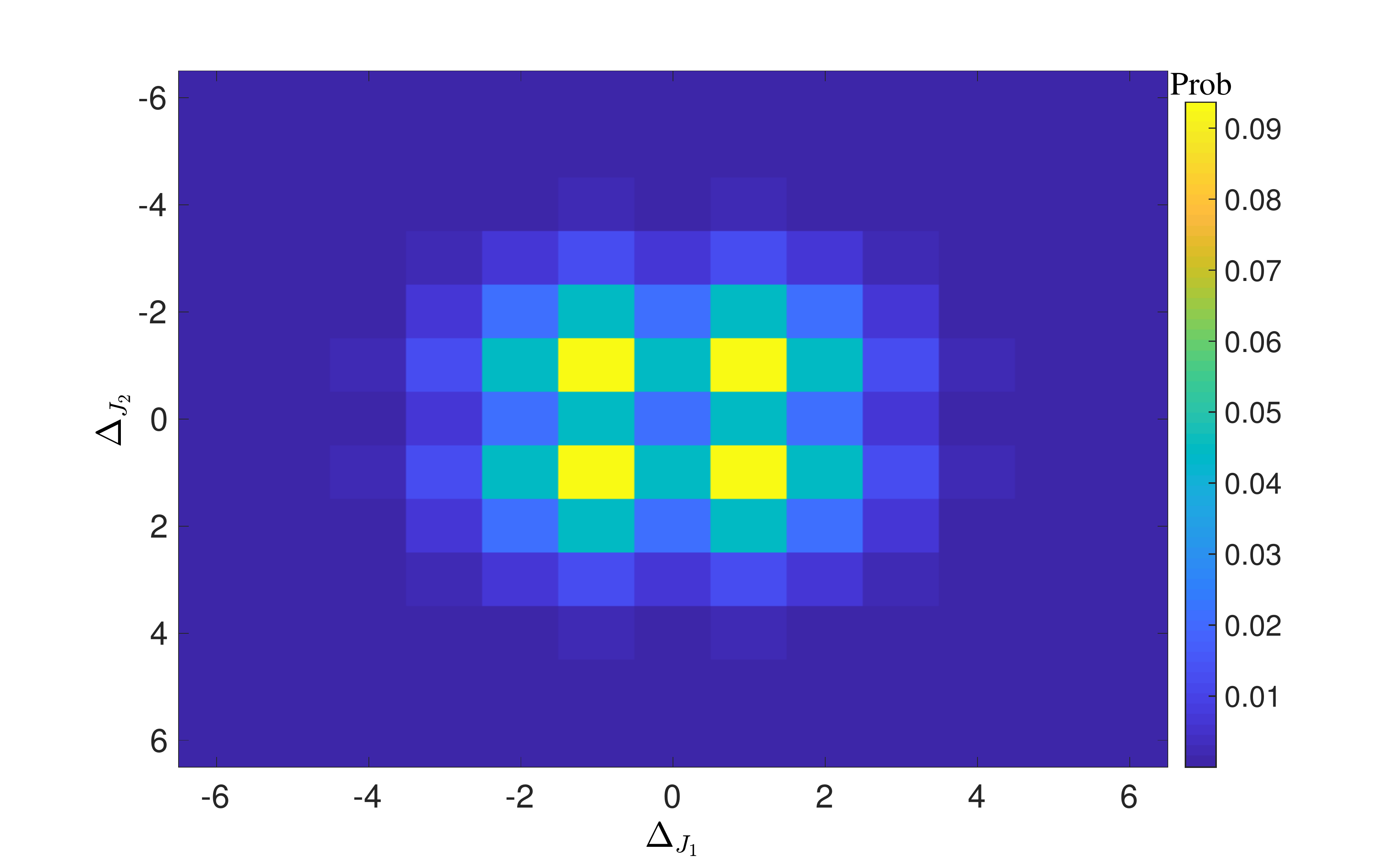}
\caption{Probability distribution for the change in the rotational quantum number, $\Delta J_1$ and $\Delta J_2$, in the resonant quenching process CO(0)+CO($v$)$\rightarrow$CO(1)+CO($v-1$), see Eq.~\eqref{eqvib}.  We choose $\sigma=$0.5 in Eq.~\eqref{eq:dist-Delta-j}, and take $\xi(0)=\xi(2)$.}
\label{figcomp}%
\end{figure}

In Eq.~(\ref{eqvib}) the coupling between different rotational states is simulated by the following distribution 
\begin{equation}\label{eq:dist-Delta-j}
\xi(\Delta J)=\frac{e^{-\sigma^2(\Delta J^2-1)}}{\sum_{i=1}^{N_J}e^{-\sigma^2(\Delta J_i^2-1)}} \ \ (\Delta J_i\ne0)\,,
\end{equation}
which represents the propensity rule for $\Delta_J$ collisions on the basis of experimental findings. The distributions for $\Delta J_{1,2}$ are shown in Fig.~\ref{figcomp}, where it is seen that the simple dipole selection rule, $\Delta_J=0,\pm1$, is not always applicable.

The accuracy of Eq.~(\ref{eqvib}) has been verified experimentally for resonant quenching transitions $v=4-10$ in CO at 100~K~\cite{Gower1975,Levon1974}. Finally, we would like to emphasize that the impact parameter approximation employed to derive Eq.~(\ref{eq12}) is justified in our case since $mvb/\hbar \gg 1$; in particular, for temperatures $\sim$50~K, $mvb/\hbar \sim 10$.

The approach of Lev-On et al.~\cite{Levon1974,Levon1973}  followed here accounts for the $V-V$ energy transfer in self-relaxation of highly harmonic molecules.  However, it does not include explicitly the role of rotational states and, more importantly, it does not include a $V-T$ transfer pathway. This may be accounted for by more involved methods such as in~\cite{Shin1971,Shin1983} or the more advanced semiclassical model in~\cite{Billing2003}. However, neither of these models have been tested below 100~K, which is the relevant temperature range of interest for our detection concept, and there is no experimental data on the vibrational relaxation rates in CO to benchmark the calculation with the exception of the work of Gower et al.~\cite{Gower1975} at 75~K. More work is needed to accurately establish the quenching rates, which could affect the precise boundary in $v'$ and $J'$ between the cascade and co-quench signals as seen in Fig.~\ref{fig:v_j_Plane}. 
 
\subsection{CO-He complex formation}

To ensure a large mean free path for the co-quench photons, we have proposed the use of a high-pressure He buffer gas. However, collisions of CO with He will eventually lead to the formation of a resonant complex COHe$^*$ as the consequence of a shape, a Fano-Feshbach, or a hybrid resonance. This complex may be further stabilized to form a COHe molecule.  Some of the CO molecules will therefore be lost as active targets for DM detection if the pressure of the He buffer gas is too high. 

Several resonances in the CO($v=0,j=0$)+He $\rightarrow$ CO($v=0,J=1$)+He cross section between 5 and 13 cm$^{-1}$ (0.62 and 1.61 meV) of collision energy have been experimentally observed and theoretically corroborated~\cite{Bergeat2015}. At 55~K, there is a $\sim$10$\%$ probability to find CO molecules and He atoms with the required relative velocity to fall into the relevant range of collision energies, and the mentioned resonances may play some role. These resonance show a width $\gtrsim$ 0.25~cm$^{-1}$ (3.1$\times$ 10$^{-5}$ eV), and the COHe$^*$ complex will therefore have a lifetime $\tau_{\text{res}}$ $\lesssim 20$~ps. This complex may be stabilized through a collision with a third body as
\begin{eqnarray}
\label{reaction}
\text{CO} + \text{He}& \rightarrow &\text{COHe}^* \nonumber \\
 \text{COHe}^* + \text{He} &\rightarrow& \text{COHe} + \text{He}\,,
\end{eqnarray}
which is known as the indirect approach for three-body recombination. It is worth emphasizing that the second reaction in Eq.~\eqref{reaction} may also lead to the dissociation of the complex into CO + He. In other words, once the complex is formed and a second collision occurs, there is a chance of stabilization and a chance of dissociation (both probabilities are unknown in our case).

To avoid the formation of COHe, the time $\tau_{\text{col}}$ for the collision $\text{COHe}^* + \text{He}$ must be smaller than $\tau_{\text{res}}$, i.e., $\tau_{\text{col}}=(\rho k_{\text{el}})^{-1}\lesssim \tau_{\text{res}}$, where $\rho$ is the density of He and $k_{\text{el}}$ is the CO+He elastic scattering rate, which has been obtained by averaging $u\sigma_{\text{el}}(u)$ over a Maxwell-Boltzmann  distribution of velocities $u$. The elastic scattering cross section as a function of the collision energy, $\sigma_{\text{el}}(u)$, is shown in Fig.~\ref{figelast}, see Sec.~\ref{xsel}. Next, taking into account the relation between density, pressure, and temperature, we find $p\lesssim (k_B T)/(k_{\text{el}} \tau_{\text{res}}) \sim \mathcal{O}$(1~bar) at 55~K, where $k_B$ stands for the Boltzmann constant. This estimate assumes that $\text{COHe}^* + \text{He} $ collisions lead to stabilization with 100$\%$ probability, leading to a very conservative estimate of the maximum allowed He pressure.  For our estimate of the mean free path of the co-quench photons, we take the He pressure to be 4~bar, which we believe is a realistic maximum He pressure before COHe occurs with a non negligible probability. 
Future calculations are required to quantify precisely the maximum allowed He pressure.  

\section{Relevant Molecular Processes in H-X Systems}
\label{HXappend1}
\subsection{Spontaneous Emission Rate}

The state-to-state Einstein $A_{v'J';vJ}$ coefficients corresponding to the decay processes $(v',J')\rightarrow (v,J)$ for the different hydrogen halides studied in the present work have been calculated following Eq.~(\ref{A}) employing the accurate potential curves of Coxon et al.~\cite{Coxon2015} and the transition dipole moment functions of Li et al.~\cite{Li2013}. The results for HF and HBr are shown in Table~\ref{TH1}, where we see that the Einstein $A$ coefficients for HF are larger than those for HBr. This is a consequence of the smaller energy spacings in HBr compared to HF, in addition to the intrinsically smaller dipole moment function of HBr compared to HF~\cite{Li2013}. Our results for HF agree well with previous theoretical calculations~\cite{Sileo1976,Zemke1991}. The Einstein $A$ coefficients for HSc have been taken from the ExoMol database~\cite{ExoMol}.

\begin{table}[t]
\centering
 \caption{Calculated rotationless Einstein A$_{v'v'-\Delta v}$ coefficients for HF(X$^1\Sigma^+$) and HBr(X$^1\Sigma^+$)  in s$^{-1}$. The upper entry is for HF whereas the lower entry is for HBr.}
 \begin{tabular}{c c c c c} 
 \hline
 & \multicolumn{3}{c}{$\Delta v$} \\
 $v'$ & 1 & 2 & 3 \\ 
 \hline
 1 & 193.7 & &  \\ 
   &  8.1& &  \\ 
 2 & 332 & 21.9 &  \\
   & 14.4 & 0.3 &  \\
 3 &  418.9& 64.5 & 1.1 \\
   &  18.9&1.04 &0.02  \\
 4 & 458.9 & 126.1& 4.2   \\
  & 21.4 & 2.52& 0.05   \\
 5&  458.1& 204.5 & 10.4 \\
  &  22.1& 4.9 & 0051 \\[1ex] 
 \hline
\end{tabular}
 \label{TH1}
\end{table}

\begin{figure*}[t]
\centering
\includegraphics[width=1\textwidth,keepaspectratio]{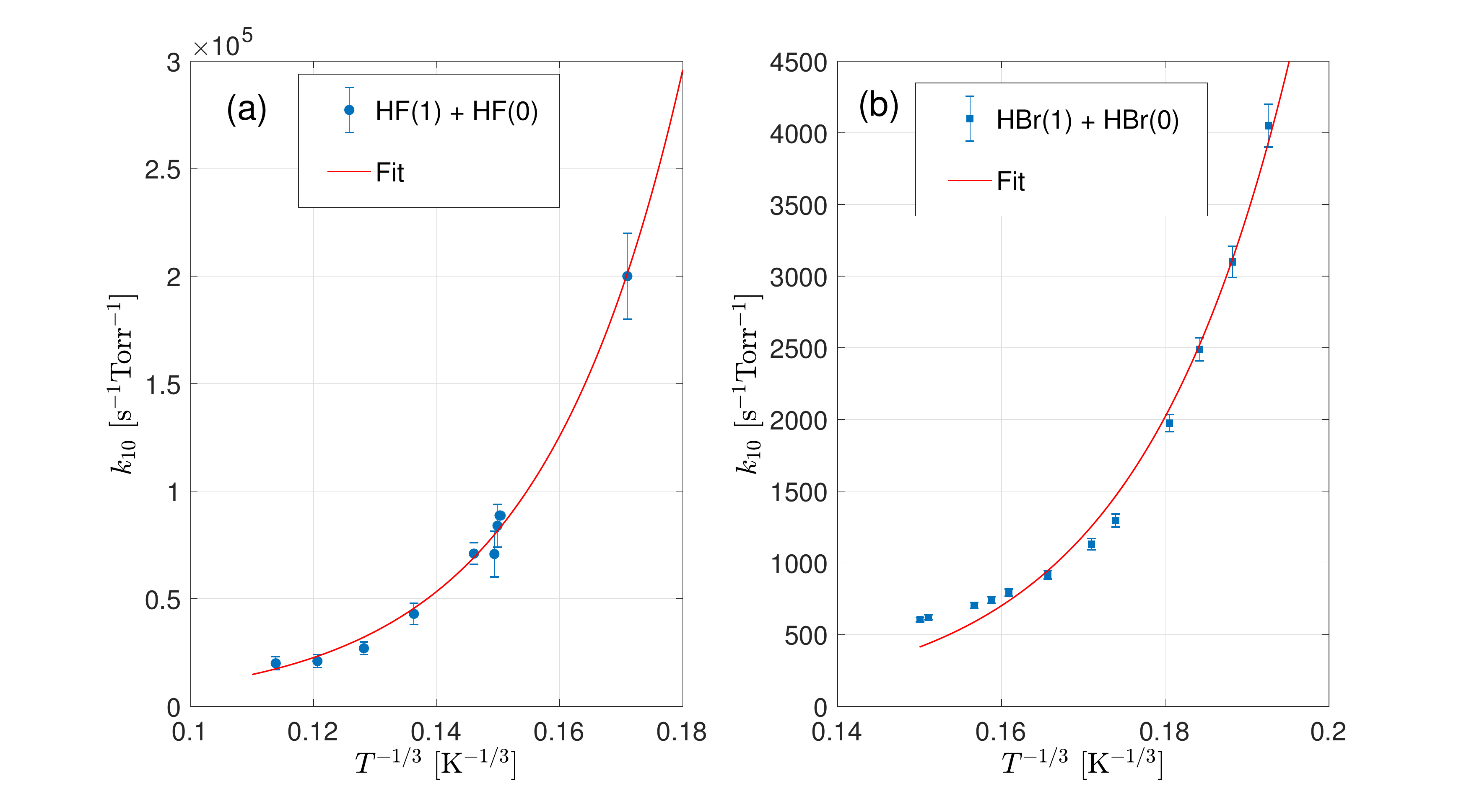}
\caption{Temperature dependence of the vibrational relaxation rate for hydrogen halides. The left panel shows the results for HF, where the blue dots are the experimental data taken from~\cite{Ahl1973,Green1973,Lucht1974,Hancock1975}.  The right panel shows the results for HBr, where the blue dots are the experimental data from~\cite{Losert1988}. In both panels the solid-red curves are fits to the data of exponential functions whose arguments depend on the temperature as $T^{-1/3}$. 
}
\label{figHX}%
\end{figure*}

\subsection{HX($v=1$)+HX($v=0$) Collisions}

The vibrational relaxation of hydrogen halides is dominated by $V-R$ and $V-T$ energy transfer mechanisms and have been studied mostly for temperatures above 200~K, with the exception of HI~\cite{Leone1982}. The relevant rates for the present work at room temperature are shown in Table~\ref{T3}. However, our proposal requires a molecular gas temperature below 100~K to avoid blackbody radiation backgrounds, which has not yet been explored experimentally or theoretically. Nevertheless, looking at the available data for the vibrational relaxation of hydrogen halides as a function of temperature~\cite{Leone1982}, we notice that between 300~K and 200~K the rates are proportional to $\exp{[T^{-1/3}]}$, as shown in Fig.~\ref{figHX} for HF and HBr. Indeed, the {\it universality} of this behavior may be related to the fact that all the hydrogen halides are dominated by the same $V-R$ and $V-T$ energy transfer mechanisms for vibrational relaxation. We extrapolate this temperature dependence to infer the vibrational relaxation rates below 100~K, the range relevant for the present proposal.  

Full quantum simulations have previously been carried out to study the vibrational relaxation of a number of molecules down to 10~K, leading to a softer dependence on the temperature instead of the exponential dependence observed at higher temperatures. In particular, the vibrational relaxation rate is almost independent of temperature for $T \lesssim 80$~K~\cite{Samantha2011,BalaCOH2} for CO-H$_2$ and H$_2$-H$_2$.  A similar behavior is predicted and observed for CO-He~\cite{Krems2002,ExpCOHe} and H$_2$-He~\cite{McGuire1975}.  Therefore, our assumptions that the vibrational rates depend exponentially on the temperature is very conservative.  Future work may indicate that we can work at much higher pressures for halides. This could dramatically improve their reach.

\begin{table}[t]
\centering
 \caption{Vibrational relaxation rate for HX$(v)$-HX(0) collisions for different vibrational states and operating temperature (in K). The rates are presented in units of 10$^6$ s$^{-1}$Torr$^{-1}$}
 \begin{tabular}{c c c c c c} 
 \hline
 Molecule & T(K) & $k_{21}$ &  $k_{31}$ &  $k_{42}$ &  $k_{21}$ \\ 
 \hline
 HF & 115 & 4.28 & 10.7 & 26.1 & 53.9  \\ 
 HCl & 83 & 6.07 & 5.55 & 6.45 & 32.5\\
 HBr &  72&  7.22 & 21.6 & 46.9 & 85.7\\
 HI & 60&  0.69 & 2.05 & 4.45 & 8.13  \\
 HSc&  46/44& 2.04/2.56 & & & \\[1ex] 
 \hline
\end{tabular}
 \label{TH2}
\end{table}

\begin{figure*}[t]
\centering
\includegraphics[width=0.47\textwidth]{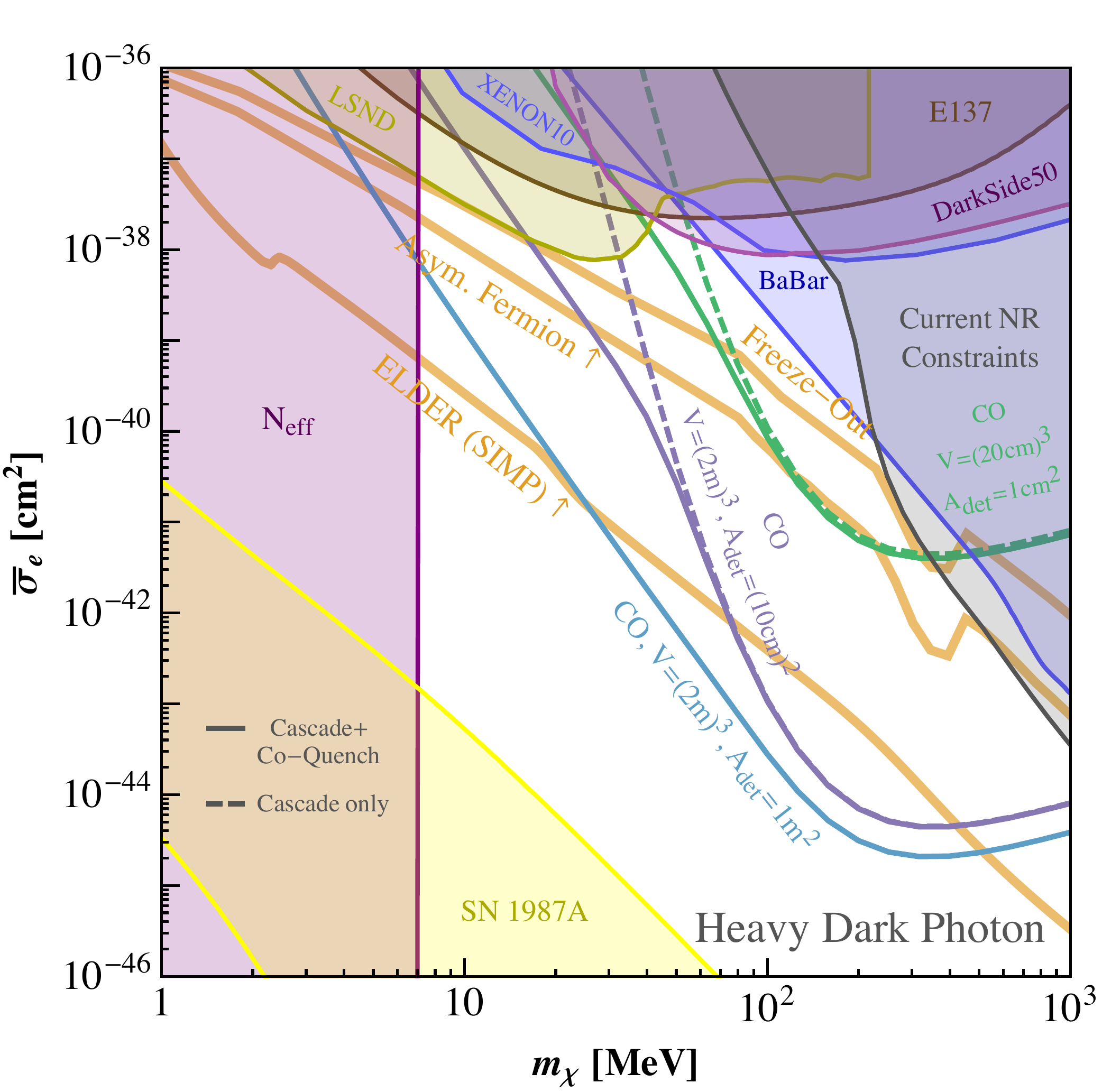}
\hspace{0.0cm}
\includegraphics[width=0.47\textwidth]{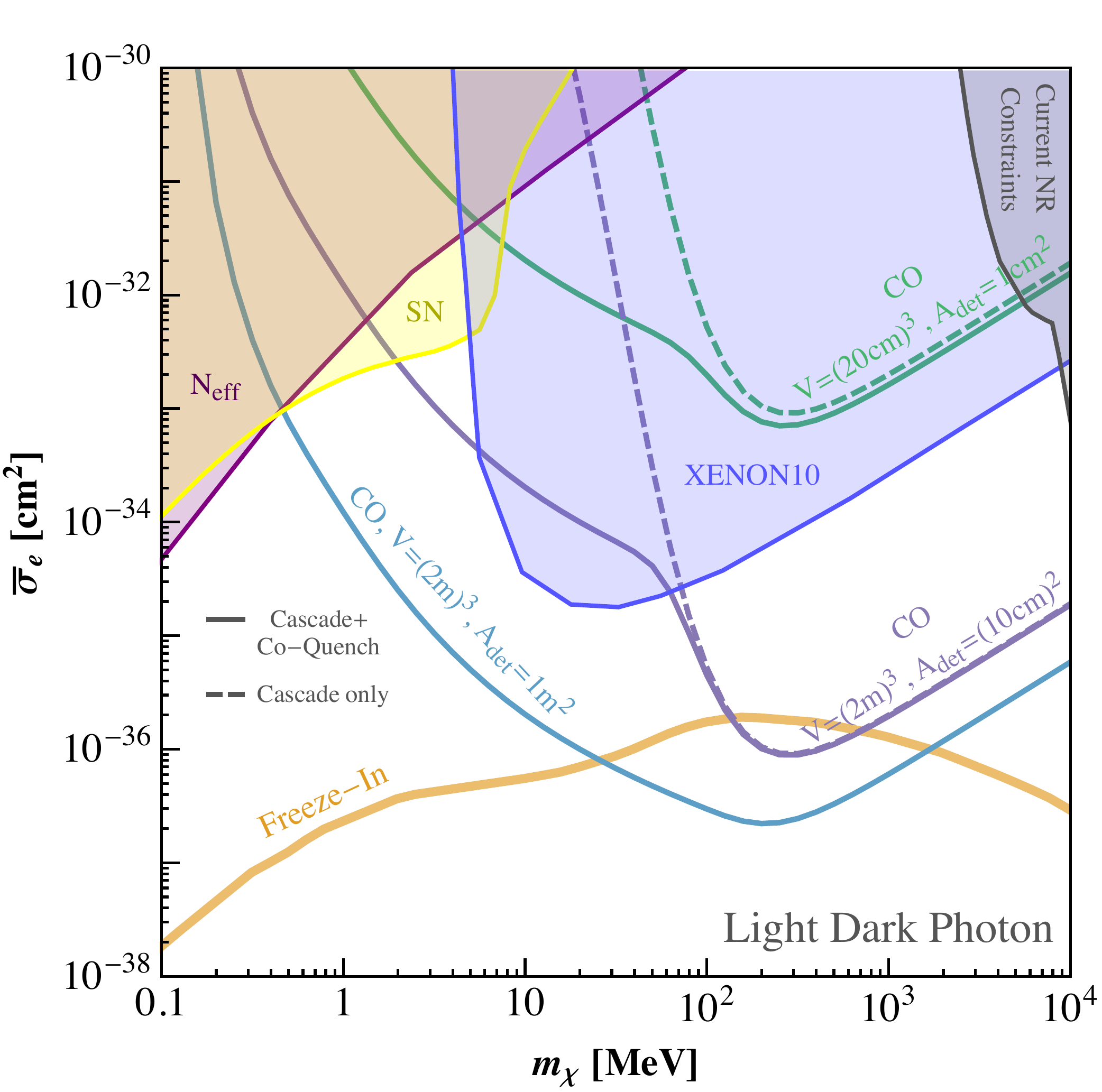}
\caption{The 95\% confidence-level sensitivity for DM spin-independent nuclear scattering off CO molecules via a dark photon mediator at a partial pressure of $5$~mbar and a temperature of $55$~K, for various tank sizes and photodetector area and for a time exposure of 1~year.  Limits are converted from $\sigmap$ into $\bar{\sigma}_e$ according to Eq.~\eqref{eq:sigman_sigmae} for comparison purposes. The signal consists of two-or-more coincident photons, and we assume zero background events. Colored-dashed curves correspond to CO cascade photons while colored solid curves correspond to the sum of cascade and co-quench photons.  Since co-quench photons are produced at a smaller energy threshold, they reach to lower DM masses. The co-quench signal assumes the addition of an He buffer gas at a pressure of $4$~bar and an internal mirror that helps to focus photons to the detector. In the {\bf left} panel, we show the prospective reach assuming a heavy dark photon with mass $m_{A'}=3m_\chi$, leading to a trivial DM form factor $\FDMsq=1$.  Also shown are existing stellar and terrestrial limits as well as the contour where freeze-out sets the correct relic abundance of the DM. In the {\bf right} panel, we show the prospective reach assuming an ultra-light dark photon with mass $m_{A'}\ll \alpha_{\rm EM} m_e$, leading to a DM form factor of $\FDMsq=\left(\alpha_{\rm EM} m_e/q\right)^4$.  Also shown are existing stellar and terrestrial limits as well as the contour where freeze-in sets the correct relic abundance of the DM. At low DM masses, the prospective reach falls sharply due to the small momentum transfer and Thomas-Fermi screening. See text for details.}
\label{fig:Sigmae_HiddenPhoton}%
\end{figure*}

HSc has not been explored in detail until now, and only some of its spectroscopic details are known.  Hence no information is known about the dynamics of its vibrational quenching. However, by studying the relation between the molecular properties and vibrational relaxation rates in hydrogen halides, it is possible to estimate the vibrational quenching rate in HSc to be $340$~s$^{-1}$Torr$^{-1}$. In particular, molecules with larger harmonicity and smaller rotational constants have smaller vibrational relaxation rates, as can be seen in Table~\ref{T3}. However, more work is required to quantify the maximum pressure for HSc gas, which could impact its ultimate utility as a DM target.  

\subsection{HX($v$)+HX($0$) Collisions}

Experimental results and theoretical predictions are available for HX($v$)+HX($0$) scattering at room temperature~\cite{Leone1982,Shin1983}. However, there is no experimental information or quantum mechanical calculations to determine the temperature dependence of the vibrational relaxation rate of highly vibrational states in halides. For HF and HCl, we assume that the temperature dependence of the self-quenching vibrational relaxation rate for $v\geq 2$ is the same as for the vibrational relaxation from the $v=1$ state, which is a very good approximation for small energy transfers~\cite{Shin1983}. HBr has been only studied up to $v=2$ and HI up to $v=1$~\cite{Leone1982}, thus in these cases, we assume that the vibrational relaxation rate as a function of $v$ follows the same trend observed in HF~\cite{Jurisch1981}. In the case of HSc, we estimate that at room temperature $k_{10}=340$~s$^{-1}$Torr$^{-1}$, as explained in the previous section.  Due to the similarities between HSc and HBr, we assume that the temperature- and $v$-dependence of the vibrational relaxation in HSc are equal to those in HBr. 

\section{Dark-Photon Mediator}
\label{subsec:Sens_Dark_Photon}
The spin-independent results presented in Sec.~\ref{sec:Projections} assume that the DM interacts with the two nuclei in the di-atomic molecule with a coupling that is proportional to the mass of the nucleus. Here we discuss the sensitivity to a model in which DM interacts with a dark photon mediator that kinetically mixes with the Standard Model photon~\cite{Holdom:1985ag,Galison:1983pa}.  We show the projected sensitivity in Fig.~\ref{fig:Sigmae_HiddenPhoton}, where the left panel assumes the dark photon mass is $m_{A'}=3m_\chi$ and the right panel assumes an ultra-light dark photon with $m_{A'} \ll \alpha_{\rm EM} m_e$. The solid and dashed-blue, purple, and green curves correspond to the choices of experimental parameters as in Fig.~\ref{fig:N_recoil_SensPlot}. The reach is plotted as a function of $\bar{\sigma}_e$, which relates to $\sigmap$ via 
\beq
\bar{\sigma}_e = \left(\frac{m_{A'}^2+q_0^2}{m_{A'}^2+\alpha_{\rm EM}^2 m_e^2} \frac{\mu_{\chi e}}{\mu_{\chi n}} \right)^2 \sigmap,
\label{eq:sigman_sigmae}
\eeq
where $\mu_{\chi e}$ is the reduced mass of the DM-electron system and $\sigmap$ is derived with $f_{P,{\rm SI}}^{(i)}=1$, $f_{N,{\rm SI}}^{(i)}=0$ and includes the effects of Thomas-Fermi screening in Eq.~\eqref{eq:Thomas_Fermi}.

Also shown in the figure are constraints on the dark-photon parameter space from various experiments. These include electron recoil results from direct detection experiments XENON10~\cite{Essig:2012yx} and DarkSide50~\cite{Agnes:2018oej}, collider and beam dump constraints from BaBar~\cite{Lees:2017lec,Essig:2013vha}, E137~\cite{Batell:2014mga}, and LSND~\cite{Batell:2009di,Izaguirre:2015yja}, constraints from an analysis of the cooling rate of supernova 1987A~\cite{Chang:2018rso}, constraints on measurements of the number of relativistic degrees of freedom ($N_{\rm eff}$)~\cite{Boehm:2013jpa,Nollett:2013pwa,Vogel:2013raa}, and constraints from conventional nuclear recoil constraints as in Fig.~\ref{fig:N_recoil_SensPlot} (converted to $\bar{\sigma}_e$). Finally, also shown as thick orange curves in the figures are values of $\bar{\sigma}_e$ for which the correct relic abundance of DM is produced in various models (see e.g.~\cite{Battaglieri:2017aum} for details). In the heavy dark photon panel, three curves are given: production via thermal freeze-out~\cite{Boehm:2003hm}, production via an initial asymmetry for Dirac Fermion DM candidate~\cite{Lin:2011gj,Essig:2015cda} and production of a Strongly Interacting Massive Particle (SIMP) via $3\rightarrow2$ DM annihilations while remaining in thermal contact with the visible sector via elastic scattering (this curve is denoted ELDER (SIMP))~\cite{Hochberg:2014dra,Hochberg:2014kqa,Kuflik:2015isi,Kuflik:2017iqs}. For the ultra-light dark photon shown in the right panel, the couplings are so small along the orange curve that the DM never thermalizes with the Standard Model sector, but the correct relic abundance can be obtained from freeze-in~\cite{Hall:2009bx,Essig:2011nj,Chu:2011be,Essig:2015cda,Dvorkin:2019zdi}.  These four curves represent specific motivated regions of DM parameter space. 

For the heavy dark photon, even a tank with volume $V=(20\text{cm})^3$ together with a small detector of area $A_{\rm det}=1\text{cm}^2$, can probe much of the freeze-out cross section. A larger tank with volume $V=(2\text{m})^3$ and detector area $A_{\rm det}=(10\text{cm})^2$ probes the entire allowed region of the freeze-out line. For the ultra-light dark photon, a large tank and large detector area can partially probe the freeze-in curve. For low DM masses, there is a significant loss of sensitivity compared to a naive translation of Fig.~\ref{fig:N_recoil_SensPlot} due to the Thomas-Fermi screening in Eq.~\eqref{eq:Thomas_Fermi}. A model without such screening, such as for a hidden scalar mediator, would result in a much stronger reach at these low DM masses.  

A DM particle that interacts with a dark photon can also scatter off the electric dipole moment of the molecule.  We leave to future work an investigation of the sensitivity to this scenario.  

\bibliography{DM}

\end{document}